\documentclass[UTF8,a4paper]{article}
\usepackage{amsmath}
\usepackage{graphicx}
\usepackage{subfigure}
\usepackage{epstopdf}
\usepackage{geometry}
\usepackage{ulem}
\usepackage{indentfirst}
\usepackage{amsfonts,amssymb,dsfont}
\usepackage{setspace}
\usepackage{threeparttable}
\usepackage{upgreek}

\usepackage[numbers,super,sort&compress]{natbib}
\setcitestyle{close={)}}
\begin{document}

\title{\bf Time Evolution and Thermodynamics for the Nonequilibrium System in Phase-Space}
\author{Chen-Huan Wu
\thanks{chenhuanwu1@gmail.com}
\\College of Physics and Electronic Engineering, Northwest Normal University, Lanzhou 730070, China}

\maketitle
\vspace{-30pt}
\rule{\textwidth}{0.3mm}
\begin{abstract}
\begin{large}

The integrable system is constrained strictly by the conservation law during the time evolution,
and the nearly integrable system or nonintegrable system is also constrained by the conserved parameters (like the constants of motion)
with corresponding generalized Gibbs ensemble (GGE)
which is indubitability a powerful tool in the prediction of thr relaxation dynamics.
For stochastic evolution dynamic with considerable noise, the obviously quantum or thermal correlations which don't exhibit the thermal behavior,
(like the density of kinks or transverse magnetization correlators),
display a asymptotic nonthermalization,
and in fact it's a asymptotic quasisteady state with a infinte temperature, therefore the required distance to the nonthermal steady state
is in a infinite time average.
In this paper,
we unambiguously investigate the relaxation of a nonequilibrium system in a canonical ensemble for integrable system or nonintegrable system, 
and the temporal behavior
of many-body quantum system and the macroscopic system, as well as the corresponding linear-coupling between harmonic oscillators.
Matrix-method in entropy ensemble is also utilized to discuss the boundary and the important diagonalization,
the approximation by the perturbation theory is also obtained.


\end {large}
\end{abstract}
\begin{large}
\section{Introduction}

The investigation of evolution of nonequilibrium system is important to the particle physics or condensed matter physics and even the cosmology 
(like the entropy of Bekenstein-Hawking black hole\cite{Eling C}),
especially in the many-body theroy prediction which by, e.g., the trapped ultracold atomic gases which have weak energy interaction with the environment
and therefore allow the observation of unitary time evolution\cite{Calabrese P2}.
For nonequilibrium system, the usualy form of glass can be blocked by the pinning field\cite{Monasson R}
and prodece a galss transition like the process of ergodic to non-ergodic,
In replica theory, since the homogeneous liquid given by replica symmetry have a inhibitory effect for entropy production, 
whereas the replica symmetry broken result in
the increase of overlap of replicas.
With the increse of degree of overlap which can be realized by enlarge the system size, 
the number of metastable states (or the hidden one) is grows exponentially, and furthermore, the entropy is grows logarithmic.

We already know that the observable chaotic classical system require the processing resource which increase exponentially with time and 
Kolmogorov entropy $h$\cite{92} due to it's
exponential sensitivity in initial state\cite{Prosen T},
while the integrable one, which is solvable by the Bethe ansatz\cite{Karrasch C}, is increse polynomially.
The time evolution of quantum entangled state may cause decoherence effect which is widely found in condensate system and 
it take a important role in quantum information processing, quantum computation and metrology, 
quantum teleporation, quantum key agreement\cite{Zhou N,Liu B}, 
and even the decoherence in neural network\cite{Tegmark M}.
The entanglement is mostly preduced by the dynamical evolution with nonlinear interaction\cite{Luo X Y} and the non-destructive measurement.
like the Dzyaloshinskii-Moriya interaction\cite{Sharma K K,Qiang Z},
and in extreme case, e.g., through the axion field\cite{Li R,Qi X L}.
Usually the quantum entanglement is studied by the two-qubit or qutrit\cite{Zhang C J,Sharma K K} system, 
in some case the tripartite system\cite{Horodecki M,Qiang Z} or even more one is consider.
In nonequilibrium and nonstationary open system, the coarse graining which connecting numerous
subsystems' degrees of freedom make more possible to realize this process\cite{Arimitsu T},
and the thermal entropy is a good measurement for the effect of coarse graining.
The quantum spaces' dimension increases exponentially with particle number due to the tensor-productor\cite{Prosen T}, similarly, 
the number of metastable which as the subsystems of the spin glasses system is increase expnonentially with size in high temperature\cite{Franz S},
phase transition and critical fluctuation occur when it from one kind of subsystem into another and the broken and restoration of symmetry is also 
affect the properties of materials\cite{Yaida S}, like the dielectric constant, etc. 

In solid-state quantum system, the spin is the best candidate among various microscopic atom intrinsic
degrees of freedom in thermal entanglement which has higher stability compared to other entanglements 
due to the spins' relatively long decoherence time\cite{Liu J} and it's in close connection with the local free erengy.
The long coherence time in many-body systems is useful to detecting the unitary dynamics, e.g., the Hubbard-typr model,
and it's important to the coherent nonequilibrium dynamics for the multiple phases transition.
Since the models that can be mapped to a spinless free fermions through Wigner-Jordan transformation and show a in-phase fermion liquid state\cite{Wang Y R}, 
have show a stationary behavior in such a equilibrium intergrability model which consider as a powerful tool to obtain the exactly solution of model\cite{Fagotti M}.
A numberical method as time-dependent density matrix renormalization group (t-DMRG) have show that the matrix produce operator 
$D(t)$ is simulation-inefficently for nonintegrable 
model which is similar to the tensor-productor, but it's efficent for integrable
and local disorded case\cite{?nidari? M}.
Except that, the method of matrix produce wave function is also a good tool to deal with this time-evolving one-dimension quantum system\cite{Vidal G}.
The time evolution on free fermions or bosons, when the time scale to infinity the thermal average of z-component spin $S^{z}$ is zero and the spin states
is half-filled\cite{Wang Y R}, in this case the interaction between particles is strongest due to the zero-polarization\cite{?nidari? M}, and the entanglement entropy 
is also increase and becomes more extensive\cite{Barthel T}.
The first implementation of using the density matrices in prediction of many-body system (equilibrium or nonequilibrium) is the Ref.\cite{Jaynes E T}.
It discuss the situation similar to the quantum irreversible process in a energy- and information-lossy system. 

A fact that the many-body quantum system will tend to equilibrium has been verified by many rencently experiments, 
like the trapped ultracold atoms in optical lattices
or the interactions with optical resonance. Whereas for the nonequilibrium system, the relaxtion and thermal entanglement and the stochastic force also
attract a lot of attention\cite{Cramer M,Behunin R O}. 
Furthermore, the system may relax to analogue of thermal state if the inital state is ground state\cite{Cramer M}.
The method of fluctuation-dissipation relation (FDR) and quantum state diffusion (QSD) 
is utilized for the evolution to steady states in integrable system whose final states are constrainted by the conserved law
(indeed, it's the scattering process of particles which constranted by conserved law)
and with a finite speed of algebraically relaxation and information transfer
under the thermodynamic limit (the large-N limit).
Note that the speed here will not bounded by the speed of light like the relativistic quauntum theory,
but bounded by a well known Lieb-Robinson group velocity\cite{Lieb E H}.
The integrable system of quantum Newton's cradle with groundbreaking is a example\cite{Kinoshita T}.
The classical system also have found the same reslut,
like the Fermi-Pasta-Ulam (FPU) theorem\cite{Berman G P} and Kolmogorov-Arnold-Moser (KAM) theorem\cite{Kollar M}.
While for some nonintegrable system, the constant of motion can be expressed by second quantized operator\cite{Manmana S R} (see below).

The collection variables are applied to investigate the evolution in studied system, except this, we also applied 
the method of density matrix and complex tensor grid
to make this paper self-contained.
For local observable system the stationary and linear value may exist (like thermal state), 
but for integrable system whose time evolution found no thermalization and it may tends to a distribution of GGE 
with a important fundamental hypothesis for statistical ensemble that has maximized entropy which is constrained by local conservation law\cite{Fagotti M2},
(e.g., the conservation quantity of momentum occupation number), hence restrict the ergodicity and can't reach the thermal state.
For a framework of macroscopic system in finite dimension is important to introduce the quantum field theory for both the equibrum and
nonequibrum state in open system\cite{Arimitsu T} to investigate its time-dependent nature and coupling (or interaction) in local and nonlocal case
as well as the dynamical fluctuation in short distance.
It's also necessary to consider a quantum field when the Hilbert space is too large to implement a well numberical simulation\cite{Steel M J}.
While the importance of entangled states for quantum computation is well understand, to 
reduce the confusion from decoherence, there is a topology way that storing the quantum information non-localized\cite{Storni M L} or through the
non-Abelian braiding statistics which support the Majorana fermions\cite{Alicea J,Chung S B} by Majorana modes in finite wire\cite{Diehl S},
and it can better solve the problem of infinety dilution of the 
stored information in local area\cite{Cramer M}.

Since for nearly integrable system, the behavior of relaxation is under the crossover effect of prethermalization and thermalization,
which is associate with the thermal correlation and the speed of information transfer,
and the prethermalized state can be well described by the GGE\cite{Marino J}, i.e., may be view as a integrable system.
Like the Ref.\cite{Kollath C} which also using the method of t-DMRG and show the nonthermalization in soft-bosons model, 
have perform the off-diagonal correlation in the two-dimension square model, and the relaxation with some fluctuation is presented in short time evolution.
The suppressed thermalization can be freed by enough perturbation to break the integrability.
This crossover effect affect both the nonintegrable system and open system. 
Through the study of this paper, we know that the recurrence will appear for large time evolution.
In the configuration which considered in this paper, part of mixed system which is of interest is coupling with the environment (not isolated),
and hence the degrees of freedom of environment system (i.e., the counterpart of the target one) can be traced out in the canonical ensemble\cite{Kollar M},
i.e., tracing over the variables outside the target region.
This provide the support on the matrix method in Sect.10.
A large number of degrees of freedom is also a important precondition to implement global relaxation with the thermodynamic limit\cite{Kollar M2}.
For nonlocal operators in equilibrium state, the dynamical parameters display a effective asymptotic thermal behavior (follow the Gibbs disturbution)\cite{Rossini D}
during equilibrium time evolution with determined temperature and decay with a asymptotic exponent law,
while the model what we focus on is towards the asymptotic quasisteady state with a infinte temperature,
which decay with a asymptotic power law\cite{Karrasch C} acted by a diffusion term (see Sect.11).
The prethermalization will shares the same properties of nonthermal steady state due to the dynamical parameters,
which makes the model after quench close to the integrable points (or superintegrable point).
But in fact, for integrable quenches, the stationary behavior for both the local and nonlocal observables can be well described by the corresponding GGE,
and the particles scattering which constranted by conserved law is purely diagonal\cite{Calabrese P3,De Grandi C2}.

This paper is organized as follows.
We introduce the model of two-coupled subsystem in Sect.2, and the bare coupling is further discussed in Appendix.A.
The evolutions in non-dissipation system is discussed in Sect.3,
and the quenching for many-body system is diacussed in Sect.4.
A system-environment partition is mentioned in these two section.
In Sect.5, we discuss the dissipation for nonlocal model.
In Sect.6, the time evolution and thermal entanglement of Heisenberg XXZ model is investigated.
In Sect.7, the correlation and transfer speed of information in quantum system is discuss where we take the one-dimension chain model
as the explicit example.
The relations between thermal behavior and the integrability is also discussed in this Section.
We discuss the nonequilibrium dynamics with strong and weak interaction in Hubbard model in Sect.8.
In this section, we investigate the phase transition of nonintegrable Hubbard model, and the relaxation of double occupation and the kinetic energy. 
We also use the method of nonequilibrium dynamical
mean-field theory (DMFT) to detect the evolution by mapping the lattice model to the self-consistent single-site problem
which can be solved numerically.
In Sect.9, we discuss the relaxation to a Gaussian state.
In Sect.10, we resort to the matrix method,
and the propertice of the boundary and the transfer speed are also discussed.
In Sect.11, we discuss the relaxation of nonequilibrium system with stochastic dynamical variables
in a free energy surface, the quantum dissipation in the damp-out process is also discussed.
The diagonal contribution to symplectic spectrum of covariance matrix is further detected in Appendix.B.
The bulk-edge-coupling type materials which is related to the spectrum gap is presented in Sect.10 and Appendix.C,
and the perturbation therory and diagonalized Hamiltonian is also discussed in Appendix.C.


\section{Model Introduction and the Coupling in Feild Theory}

We begin with the perturbation theory with space-time dimension, which is important to consider in the strong coupling case\cite{Kollar M}, 
weak-perturbation limit of nonintegrable
system, and even the breaking of ergodicity\cite{Monasson R}.
In dimension of $(d+1)$ in sapce-time, since the particles obtain mass from the broken of non-Abelian gauged symmetry, 
the couping constant $g$ is dimension-dependent,
except the bare couping $g^{b}$ which vanish in $d+1=4$ limit \cite{McKeon D G C}.
the broken translational symmetry also make the spin liquid state rapidly solidified and turn into the crystalline sturcture \cite{Yaida S,Elcoro L}.
Then we define two $d$-dimension system $\psi_{i}$ and $\psi_{j}$ with potential $\phi_{i}$ and $\phi_{j}$, respectively.
In weak couping condition which suitable for the perturbative calculation \cite{Yaida S}, there is exist a spin density wave (SDW) which in a Fourier expression
is $\psi_{i}=L^{-d}\sum_{i}e^{-iqr_{i}}\phi(x-r_{i})$, and $\psi_{j}$ is as the same form.
Although the $L$ here is constrained by the model dimension $d$, but $L$ itself could be dimensionless and with dimensionless length scale and time scale (see Ref.\cite{Drummond P D}).
The $\phi$ here describe the fluctuation as a function in arbitrary dimension,
and it's also useful for quantum fluctuation or even the vacuum fluctuation.
The dimension of $\phi$ may even up to ten according to D-branes of string theory \cite{Ho?ava P}.
In the space dimension of $d=3$, the kitaev model despict a triangular parameter space with different degrees of coupling in three direction x y and z,
and the small triangular area which connecting the three midpoints of three side is gapless phase region \cite{Baskaran G}. In this model, I set coupling in these
three direction in a range of 0 to n, for which the top value $n$ is $n=2^{d/2}N$ in $SO(d)\times SU(N)$ system \cite{Pelissetto A}.
So a continuous phase transition with weak coupling pertubative RG under the time evolution can be expressed by 
$S=\int d^{d}x \mathcal{L}$ which is a exponent appear in the imaginary-time path integral 
$Z=\int D\psi_{i}^{\dag}D\psi_{i}D\psi_{j}^{\dag}D\psi_{j}e^{-S}$ \cite{Machida M}. 

The nonrelativistic Lagrange $\mathcal{L}$ is\cite{McKeon D G C,Ho?ava P}

\begin{equation}   
\begin{aligned}
\mathcal{L}=\int^{\tau}_{\tau'} d\tau [(i\psi_{i}^{\dag}\partial_{\tau}\psi_{i}+\frac{1}{2\mu}\psi_{i}^{\dag}\nabla\psi_{i}-\mu \psi_{i}^{\dag}\psi_{i})
+(i\psi_{j}^{\dag}\partial_{t}\psi_{j}+\frac{1}{2\mu}\psi_{j}^{\dag}\nabla\psi_{j}-\eta \psi_{j}^{\dag}\psi_{j})],
\end{aligned}
\end{equation}
where $\tau$ and $\tau'$ is initial and final time, $\nabla$ is the Laplace operator, and $\eta$ is the chemical potential.
This time evolution Lagrange ignore the interactions, e.g., the inpurity induced long range order \cite{Ho?ava P,Matsumoto M}.
Since the half integer spin correponding to the gappless area which mentioned above, the fermion system in this area can be written as
$H=\sum_{a=x,y,z}J_{a}\sum_{\langle i,j \rangle_{a}}\psi_{i}\psi_{i}^{a}\psi_{j}\psi_{j}^{a}$, ($a=x,y,z$), with $\psi_{i}\psi_{i}^{a}=s_{i}/i$ and 
$\psi_{j}\psi_{j}^{a}=s_{j}/i$, then we have $H=-\sum_{a=x,y,z}J_{a}\sum_{\langle i,j \rangle_{a}}s_{i}s_{j}$.
In Eq.(1) we take the imaginary time approach which the quantum Monte Carlo (QMC) method is utilized \cite{De Grandi C}, 
the differential symbol $\partial_{\tau}$ has the below relation according to the definition of Bernoulli number\cite{Bender C M}

\begin{equation}   
\begin{aligned}
n\partial_{\tau}(\frac{\tau^{1-z}}{1-z})=\frac{\tau^{1-z}}{1-z}\sum^{\infty}_{n=0}B_{n}\frac{(-\partial_{\tau})^{n}}{n!},
\end{aligned}
\end{equation}
and the differential symbol for mass is as the same form
\begin{equation}   
\begin{aligned}
n\partial_{\mu}(\frac{\mu^{1-z}}{1-z})=\frac{\mu^{1-z}}{1-z}\sum^{\infty}_{n=0}B_{n}\frac{(-\partial_{\mu})^{n}}{n!}.
\end{aligned}
\end{equation}

The Gardner transition which the critical dimension $d_{c}=3$ is a important object in study of properties of amorphous solids \cite{Berthier L}.
In (3+1) space-time dimension using the renormalized coupling, since the bare coupling is absent in the dimension $d+1=4$, 
the resulting dimensionless bare action with unbroken Quantum electrodynamics (QED) symmetry is 
\begin{equation}   
\begin{aligned}
S=\int dx\left\{\frac{1}{2}\sum_{x,y=0}^{n}[(\partial _{\mu}\phi_{xy})^{2}+r\phi^{2}_{xy}]-\frac{1}{3!}(g_{i}^{b}\sum_{x,y=0}^{n}\phi_{xy}^{3}+g_{j}^{b}\sum_{x,y,z=0}^{n}\phi_{xy}\phi_{xz}\phi_{yz})\right\},
\end{aligned}
\end{equation}
and the action of Landau-Ginzburg-Wilson (LGW) Hamiltonian with N-component O(N) symmetry and noncollinear order is\cite{Kawamura H}
\begin{equation}   
\begin{aligned}
S=&\int d^{d}x \int_{\tau}^{\tau'} d\tau
\bigg\{
    \frac{1}{2}     \sum_{x,y=0}^{n}[(\partial _{\mu}\phi_{xy})^{2}+r\phi^{2}_{xy}]\\
  &+\frac{1}{4!}[g_{i}(\sum_{x,y=0}^{n}\phi_{xy}^{2})^{2}+g_{j}\sum_{x,y,z=0}^{n}\overline{[(\phi_{xy}\phi_{z})^{2}-\phi_{xy}^{2}\phi_{z}^{2}]}]
\bigg\} 
\end{aligned}
\end{equation}
The summation index $xyz$ range from zero to $n-1$ corresponding to the parameter space setted above, and the average term 
$\sum_{x,y,z=0}^{n}\overline{[(\phi_{xy}\phi_{z})^{2}-\phi_{xy}^{2}\phi_{z}^{2}]}$
exhibit the correlation between these two fluctuation functions.
Using the method of time dependent density matrix RG which have been proved valid for particles at a fixed evolution time \cite{?nidari? M},
this fermion system shown as $T_{ij}\delta_{ij}={\rm Tr}\{\sigma^{i}\sigma^{j}\}$ where $T_{ij}$ is the interaction tensor, the $\sigma^{i}$ and $\sigma^{j}$ are the 
matrices of $\psi_{i}$ and $\psi_{j}$ respectively and $\delta_{ij}=\{c_{i}c_{j}^{\dag}\}$.
This expression is indeed take the diagonal part of $T_{ij}$.
Ref.\cite{McKeon D G C} put forward a valuable view that connecting the bare coupling to the renormalized coupling
by a infinte cutoff, and then the mass-independent bare coupling can be shown as \cite{McKeon D G C}

\begin{equation}   
\begin{aligned}
g^{b}=\mu^{3-d}\bigg\{g&+\delta_{11}\frac{g^{3}}{3-d}\\
                       &+\delta_{21}\frac{g^{5}}{3-d}+\delta_{22}\frac{g^{5}}{(3-d)^{2}}\\
                       &+\delta_{31}\frac{g^{7}}{3-d}+\delta_{32}\frac{g^{7}}{(3-d)^{2}}+\delta_{33}\frac{g^{7}}{(3-d)^{3}}+O(g^{9})\bigg\},            
\end{aligned}
\end{equation}
which is satisfactory consistent with the series expansion of $\beta$ function given in Ref.\cite{Banks T}

\begin{equation}   
\begin{aligned}
\beta(g)=-\beta_{0}\frac{g^{3}}{16\pi^{2}}-\beta_{1}\frac{g^{5}}{(16\pi^{2})^{2}}-\beta_{2}\frac{g^{7}}{(16\pi^{2})^{3}}-O(g^{9}).
\end{aligned}
\end{equation}
This $\beta$-function is series expand to the seven-order of coupling, i.e., the three-loop level for the gauge field.
The specific quantitative analysis of $\beta$-function is presented in the Appendix A.
Fig.1 show the $\beta(g)$ as a function of $g$ in SU(3) system (i.e. $C^{(2)}_{ij}=3$ (see Appendix.A) with different number of fermion multiplets $m$,
We set the $m$ from 0 to 20. It's obvious to see that the curves shows a drastic non-linear change,
and the $m$-dependent interaction tensor $T_{ij}$ also played a decisive role
in the relation between $\beta(g)$ and $g$.

\section{Evolution Behavior in Non-Dissipation System}

Since the long time scales exist in the metastable states which the quantity grows exponentially with system size \cite{Yaida S}, 
e.g., the single positive charge state in p-type material\cite{Paudel T R} or the p-spin model\cite{Kirkpatrick T R}.
the imaginary-time path integral can be expressed by the trace of time evolution operator $Z={\rm Tr}(e^{-\beta H})$ with the evolution propagator 
$U=e^{-\beta H}={\rm Tr}(\sigma^{i}_{1}\sigma^{i}_{2}\cdot\cdot\cdot\sigma^{i}_{n}\sigma^{j}_{1}\sigma^{j}_{2}\cdot\cdot\cdot\sigma^{j}_{n})$, 
where $\beta$ is inverse temperature $1/T$ which we use the unit of Boltzmann constant $k=1$
Note that the spin pauli matrix here is contain all the component in finite dimension of Hilbert space and 
$H_{ij}$ is the nearest neighbor Hamiltonian and can be decomposed using the Trotter-Suzuki method 
which mapping the one dimension quantum system into two dimension\cite{Xiang T} and the path integral is becomes 
$Z={\rm Tr}(\Pi_{i,j}e^{-\beta H_{ij}})$,
In this way, the long range interaction can be treated locally as a nearest-neighbor pair in this spin isotropic system through a single two-qubit exchange gate 
$U_{i,i+1}=e^{-H_{i,i+1}\delta\tau}$ due to the iterative nature and acting on two adjacent site with single time step $\delta\tau$ evolution, 
it is also meets with the realignment criterion\cite{Zhang C J},
that is, the local field effect. Then we have

\begin{equation}   
\begin{aligned}
e^{-\beta H_{i,i+1}}=\prod_{i}U_{i,i+1}.
\end{aligned}
\end{equation}

Except the Andenson localization, the local length may strongly increase obey the logarithmic law\cite{Shepelyansky D L}.
The Hamiltonian here was divided by the partition function Z through the temperature interval or the external magnetic field $h$\cite{Pelissetto A}.
By investigate the asymptotic behavior of Z, when $\beta\rightarrow\infty$, i.e., the temperature decrease with the imagnary time evolution, 
the $Z\rightarrow 0$, and the system tends to the ground state which is 
$|\psi(0)\rangle=|\psi^{i}_{1}\rangle\otimes|\psi^{i}_{2}\rangle\cdot\cdot\cdot\otimes|\psi^{i}_{n}\rangle
                  \otimes|\psi^{j}_{1}\rangle\otimes|\psi^{j}_{2}\rangle\cdot\cdot\cdot\otimes|\psi^{j}_{n}\rangle$,
and denote the $\varepsilon_{n'}$ is the energy of the $n'$th level ($n'<n$) in this system above the ground state, $\varepsilon_{n'}=E_{n'}-E_{n'-1}$.
Then the pauli operator $\sigma_{n'}^{i/j}$ within the evolution propagator $U$ is $\sigma_{n'}^{i/j}=\sigma^{0\otimes n'}\otimes\sigma^{i/j}_{n'}\times\sigma^{0\otimes(n-n')}$\cite{Prosen T}
Before that happen, the entanglement between paiticles which depending on time rapidly reaches the maximum value, 
which make the method of time dependent density matrix RG invalid due to the too large growth speed of entanglement entropy. 
The evolution by the evolution propagator $U$ is 

\begin{equation}   
\begin{aligned}
|\psi(\beta)\rangle=U|\psi(0)\rangle,
\end{aligned}
\end{equation}
and specifically, in the form with imaginary-time analogue $e^{\tau H(\tau)}$ it has\cite{Vidal G,Jordan J}
\begin{equation}   
\begin{aligned}
|\psi(\tau)\rangle=\frac{e^{\tau H(\tau)}|\psi(0)\rangle}{||\ e^{\tau H(\tau)}|\psi(0)\rangle\ ||},
\end{aligned}
\end{equation}
where we define the  imaginary-time as $\tau=t+i0^{+}$,
while for the evolution Hamiltonian is $H(\tau)=e^{\alpha H}He^{-\alpha H}$ where $\alpha=\beta+i0^{+}$.
Since $\partial_{\beta}\psi(\beta)=H\psi(\beta)$,
we have $\beta \propto (\partial_{\tau})^{n}$, which is also shown in the Eq.(2).  
For thermal average of a imaginary-time-dependent quantity $\mathcal{F}$, its expectation value which describe the ensemble average can be written as 

\begin{equation}   
\begin{aligned}
\langle\mathcal{F}_{\tau}\rangle=\frac{\langle \psi(\tau)|\mathcal{F}|\psi(\tau\rangle)}{\langle \psi(\tau)|\psi(\tau\rangle)},
\end{aligned}
\end{equation}
where $\langle \psi(\tau)|\psi(\tau\rangle)$ is the partition function here,
and the accurate value of $\langle \psi(\tau)|\psi(\tau\rangle)$ and $\langle \psi(\tau)|\mathcal{F}|\psi(\tau\rangle)$ can be determined
by the method of tensor RG.
The cumulative effect is efficiently in this averaging process\cite{Cramer M} and often do a cumulant expansion at the expectation value for simplified result
whose truncation depends on the detail of dissipation\cite{Stockburger J T}.
Through this, a world-line tensor grid RG can be formed by taking coordinate as the horizontal axis, and the time (or temperature) as vertical axis,
i.e., form a tensor network.
The tensor network separated by the inverse temperature $\beta$ have the spacing $\zeta=\beta/M$ where $M$ is the total number of lattices in the network
(also called the Trotte number\cite{Xiang T}).
Such a method which utilize the evolution of time and phase also called Trotterization \cite{Babbush R}.
Through the theory of t-DMRG, 
the $\mathcal{F}$ can be treated as a matrix product operator which depends on the time-evolution, 
$\mathcal{F}_{\tau}=U(\tau)\mathcal{F}U^{\dag}(\tau)$, here $\mathcal{F}_{\tau}$ and $\mathcal{F}$ base on different basis.
With the nonequilibrium time evolution, the integrable system which has the important feature of localization will relax to the stationary state after quantum quenches, 
i.e., the suddenly
change of interaction strengh\cite{Cramer M},
and the 
density matrices which constraint by the expectation value will leads to a maximum entropy ensemble\cite{Barthel T}. 
Usually we model the integrable (superintegrable) model by choosing the special initial state,
typically, like the XY spin model,
and it can be affected deeply by the constants of motion in the integrable (superintegrable) points like reach the nonthermal steady state and so on.
The density matrices here is denpends exponentially on conserved quantity and the Hamiltonians which related to the initial state.
For the matrix-product operators which describe the quantum states,
the minimal rank $D$ is requied to the maximal one of the
the reduced density matrix of bipartition system\cite{Prosen T}
(i.e., bipartition of the target one and its enviroment)
and it needed to truncated 
by the method of singular values decomposing to keep the size of $D$ polynomial increase which is local and time-computable, 
and we keep only the largest singular values after the truncation, i.e., only keep the basis states\cite{Jiang H C}.
In fact, for dissipation system, the linear or nonlinear dissipation coupling accompanied by the phase noise\cite{Demir A}
(like the Wiener noise (see Sect.11)) as well as the white noise or colored one\cite{Jumarie G} also have inhibition on the exponential increase. 

In Schr\"{o}dinger picture, the observables of thermal states are achieved by carry the integrable system into the nonintegrable one
(by perturbations) and in the mean time the energy-level spacing disturbution is evolves from the Poisson distribution with diagonal matrices to the Gaussian one 
(i.e., the wigner-Dyson type one) with level repulsion and random symmetric matrices\cite{Hsu T C} 
(there are also symmetrically ordered operators in quantum dynamics
by Wigner representation\cite{Drummond P D}).
It's possible to back to the Possion distribution by applying a series of single gate which prevent the
exponential increase of rank $D$ but
introduces the norm error\cite{?nidari? M}
\begin{equation}   
\begin{aligned}
\eta=\sum_{i=0}^{n-1}(1-\sum_{j=0}^{D-1}\lambda_{j}^{2}(U_{i})),
\end{aligned}
\end{equation}
where $U_{i}$ is a single gate and $\lambda_{j}(U_{i})$ is the decreasing ordered singular values after removing the maximum one,
and the maximum entropy is accessible through the local relaxation and the same as the entanglement.
Although for nonintegrable system the growth of $D$ is founded to be exponential,
there exist methods like the diagonalization which keep the size of matrix always proportional to the time (or the system size),
like Bogoliubov rotation (see Appendix.C).
The procedure of eliminating the small singular values result in a low-rank matrix,
and this is also to keep the local free energy 
\begin{equation}   
\begin{aligned}
E_{{\rm free}}=-\frac{1}{\beta}{\rm ln}(\sum_{i}\lambda_{i}^{2})
\end{aligned}
\end{equation}
smallest ($\lambda_{i}$ is the singular values),
and also to enhance the equilibrium characteristics which treated as a thermodynamics anomaly in glass system\cite{Odagaki T}.
This equation also explicitly show the measurement of erengies in units of (inverse) temperature.
To solve the problem of density matrix in the t-DMRG,
one introduce a way to solve the rank minimization problem which make this method valid even for the low rank matrices (see Ref.\cite{Cai J F}),
and it's help to reducing the error and keep computational cost low at the same time.
On the other hand, that also provide the convenience that 
make the matrix nondecreasing and so that the maximum rank is always appear in the final step of the algorithm.

Since we have implement the system-environment partition, in a full quantum dynamics, we can 
yield a well approximation in the weak-couping regime by the low-order truncation, 
e.g., the wigner truncation approximation which truncate in the power of one-order\cite{Drummond P D}.
In such a phase space, the coupled two subsystem have the relation
$\sum_{k_{i},k_{j}}(-k_{i}!/A^{k_{i}})g^{k_{i}}(-k_{j}!/A^{k_{j}})g^{k_{j}}=\sum_{k}(-k!/A^{k})g^{2k}$\cite{Yaida S},
where $k$ is the number of powers of truncation in phase space (e.g., $k=1$ when truncate in the first-order) and $A$ is the angles which dominate the 
series expansion of the dimensionless coupling $g$ (see Sect.2).

From the discussion on this Setction, we can see that the imaginary-time propagation has the similar behavior with the real-time one,
it will provide us another way to detect the decaying progress including the die out of excitations, and it's available for similar real-time setups\cite{De Grandi C},
or application to the nonequilibrium problem with stochastic series expansion in integrable system without the constraint of local conservation law.
Therefore it's more feasible to detect the asymptotics phenomenon in time evolution, expecially for the low-order perturbation theroy with extended potential.

\section{Quenching in Many-Body Local System}
For integrable open system, we imagine the bipartion of the Hilbert space and into the two formulated finite-dimension linear space (two associated configuration)
$V_{i}$ and $V_{j}$ which assumed have same spectrum and their reduced density matrices are

\begin{equation}   
\begin{aligned}
\mathcal{J}_{i}=\sum^{R}_{R=1}\lambda_{R}|\psi_{R}^{i}\rangle\langle\psi_{R}^{i}|,
\end{aligned}
\end{equation}
\begin{equation}   
\begin{aligned}
\mathcal{J}_{j}=\sum^{R}_{R=1}\lambda_{R}|\psi_{R}^{j}\rangle\langle\psi_{R}^{j}|,
\end{aligned}
\end{equation}
where $\lambda_{R}$ is the Schmidt coefficients (the decresing singular values).
The bipartite state $|\psi\rangle\in\mathbb{C}^{d_{i}}\otimes\mathbb{C}^{d_{j}}$ which realized through the Schmidt decomposition via singular value decomposition, 
and the Schmidt rank is ${\rm min}[d_{i},d_{j}]$\cite{Cubitt T}. For inseparable case, the reduced density matrix $\mathcal{J}_{i}^{'}$ 
(if it's pure state density matrix with feature of unitarily invariant) can be obtained by
tracing over the the pure state in its extended subsystem (i.e., $\mathbb{C}^{d_{j}}$),
and the product space which form by two subsystem is $V_{i}\otimes V_{j}$.
This bipartition can be used in most of the quantum many-body model, like the Ising transverse field model, XXZ model, and kitaev model, ect.

Integrability is usually relies on the localization, especially the superintegrable one (like the XY spin model) which are fully relies on the localization\cite{Fagotti M2}.
For a concrete example, we consider a XY spin two-chain model without the magnetic field, which the bulk Hamiltonian is\cite{Bracken A J}
\begin{equation}   
\begin{aligned}
H_{i,i+1}=\sum_{i=0}^{N-1}\frac{1}{2}(\sigma_{i}^{x}\sigma_{i+1}^{x}+\sigma_{i}^{y}\sigma_{i+1}^{y})\cdot {\rm exp}[\frac{J}{4}\sum_{j=0}^{N-1}(\sigma^{2x}_{i+\Uptheta(j-i)}+\sigma^{2y}_{i+\Uptheta(j-i)})],
\end{aligned}
\end{equation}
where $J$ is the coupling, $i,j$ stands for the different chains, and $\Uptheta(j-i)$ is a step function.
The correlation in such a system is\cite{Franz S}
\begin{equation}   
\begin{aligned}
\langle s_{i},s_{j}\rangle=\frac{1}{N-1}\sum_{ij}s_{i}s_{j}=\frac{1}{2} q_{ij}(N-1),
\end{aligned}
\end{equation}
where $q_{ij}$ is the overlap between these two spin configuration.
The local quantum integrability in the bounded bulk model can be deriving by the explicit form of the quantum R-matrix as well as 
the boundary transfer matrices, e.g, see Ref.\cite{Zhou H Q,Bracken A J,Fagotti M2}.

For quench behavior due to the perturbation from local operators, which for the out-of-equilibrium protocol is striking,
the amplitude from initial state
to instantaneous $n$ state is\cite{Gritsev V}

\begin{equation}   
\begin{aligned}
A_{n}(t)=-\int_{t_{i}}^{t_{f}}dt\langle n|\partial_{t}|0\rangle {\rm exp}[i(\varphi_{n}(t)-\varphi_{0}(t))]\ \ (0\le i\le n),
\end{aligned}
\end{equation}
where $\varphi(t)$ is the dynamics phase.
Such a amplitude is also the eightvalue of density matrices in entropy ensemble with the specific heat $\sum_{n}[E_{n}-E_{0}|A_{n}(t)|^{2}$.
The sum of square of amplitudes is the excitation probability $P_{ex}=\sum_{n}|A_{n}(t)|^{2}$
for electrons, particles, or holes,
i.e., quenched away from initial (ground state) to new state.
Here we suppose the quench is very fast that the initial state $\psi_{0}$ and the quenched state $\psi_{n}$ are amlost exist at the same time $t_{i}$.
Then using the evolution propagator $U(t)$,
we obtain the amplitude\cite{Tong D}
\begin{equation}   
\begin{aligned}
\langle n|U(t)|0\rangle&=-i\langle n|\int^{t_{f}}_{t_{i}}dt H(t)|0\rangle\\
                          &=-i\langle i|H_{{\rm int}}|0\rangle\int^{t_{f}}_{t_{i}}dt'{\rm exp}[i(E_{n}-E_{0})t']\\
                          &=-\langle i|H_{{\rm int}}|0\rangle\frac{{\rm exp}[i(E_{n}-E_{0})t]-1}{E_{n}-E_{0}},
\end{aligned}
\end{equation}
where $E_{0}$ is the energy in the initial state $\psi_{0}$
Through the fermi golden rule, where $H_{{\rm int}}$ is the interaction Hamiltonian with scattering amplitude $A_{i}$, which is
\begin{equation}   
\begin{aligned}
H_{{\rm int}}=\frac{U(t)(E_{n}-E_{0})}{\sqrt{2-2{\rm cos}[(E_{n}-E_{0})t]}}.
\end{aligned}
\end{equation}
For further detect the perturbation from local operators, we present in
Fig.2 (a) the energy difference between the excited state and initial one 
with different staggered magnetic field $h_{s}$ in different dimension $D$ of a quantum lattice model, and
(b) the excitation probability as a function of the temperature, it's clearly that the probability distribution obey a Gaussian form.
Since the quantum noise comes from the random initial state, we define a Gaussian in the initial state white noise which have a zero mean and therefore the initial 
probability distribution is Gaussian.
Then the probability distribution in the process of relaxation is\cite{Marino J2}
\begin{equation}   
\begin{aligned}
P=\sum_{m,n}\delta[\Delta E-(E_{m}(t')-E_{n}(t_{i}))]|\langle\psi_{m}(t')|U(t)|\psi_{n}(t_{i})\rangle|^{2}|\langle\psi_{n}(t_{i})|\psi_{0}(t_{i})\rangle|,
\end{aligned}
\end{equation}
where $\delta$ is the amplitude of the Gaussian (see Sect.11) and $\Delta E$ is the energy-difference between the initial and final state of relaxation.
In phase space,
such a relaxation can be expressed by the density matrix 
\begin{equation}
\begin{aligned}
\mathcal{J}(t)=\sum_{{\bf k},{\bf k+q}}{\rm exp}[-\phi({\bf k})t]\mathcal{J}(0)
=\sum_{{\bf k},{\bf k+q}}{\rm exp}[-(E_{\bf k+q}-E_{\bf k})t]\mathcal{J}(0),
\end{aligned}
\end{equation}
where ${\bf k}$ and ${\bf k+q}$ are two spectral parameters. For slow quench which the time scale to infinity, 
the non-diagonal contribution to $\mathcal{J}(t)$ (i.e., the part of ${\bf q}\not= 0$) is 
vanish due to the fast oscillation of Fourier kernel ${\rm exp}[-(E_{{\bf k+q}}-E_{\bf k})t]$. 


In fact, the non-diagonal contribution to the mean-field-representation (or the second moments of the disturbution of momentum\cite{Cramer M})
$\langle c_{i}c_{i+1}^{\dag}\rangle=\int d^{n}k f(k){\rm cos}(\varphi(k)t)$ is 
asymptotically to a fixed value with the time evolution\cite{Kollath C}.
When
a external perturbing field is considered in the free energy landscape, 
a perturbing term should be added to the local free energy,
and since the perturbation is bad for the conservation of energy, the quantum system under the influence of noise variables
will not completely isolated even for the closed quantum system.
The coupling between this perturbing field and the Hamiltonian is beneficial to
enhance the system ergodicity by increase the coupling of metastates.
For closed system which have total energy conservation,
the ergodicity for observables under the long-time limit can be large enough to expect the time average to the thermal average\cite{Manmana S R},
but there are restriction on the observables like the bound of the von Neumann entropy, and hence prevent it closing the thermal state.
(Note that here the correlation between each distinguishable particle and the environment is still localized.)
The entropy of pinning field is increase with the overlap in a metastate, can associate with the hidden glass states, and it's confirmed
equal to the mean field potential of glass system\cite{Kagan D M}.
Both the entropy $S_{{\rm hidden}}$ (not the diagonal one) and 
its free energy as well as the non-diagonal contribution vanish in the final of the process of relaxation to steady equilibrium state, e.g., the 
commensurate superfluid state.

Since for the integrable system, most solvable Hamiltonian can be mapped to the effective noninteracting Hamiltonian\cite{Kollar M}
\begin{equation}   
\begin{aligned}
H_{{\rm eff}}=\sum^{N-1}_{i}\epsilon_{i}P_{i}
\end{aligned}
\end{equation}
with the eigenenergy $\epsilon_{i}$ and conserved quantity $P_{i}$,
and the maximum entropy ensemble after quench with local conserve-law can be written using the density matrix as

\begin{equation}   
\begin{aligned}
\mathcal{J}_{{\rm quenched}}=\frac{1}{Z}{\rm exp}(-\sum_{i}P_{i}Y_{i}),
\end{aligned}
\end{equation}
where the conserved observable quantity $P_{i}$ has the form $P_{i}=a^{\dag}_{i}a_{i}$ where $a_{i}$ is the annihilation operator of bosons or fermions
and has commute relation $[H,P_{i}]=[P_{i},P'_{i}]=0$, the $Y_{i}$ is a initial state-dependent quantity.
The partition function $Z={\rm Tr}[({\rm exp}(-\sum_{i}P_{i}Y_{i}))]$. 
This is in fact only a local steady state but not canonical steady states for the full system\cite{Barthel T}.
For integrable system begin with the maximal entropy in GGE, 
the $Y_{i}$ here can be replaced by a Lagrange multiplier set $\{\lambda_{i}\}$\cite{Fagotti M2,Fagotti M3,Kollar M,Rigol M},
(which is\cite{Cassidy A C} $\lambda_{i}={\rm ln}[(1-\langle \psi(0)|P_{i}|\psi(0)\rangle)/\langle \psi(0)|P_{i}|\psi(0)\rangle]$
and constrained by $\langle n\rangle_{GGE}=\langle\psi(0)|c^{\dag}c|\psi(0)\rangle={\rm Tr}(\rho n)$ where $n$ is the conserved number of particles).
For integrable systems which are exact solvable (i.e., all the eigenvalues and eigenfunctions can be obtained),
since the $\epsilon_{i}$ is linear eigenenergy, for a simplest conserved quantity, the number of particles $n_{i}$,
the number eigenstate can be treated as the energy eigenstate $E=\sum_{i}\epsilon_{i}n_{i}$ which on the eightbasis of $\{n_{i}\}$\cite{Drummond P D}.

Within the scheme of adiabatic perturbatic $\langle{\bf k}\cdot{\bf p}\rangle$ theory, the asymptotic behavior be manipulated by the velocity and acceleration of tuning parameter
in quench dynamic\cite{Gritsev V}. The tuning-dependent Hamiltonian $\psi(\lambda(t))$ ($\lambda(t)$ is the time-dependent tuning parameter) 
can also take effect in the adiabatic excitation of system in ground state which is
similar to $\psi(t)$, and recover due to the asymptotic effect of time evolution\cite{De Grandi C}.
The asymptotic freedom of system will preserved until the number of fermion species is too large\cite{Banks T}, so this asymptotic state with the 
scaling theory is depend only on the configuration, e.g., the fluctuation of system \cite{De Grandi C,Tissier M,Franz S}, and scales show a collection of the effects
from fluctuation and tend to Gibbs value when the momentum vector ${\bf q}\rightarrow 0$\cite{Tissier M}. One of the reflection is the equilibrium  Gibbs free energy
as below\cite{Monasson R} (without restrictions)

\begin{equation}   
\begin{aligned}
E_{{\rm Gibbs}}=-\frac{1}{\beta}{\rm ln}\int dt e^{-\beta H(t)},
\end{aligned}
\end{equation}
and since the Hamiltonian here is often the potential field-characterized, the free energy also treated as a potential function
with determined weigh (probability distribution).

\section{Dissipation in Nonlocal Model}
For nonlocal model, there is a large different compare to the local one. 
The nonequilibrium long-range force is also usually unobservable in localiaed interaction models\cite{Dmitra?inovi? V}.
Consider the Yang-Mills theory, the action of field can be expressed as 

\begin{equation}   
\begin{aligned}
S=\frac{1}{4}\int d^{d}x\int^{t'}_{t}dt F^{\mu\nu}_{i}F_{\mu\nu}^{i},
\end{aligned}
\end{equation}
where $F^{\mu\nu}_{i}$ is the field strength tensor (see Appendix.A),
$F^{\mu\nu}_{i}=\partial_{\mu}A_{\nu}^{i}-\partial_{\nu}A_{\mu}^{\nu}-gC_{iab}A_{\mu}^{a}A_{\nu}^{b}$\cite{Gross D J}, where $A_{\mu}^{a}$ and $A_{\nu}^{b}$ 
are the vector potential of the field and here $C_{iab}$ is for intruduce a SU(3) structure factor which
is $C_{iab}=\gamma^{iab}F^{a}F^{b}$, where $\gamma^{iab}$ is the SU(3) structure constant and $F^{a}$ is the group generator. 
The relation between the Lie group structure constant $C$ and quadratic Casimir operator is $\sum_{ab}C_{iab}C_{jab}=C^{(2)}_{ij}\delta_{ij}$\cite{35}.

The dissipative effect which derived from the macroscopic entangled system give arise the reservior problem 
and accompanied by a process of coarse-graining by the isometry
that integrating the degrees of freedom of subsystems\cite{Behunin R O} and with a dimension smaller than the maximum dimension of Hilbert space\cite{Vidal G2}.
The nonlocal correlation between the nearest neighbors can be treated locally by using the matrix product operator with determined rank
and the unitary transformation with time-evolution oparator (see below).
For the localized interaction with nearest neighbor spin accompanied by the local field effect, since the relatively large
coupling constant and long time configuration, it's priority to use the nonperturbative method\cite{Stockburger J T},
but the quantum dissipation which is nolinear is more acceptable to use the perturbative RG, and the reservoir interaction is also perturbed,
The Gaussian probability disturbution which exist in the linear case is not exist in the nolinear case anymore,
and the dimension of density matrices is also grows non-linearly with time\cite{Prosen T}.
But there still exist some linear relation, e.g., the 
entanglement entropy is change linearly with the evolution of time with a straggered magnetic field in the disorded case \cite{?nidari? M}.

In a open quantum system, the thermal average of observable $\mathcal{F}$ can be written as (here $\tau$ is the complex-time for propagators) 

\begin{equation}   
\begin{aligned}
\langle\mathcal{F}\rangle_{\tau}=\frac{{\rm Tr}(e^{-\beta H}e^{-H\tau} \mathcal{F}e^{H\tau})}{{\rm Tr}(e^{-\beta H})}.
\end{aligned}
\end{equation}
For integrable system, this equation which describe the thermal average in Gibbs ensemble\cite{Fagotti M} is equal to the energy of inital state of relaxation process after quench 
which evolution with time $\tau$. Such thermal average is also meaningful in thermodynamics description for quasi-equilibrium state\cite{Odagaki T}.
Base on the Eq.(8) and using the second order Trotte-Suzuki formation, the evolution propagator can be decomposed as 
$e^{-\beta H}=e^{-\beta H_{x}}e^{-\beta H_{y}}e^{-\beta H_{z}}+O(\tau^{2})$\cite{Jiang H C}, and the Eq.(8) can be rewritten as 

\begin{equation}   
\begin{aligned}
e^{-\beta H_{ij}}=\prod_{i=0,j=0}^{n}U_{i,i+a;j,j+a}\ \ \ (a=x,y,z).
\end{aligned}
\end{equation}
To study the dissipation of the remaining degrees of freedom in subsystems after coarse granulation in such a no-spacing-interaction macroscopic model, 
the reservoir is very important. To introducing FDR to the steady state, we rewrite the Eq.(27) by the method of path integral as
\begin{equation}   
\begin{aligned}
\langle\mathcal{F}_{\tau}\rangle=\int D\psi(\tau) e^{\tau H}
\frac{\langle \psi(0+\varepsilon^{+})|\mathcal{F}(0)|\psi(0+\varepsilon^{-})\rangle}{\langle \psi(\tau+\varepsilon^{+})|\mathcal{F}(\tau)|\psi(\tau+\varepsilon^{-})\rangle},
\end{aligned}
\end{equation}
with $\varepsilon\rightarrow 0$, and $\psi(\tau)=U\psi(0)$ where $U$ the time-evolution oparator $U=T_{\tau}{\rm exp}(-\int^{\tau'}_{\tau}d\tau H(\tau))$.
For statistical linear dissipation system, the correlation between reservoirs $\langle R_{i}R_{j}\rangle\not= 0$,
the method of unperturbed linear dissipation is also suitable for perturbed macroscopic model if the 
perturbation Hamiltonianis linear with reservoir $H_{p}=\sum_{i} f(i)R_{i}$ where $f(i)$ is a linear term 
and therefore the collective response to perturbation is mostly linear\cite{Stockburger J T}.
This form of $H_{p}$ is suitable for all the integrable or nonintegrable linear dissipation model.
While for the non-linear dissipation case, since the reserviors in different subsystems is independent with each other, so we constraint the reservoir states in
the Liouville spaces, and have $\langle R(0)|H_{SR}|R(\tau)\rangle=0$\cite{Arimitsu T}, where $H_{SR}$ is the interaction term between system and reservoirs and 
there exist shared influence function for all constituent\cite{Behunin R O}. 

For non-dissipation system, the propagation along time scale can be expressed by the initial Hamiltonian and the observable conserved quantity 
(i.e., Eq.(24)),
whereas for linear dissipation, since it need a stochastic term to compensate the lost energy,
and it has a history-independent potential term $\partial_{\tau} \psi(\tau)=H_{0}(\tau)\psi(\tau)-\sum_{i}f(i)q(i)\psi(\tau)$,
where $q(i)$ is the stochastic force or the noise. 
For nonlinear-dissipation system, the state of reservoir variables is span only in the Liouville space\cite{Arimitsu T}.
Both the linear-dissipation and nonlinear-dissipation contain a friction force term but the nonlinear-dissipation have a complex memory which it's
obvious from the feature of history-dependent\cite{Laio A} in evolution while the linear one haven't.

\section{Time Evolution and Thermal Entanglement in Integrable Heisenberg XXZ Model}

We already know that for non-dissipation system the antiferromagnetic Ising chain\cite{Prosen T}, XY spin chain\cite{Bracken A J,Prosen T} and the bulk model\cite{Bracken A J}
is integrable and can be exactly solved.
The Heisenberg XXZ model is also suggested integrable and own the local conserved quantity, e.g., the observable microscopic quantity like the $S^{z}$
or the observable macroscopic one like energy or number of particles.
To investigate the imaginary-time evolution in Heisenberg XXZ model, 
we firstly need to use a c-number representation which depict a shift of $-i\hbar \alpha$ in the axis of ${\rm Im}\tau$
(see, e.g., Ref.\cite{Stockburger J T}). 
Then we introduce the Heisenberg XXZ model with spin 1/2 antiferromagnetic free fermions interaction,
the n-component anisotropy Heisenberg Hamiltonian of this system contain a homogeneous external field $h$ is 

\begin{equation}   
\begin{aligned}
H=\sum^{n-2}_{i=0}(JS_{i}^{x}S_{i+1}^{x}+JS_{i}^{y}S_{i+1}^{y}+J_{z}S_{i}^{z}S_{i+1}^{z})
  +\sum^{n-1}_{i=0}(h_{i}S_{i}^{z}),
\end{aligned}
\end{equation}
where $J$ and $J_{z}$ are the coupling, and $S^{\alpha}_{i}=\frac{1}{2}\sum_{i}\sigma^{\alpha}_{i}\ (\alpha=x,y,z)$ is the total spin in the $\alpha$-component.
The important coupling ratio can be defined as

\begin{equation}  
\frac{J_{z}}{J}=
\left\{
\begin{aligned}
{\rm cos}\ \gamma,& \ \ J_{z}\le J\\,
{\rm cosh}\ \mu,&   \ \ J_{z}> J,
\end{aligned}
\right.
\end{equation} 
where the tilted angle $\gamma$ and $\mu$ is enlarge with the increase of degrees of anisotropy.
We focus on the $J_{z}/J={\rm cos}\ \gamma$ case.
In the case of $J_{z}=0$, 
i.e., becomes the noninteracting spinless fermion system with strongly correlated electronic characteristics 
under the Wigner-Jordan (WJ) transformation 
which turns the regular integrable system terms into the chaotic one\cite{Prosen T}.
In this case, the fermion representation of the gapless bilinear fermionic system is

\begin{equation}   
\begin{aligned}
H_{{\rm bf}}=\sum_{i}(c_{i}c_{i+1}^{\dag}+c_{i}^{\dag}c_{i+1}+h_{i}n_{i}),
\end{aligned}
\end{equation}
with $\Delta_{i}=\langle c_{i}c_{i+1}^{\dag}\rangle$ stands for a mean-field and also represent the covalent bonding of WJ fermions\cite{Wang Y R},
and this is also the tight-bingling fermionic model with dispersion relation $\kappa=\pm 2{\rm cos}\ k$\cite{Barthel T} in $\pi$-phase
(the phase difference between neighbor site is $\pi$).
In this case, this Heisenberg Hamiltonian becomes a strongly correlated electronic system with a finite entropy (will saturation)\cite{Bracken A J,?nidari? M}.
The operator of number of the spinless particles is $n_{i}=c^{\dag}_{i}c_{i}$, and the electron correlation is $J_{z}n_{i}n_{i+1}$.
To investigate the nonlinear-dissipation in this spinless fermions chain model, 
we need to introduce the master equation with system density matrix $\mathcal{J}$
\cite{Steel M J},

\begin{equation}   
\begin{aligned}
\partial_{t}\mathcal{J}=-i[H,\mathcal{J}]+\mathcal{K}\sum_{i}[O_{i}\mathcal{J}O_{i}^{\dag}-\frac{1}{2}(O_{i}^{\dag}O_{i}\mathcal{J}+\mathcal{J}O_{i}^{\dag}O_{i})]\equiv\mathcal{L}\mathcal{J},
\end{aligned}
\end{equation}
where $\mathcal{J}$ corresponds to the pure state or mixed state and $O_{i}$ is the Lindblad operator describing the bath coupling. 
The right-hand side of this equation contain two terms,
the first one is the unitary part of the Liouvillean, while the second one is the disspative term and 
$\mathcal{K}$ is the coupling strengths within the dissipation scenario.
We consider the damping here due to the nonlinear-dissipation. The Gaussian area arrived in time evolution have $\partial_{t}\mathcal{J}=0$,
in this case $\mathcal{K}$ is almost vanish and produce a zero dissipative area, that suggest that the observables
exponential fast approach to the steady state\cite{Diehl S},
while the entries of the density matrix is close to the main diagonal.

To introduce the thermal entanglement in the evolution, 
we define the generate and annihilate operator for $i$ sites as
\begin{equation}   
\begin{aligned}
c^{\dag}_{i}=e^{i\varphi_{i}}S^{+}_{i}\ ,\ c_{i}=e^{-i\varphi_{i}}S^{-}_{i}.
\end{aligned}
\end{equation}
The operators obey commutation relation $[c_{i},c_{j}^{\dag}]_{\alpha}=\delta_{ij}$ (boson operator and fermion operator for $\alpha=1$ and $-1$, respectively),
and $c_{i}^{\dag}c_{j}+c_{j}c_{i}^{\dag}=\delta_{ij}$ ($\alpha=-1$) under the WJ transformation that treat $c_{i}$ as operator field\cite{t Hooft G}.
The time-involve phase $\varphi_{i}$ have

\begin{equation}   
\begin{aligned}
\varphi_{i+1}-\varphi_{i}=c n_{i},
\end{aligned}
\end{equation}
where $c$ is a c-number-correlated factor which defined as the imaginary part of ${\rm In}(\tau'-\tau)$, i.e., the scale of imaginary-time, and the phase function
$\varphi_{i}=\sum_{i}c^{\dag}_{i}c_{i}c$.
Then the
Hamiltonian (Eq.(30)) can be represented as 

\begin{equation}
H=\begin{bmatrix}
J_{z}/2+h & 0     & 0     & 0   \\
0       &-J_{z}/2 & J     & 0   \\
0       &J      &-J_{z}/2 & 0   \\
0       &0      &0      & J_{z}/2-h\\
\end{bmatrix},
\end{equation}
when $|J|<h-J_{z}$, the ground state is disentangled state $|0,0\rangle$ which have the eightvalue $J_{z}/2-h$; when $|J|>h-J_{z}$, 
the ground state is entangled state $\frac{1}{\sqrt{2}}(|0,1\rangle-|1,0\rangle)$ for $J>0$ or $\frac{1}{\sqrt{2}}(|0,1\rangle+|1,0\rangle)$ for $J<0$ 
which have the eightvalue $-J_{z}/2-|J|$, and this entangled state will goes to maximal with time-evolution.
Thus, the entanglement increase with the enhancement of coupling $J$ and $J_{z}$
no matter they are both greater than zero (ferromagnetic) or both less than zero (antiferromagnetic),
but it's always symmetry compare to the case of inhomogeneous magnetic field.
We can obtanin the relaxations in long-time scale after the sudden quench of $J$ and $J_{z}$,
and regulate the entanglement by the quench of magnetic field $h$.
In equilibrium case, the density matrix of this thermal state can be written as\cite{Zhang G F}

\begin{equation}
\mathcal{J}=\frac{1}{Z}{\rm exp}(-\beta H)=\frac{1}{Z}\begin{bmatrix}
e^{-(J_{z}/2+h)/T} & 0                          & 0    &0\\
0                 &e^{J_{z}/2T}{\rm cosh}(|J|/T)     & -s   &0\\
0                 &-s      &e^{J_{z}/2T}{\rm cosh}(|J|/T)   &0\\
0                 &0       &0                         &e^{-(J_{z}/2-h)/T}\\
\end{bmatrix},
\end{equation}
where $Z=e^{-(J_{z}/2+h)/T}(1+e^{2h/T})+2e^{(J_{z}+h)/T}{\rm cosh}(|J|/T)$ and $s=Je^{J_{z}/2T}{\rm sinh}(|J|/T)/|J|$.
Usually, we can creating strong entanglement by raising the ratio of $J_{z}/J$, or raising the degree of inhomogeneity of magnetic field $h$, or 
properly lower the temperature through the previous study\cite{Zhang G F,Liu J,Feiguin A E}.
Sometimes the lower temperature which can be implemented by increase the system size\cite{Feiguin A E} can
decrease the eigenvalue of density matrix (Eq.(37)).

\section{Correlation and Transfer Speed in One-dimension Chain Model}
In this section, we focus on the two-point spin correlation in $S=1/2$ Heisenberg chain and $S=1$ Ising chain,
and define that the $J_{1}$ ad $J_{2}$ as the nearest neighbor coupling and next-nearest neighbor coupling in the chain, respectively.
The $\beta$(inverse temperature)-dependent magnetic susceptibility can be written as $\chi(\beta,t,i)=\beta 2^{-n}\sum^{n-1}_{i=1}\langle S^{z}_{0}S^{z}_{i}\rangle$
for a n-qubit chain,
the latter term in this expression is the spin-spin correlation function for the Heisenberg model\cite{?nidari? M}.
The Fig.3 shows the spin correlation $C$ and inverse correlation length $\xi^{-1}$ for (a) $S=1$ Ising spin chain and (b) $S=1/2$ Heisenberg chain
with different $J_{2}$ at different site $i$.
We show that the nonlocal order parameter decay exponentially due to the perturbations from long-range spin-spin interaction which breakig the integrability
and therefore exhibit a effectively asymptotic thermal behavior, though the latter one is
exactly solvable (i.e., all eigenvalues can be obtained by the method of Bethe ansatz in thermodynamic Bethe ansatz (TBA)\cite{Klumper A}) beforce the perturbation.
Such an exponentially decay for the nonlocal operators in nonintegrable model has been widely observed, 
e.g., the order parameter in transverse field Ising chain for 
ferromagnetic state or paramagnetic state\cite{Calabrese P} or the number of quasiparticles in the time evolution in a quantum spin chain\cite{Marcuzzi M}, etc.
We also can see that the $\xi^{-1}$ is tends to saturated with the increase of distance which is obey the equilibrium law,
and in fact it's equivalent to the coherent state with coherent amplitude in terms of a exponential form,
and therefore the phase coherence rate will display a similar behavior with the correlation length.
Fig.4 shows the spin correlation for $S=1/2$ Heisenberg chain as a function of temperature with different $J_{2}$.
We can see that, with the increase of $J_{2}$, the spin correlation is increase.
We also make the comparision for the spin correlation $C$ at different temperature for $S=1$ Ising chain and $S=1/2$ Heisenberg chain in the Fig.5.
It's obviously that the $S=1/2$ Heisenberg chain is earlier becoming saturated compare to the Ising one.
Furthermore,
we present the correlation (a) for $S=1/2$ Heisenberg chain which is obtained by the by the method of Bethe ansantz 
and make a comparison on the results of correlation in low-temperature for $S=1/2$ Heisenberg chain 
between the methods of Bethe ansatz and renormalization group (b) in Fig.6.

Since the equal time spin correlation $C$ have the relation
\begin{equation}   
\begin{aligned}
C(r,t)=\langle S(0,t)S(r,t)\rangle\propto{\rm exp}(-r/\xi),
\end{aligned}
\end{equation}
which is consistent with the expression of correlation length $\xi^{-1}=-\lim_{L\rightarrow \infty}{\rm ln}\langle S_{i}S_{i+L}\rangle$
in Ref.\cite{Xiang T}, here the distance $r$ can be quantified as $i$ which stands for number of
position in spin chains and $\xi$ is the correlation length.
Note that this expression for equal time two-point correlation is well conform the ordered phase in the long-time limit,
while for disordered phase the $\xi$ has more complicated form\cite{Calabrese P}.
Now that this spin correlation function display a effective asymptotic thermal behavior as introduced in Sect.1,
and correlation length $\xi$ is related to the quantum quench protocol\cite{Marino J},
the thermal behavior for a nondissipation system after quench can also have the relation which mentioned above (Eq.(38)),
but note that although this spin correlation here is in a exponential form,
the correlation length is not follow the thermal distribution but a nonthermal distribution\cite{Marino J} and guided by GGE.
This is because the correlation length is local quantity which behave nonthermally.
Similar behavior appear in the correlators like the transverse magnetization and so on.
We still need to note that though for infinite system which follow the effective thermal distribution is mostly nonintegrable,
but the initial state of integrable system which dictated by the noninteracting Hamiltonian may still follow the thermal disturbution\cite{Cazalilla M A}
since without the affect of interactional quench Hamiltonian.
Further,
if we mapping to the Fourier space, the equal time correlation (Eq.(38)) 
for the spin-1/2 square lattice model has a more specific form\cite{Ghaemi P}
\begin{equation}   
\begin{aligned}
\langle S_{i}(0)S_{j}(r)\rangle e^{ikr}\propto\frac{e^{-r/\xi}}{r^{4}}(1+\frac{r}{\xi})\delta_{ij},
\end{aligned}
\end{equation}
which follow the power law decay when $r\ll\xi$ and exponentially decay when $r\gg\xi$.

Since the pinning field play a important role in the process of ergodic to non-ergodic transition which plug the correlation between subsystems and even the velocity
of spin wave $v_{s}$\cite{Eggert S}, which associate with the slope of the dispersion relations in momentum space.
For the case of $J_{z}/J={\rm cos}\ \gamma$, $v_{s}$ can be written as\cite{Eggert S}

\begin{equation}   
\begin{aligned}
v_{s}=\frac{J\pi}{2}\frac{{\rm sin}\ \gamma}{\gamma}
\end{aligned}
\end{equation}
which is consistent with slope of dispersion relation $\partial_{k}\kappa=\mp{\rm sin}\ k$.
Then a question is arisen that if the speed of information transfer which govern the relaxation time of a post-quench state relate to the speed of spin wave
in a spin sysytem?
The answer is yes.
A direct evidence is the Lieb-Robinson type boundary (the details in a Bose-Hubbard model is presented in the next section). 
In fact the spin wave is also related to the 
momentum transfer\cite{Plihal M}
and even the damping of oscillation of superfluid regime (see Sect.8 and Ref.\cite{Halperin B I}).
We know that the missing of symmetry is related to the influence of initial states, 
and the collapse of physical phenomenas like the interference pattern\cite{Kollath C} or the collective excitation\cite{Li G Q,Jin D S} by inhomogeneous oscillation
in condensate with a density wave order
which act like a single phase wave or standing wave\cite{Chiquillo E}, 
is revives in the latter time of relaxation.
the transfer of correlation with a finite velocity also construct a line-cone which well describe the relaxation behavior.

\section{Double Occupation and The Interaction Quench in Nonintegrable Hubbard Model Near The Phase Transition Point}
Since the time evolution operator is dependents on the Hamiltonian (like Eq.(8)),
we next construct the Bose-Hubbard lattice model 
as a explicit example

\begin{equation}   
\begin{aligned}
H=-\mathcal{P}\sum^{n-2}_{i=0}(b_{i}^{\dag}b_{i+1}+H.c.)+U\sum_{i=0}^{n-1}\frac{n_{i}(n_{i}-1)}{2}-\mu_{i}\sum_{i=0}^{n-1}n_{i}
\end{aligned}
\end{equation}
where $\mathcal{P}$ is the hopping constant, $U$ is the chemical potential and $\mu_{i}$ is the local potential of each particles.
The interaction between the next-nearest nerghbor is assumed zero in this model,
and so that this model is integrable,
i.e., the second term of above equation can be replaced as $U\sum_{i=0}^{N-1}n_{i}n_{i+1}$.
A dimensionless reduced coulping
is defined as
\begin{equation}   
\begin{aligned}
g_{{\rm red}}=\frac{UN}{\mathcal{P}}
\end{aligned}
\end{equation}
where $N$ is the number of interactional particles.
We can implement the phase transition from Mott-insulator to the condensed state or the superfluid by modulating $g_{{\rm red}}$,
and it has been implemented experimentally\cite{Greiner M,Greiner M2,Cramer M2}.
Even for systems which without hopping at all (i.e., $\mathcal{P}=0$), 
the phase transition of metallic state and the Mott insulator are also realizable by the interaction quench
of $U$, and in this case the osillations with the collapse-and revival are periodic with period $2\pi U/\hbar$ \cite{Kollar M2}
(The Table.A shows the time scale of relaxation and the period of collapse and revival for several models).
In fact, most mang-body system can exhibit different quantum phase with different entanglement structure in the complex mixed dynamical,
and it's usually realizable by tuning the strenght of this competing interaction\cite{Luo X Y}.
The fluctuation of correlation amplitude due to the fast oscillation of phase factor are related to the distribution of initial state,
and the short-range correlation also shows distinguishable differences for different configuration of initial states.

In this model we next define the hopping-determined operator $\mathcal{R}:=it\mathcal{P}$,
this periodic-time-dependent evolution operator for a single-site can be expanded as\cite{Cramer M}

\begin{equation}   
\begin{aligned}
e^{\mathcal{R}}:=e^{it\mathcal{P}}=\sum_{k\ge d_{r}}\frac{(it\mathcal{P})^{k}}{k!}\le\sum_{k\ge d_{r}}\frac{(6\mathcal{P}t)^{k}}{k^{k}}
\end{aligned}
\end{equation}
where $k$ denotes the unit vector in phase space and $d_{r}$ is the distance between site $i$ and $i+r$.
There exist a upper bound for $d_{r}$ as $d_{r}<6\mathcal{P}t/e$ where $e$ is the natural constant,
since it's a insurmountable maximum speed 
for information ransfer in this model.
The summation of all the other places which beyond the distance $d_{r}$ have the above relation.
Thus we also have

\begin{equation}   
\begin{aligned}
e^{\mathcal{R}}\le \frac{(6\mathcal{P}t)^{d}}{d^{d}-6\mathcal{P}t\cdot d^{d-1}},
\end{aligned}
\end{equation}
which requires $d_{r}>6\mathcal{P}t$ while the critical distance $d_{c}$ which corresponds to the upper bound is nearly equals to $6\mathcal{P}t$.
If we relate the conserved particles-number $\mathcal{P}$ to a matrix, then it has operator norm 
$\|\mathcal{P}\mathcal{P}^{*}\|_{{\rm op}}=1$ and $\mathcal{P}^{\dag}\mathcal{P}=\mathcal{P}\mathcal{P}^{\dag}={\bf I}$ where ${\bf I}$ is a identity operator.
This is related to the case mentioned in the Ref.\cite{Kollar M2} that $n_{i}$ only have the two eightvalues 0 and 1,
and here the maximal eightvalue 1 is nondegenerate for our scenario, while other eightvalues approaches to 1 smoothly in the long-time limit.

Since the in long-time limit the relaxation will removing the non-diagonal part of the density matrix,
the differece between the density matrices and its diagonal one 
is $\Delta\mathcal{J}=\mathcal{J}(t)-\mathcal{J}_{G}$,
thus for the hopping matrix which mentioned above, its trace norm has
\begin{equation}
\begin{aligned}
\frac{(6\mathcal{P}t)^{d_{c}}}{d_{c}^{d_{c}}-6\mathcal{P}t\cdot d_{c}^{d_{c}-1}}>\|\Delta\mathcal{J}\|.
\end{aligned}
\end{equation}
Note that here the critical value $d_{c}$ is independent of the size of system.

We have present the upper bound of of speed of information transfer by a form of suppressed exponent.
Since the nondiagonal contribution won't vanish until $t\rightarrow\infty$ (which corresponds to $\Delta\mathcal{J}=0$),
and it's decay in a time scale as $1/t$\cite{Manmana S R,Barthel T}, i.e., the dephasing process, (note the for large-size system, 
the inequality of Eq.(45) will becomes more obvious,
and the vanished nondiagonal contribution will reappear if the size is large enough, which called ``rephasing''),
the phase can be expressed as $\varphi({\bf k})=\varphi(0)+{\bf q}^{\ell}+O({\bf q}^{\ell +1})$\cite{Barthel T}
where $\ell$ is a tunable parameter in phase space.
The contribution in such a dephasing with scale $1/t$ in phase space is

\begin{equation}   
\begin{aligned}
{\bf k}^{\ell}=\int d{\bf k}^{\ell}e^{i\varphi(k)}\frac{{\bf k}^{1-\ell}}{\ell}\int d^{d-1}{\bf k}f(k),
\end{aligned}
\end{equation}
where $\varphi(k)=\varphi_{0}+{\bf k}^{\ell}$.


Next we form the the Bessel formula to show the reducing property of the evolution operator $e^{i\mathcal{P}t}$
which with large size $N$ and can be viewed as the Riemann sum approximation of the following function with phase number $\alpha$ \cite{Cramer M},

\begin{equation}   
\begin{aligned}
J_{\alpha}(x)=&\frac{1}{2\pi i^{\alpha}}\int^{2\pi}_{0}{\rm exp}[i(\alpha \varphi+x\ {\rm cos}\varphi)]d\varphi\\
=&\frac{1}{2\pi}\int^{2\pi}_{0}{\rm exp}[i(\alpha \varphi-x\ {\rm sin}\varphi)]d\varphi,
\end{aligned}
\end{equation}                     
which is shown in the Fig.7.
Through this, 
the maximum rate for the system to relaxation to the Guassian state is obtained as $(2\mathcal{P}t)^{-N/3}$ for a $N$-site system.

For this one dimension bosonic system what we are discussing, the Mott gap $U-U_{c}$ is allowed to exist during the relaxation process\cite{Kollar M2},
(for a experiment, see Ref.\cite{J?rdens R}).
For coupled bose-lattice model, one
forms the time-dependent continuous variable $n(t)$ to describe the quasiperiodic decaying\cite{Polkovnikov A},
the semiclassical motion equation which in a continuum bath of harmonic potential and additively applied a confining parabolic potential, is

\begin{equation} 
\begin{aligned}
\frac{d^{2}n}{dt^{2}}+4n+4g_{{\rm red}}n\left[{\rm cos}(\varphi(0))+\frac{g_{{\rm red}}n^{2}}{2}\right]=0,
\end{aligned}
\end{equation}
where $\varphi(0)$ is the initial phase. Thus the double occupation $\langle n^{2}(t)\rangle$ (also the double momenta occupation number in momentum space)
under the quenches from different Mott insulator initial state (with different initial phase) to weak interaction one 
(with weak $g_{{\rm red}}$) is\cite{Polkovnikov A} (ignore the influence of high-order $U$)
\begin{equation} 
\begin{aligned}
\langle n^{2}(t)\rangle\approx n^{2}(0)-\frac{1}{2\pi}\int^{\pi}_{0}{\rm sin}^{2}\varphi(0){\rm cos}[4t\sqrt{1+g_{{\rm red}}{\rm cos}\varphi(0)}]d\varphi(0),
\end{aligned}
\end{equation}
where $n^{2}(0)=1/4$ here as a effective approximation for two uncouping system in semiclassical theory. 
The $n^{2}(t)$ with weak $g_{{\rm red}}(<1)$ according to Eq.(49) is shown in the fig.8, note that
since the critical value of interaction for superfluid-to-Mott insulator phase transition in the Bose-Hubbard lattice model 
requires $U/\mathcal{P}\approx 16.7$ \cite{Elstner N},
and the reduced coupling $g_{{\rm red}}\sim N^{2}$, so the ground state of this system will keep this superfluid regime
in a large range of $g_{{\rm red}}$ if without excitation like the quench behavior.
But this expression doesn't works for the region of $g_{{\rm red}}>1$, e.g., see (d) and (e) in Fig.8.
The long-time behavior with very weak $g_{{\rm red}}$, the asymptotic behavior of Eq.(49) is
\begin{equation} 
\begin{aligned}
\langle n^{2}(t)\rangle\approx n^{2}(0)-\frac{1}{\sqrt{16\pi g_{{\rm red}} t}}[{\rm cos}(4t\sqrt{g_{{\rm red}}+1}+\frac{\pi}{4})+{\rm cos}(4t\sqrt{1-g_{{\rm red}}}-\frac{\pi}{4})],
\end{aligned}
\end{equation}
which is presented in Fig.9.
We can see that the amplitude fluctuation is increse with the reduction $g_{{\rm red}}$,
and in long-time limit the undulate of oscillation becomes more flat but no completely governed by the time-independent Hamiltonian.
This corresponds to the superfluid regime with obvious amplitude fluctuation and the recurrences and interference pattern will occur (not shown).
For the case of initial $g_{{\rm red}}=N$, when the quenched $g_{{\rm red}}\gtrsim 7N$,
this nonequilibrium system will into the nonthermal steady state though it's a nonintegrable system according to the results shown in Ref.\cite{Kollath C}.


For one-dimension nonintegrable case of hard-core bosons (which generalized eigenstate thermalization occurs\cite{Vidmar L}), 
a typical model of $1/r$ Hubbard chain also have the feature of collapse-and-revival oscillations\cite{Kollar M}
like the nonintegrable one, but it's dispersion-linear, i.e., it can be effectively solved by Eq.(23) while the nonintegrable one can not.
Now we consider the large $g_{{\rm red}}$ into strong-couping perturbation in a two-dimension version of $1/r$ Hubbard model, 
the lattice fermions Hubbard model, the double occupation $d(t)=\langle n_{\uparrow}n_{\downarrow}\rangle/N$ can be written as\cite{Eckstein M2}
\begin{equation} 
\begin{aligned}
d(t)= d(0)+\sum_{i=0}^{N-1}\frac{1}{g_{{\rm red}}}\langle c^{\dag}_{i}c_{i+1}(n_{i}-n_{i+1})^{2}\rangle+O(\frac{V^{2}}{U^{2}}).
\end{aligned}
\end{equation}
whose graphs have been presented in the Fig.2 of Ref.\cite{Eckstein M2}.
This is corresponds to the state of Mott insulator with strong interaction and have
\begin{equation} 
\begin{aligned}
\mathcal{P}\langle c^{\dag}_{i}c_{i+1}(n_{i}(0)-n_{i+1}(U))^{2}\rangle=2\sum_{i}[\kappa_{i}(n_{i}(0)-n_{i+1}(U))],
\end{aligned}
\end{equation}
where $\kappa_{i}$ is the dispersion relation related to the kinetic energy $T_{{\rm kin}}$.
The prethermalization regime is also exist in this case for one-dimension or two-dimension Bose-Hubbard model\cite{Kollath C},
but this prethermalization regime as well as the general collapse-and-revival oscillations vanish in a little range before the critical value $U_{c}$
which origin from the discontinuity momentum distribution in Fermi surface under the quenching.

We show the bandwidth-dependent kinetic energy of $1/r$ Hubbard chain in Fig.10.
with different bandwidth: $W=1$, $W=4$, and $W=1/2$ which have been obtained by the method of local density approximation (LDA)\cite{Mattheiss L F}.
It's obviously to see that the amplitude of hopping is increases as the bandwidth $W$ increases (inset),
and the $T_{{\rm kin}}$ decay rapilly with the increase of distance along the chain.
When quenches to large $U$ , the oscillations of Eq.(51) makes a difference\cite{Eckstein M2} $\Delta d=\mathcal{P}\pi(1-2n/3)/U$ 
which is halved for Falicov-Kimball model
in nonequilibrium dynamical mean-field theory (DMFT)
due to the vanishing of $\mathcal{P}$ for one of its two spin species and therefore only one
spin specie contributes to kinetic energy.
In DMFT, this kinetic function due to the considerable noise (see Setc.10, Appendix.C) yields a single-site Green's function
\begin{equation} 
\begin{aligned}
G(t,t')=i\langle c(t)c^{\dag}(t')\rangle,
\end{aligned}
\end{equation}
where the contour-order correlation $\langle c(t)c^{\dag}(t')\rangle$ has
\begin{equation} 
\begin{aligned}
\langle c(t)c^{\dag}(t')\rangle=\frac{{\rm Tr}[e^{\beta H_{G}}T_{\mathcal{C}}e^{S}c(t)c^{\dag}(t')]}{{\rm Tr}[e^{\beta H_{G}}T_{\mathcal{C}}e^{S}]},
\end{aligned}
\end{equation}
where $T_{\mathcal{C}}$ is the contour-order temperature, and the single-site action\cite{Eckstein M4}
\begin{equation} 
\begin{aligned}
S=\int_{\mathcal{C}}dtdt'c^{\dag}(t)\Lambda(t,t')c(t')+\int_{\mathcal{C}}dtV(t),
\end{aligned}
\end{equation}
where $\Lambda(t,t')$ is a hybridization of site with fermion operators and the rest of the lattice,


By the nonequilibrium DMFT, which well describe the time evolution of an interacting many-body system (fermions lattice Hubbard model here), 
we can map the lattice model to the single-site impurity model as shown in above.
Unlike the Eq.(51), the method of DMFT is nonperturbative,
but since we consider the perturbation from noise into the Green's function,
the resulting Green's function is 
\begin{equation} 
\begin{aligned}
G(t,t')=G_{0}(t,t')+G_{0}(t,t_{i})\Sigma_{ij}G(t_{j},t'),
\end{aligned}
\end{equation}
where $G_{0}$ is the unperturbed Green's function,
and it has\cite{Gull E}
\begin{equation} 
\begin{aligned}
\frac{e^{V}-1}{e^{V}-iG_{0}(e^{V}-1)}*G_{0}(t,t')=\Sigma *G(t,t'),
\end{aligned}
\end{equation}
where $V=H-H_{G}$ is the non-Gaussian part of the Hamiltonian,
i.e., the interaction term $U(t)n_{\uparrow}n_{\downarrow}$
which is noncommuting\cite{Rombouts S M A}.
So to linearize the rest part of the Hamiltonian, 
we need to tend the partial function which is the denominator of Eq.(54) into interaction representation
with decomposed Boltzmann operator using the method of Hubbard-Stratanovich transformation
which require the convergency of the gaussian integrals\cite{Fyodorov Y V}.
Since this partial function select all the possible configuration of single-site along the contour $\mathcal{C}$,
which make it possible to be decouped by a auxiliary-field quantum Monte Carlo methods\cite{Rombouts S M A,Eckstein M4},
(Note that the integrable lattice model for soft-core bosons
, the nonGaussian disturbution is origin from the off-site hopping\cite{Fisher} 
term unlike the case what we are talking).
the single-energy variables $s_{i}$ along the contour $\mathcal{C}$ have \cite{Rombouts S M A}
$e^{V}_{\sigma}={\rm diag}(e^{\gamma\sigma s_{1}},e^{\gamma\sigma s_{2}},\cdot\cdot\cdot,e^{\gamma\sigma s_{i}})$
where $\sigma$ denote the spin order $\sigma=\pm 1$ and $\gamma$ here is a temperature- and interaction-dependent parameter.
This equation means that eighvalues (which can be specificized as the band energy $\epsilon_{k}$ in Hubbard model) 
of hopping matrix $V$ can be diagonalized by the diagonal matrices which shown in the bracket of this equation.


Since the total Hamiltonian must be conserved in the evolution,
the kinetic energy of $1/r$ Hubbard chain is suppressed by the term $E_{{\rm pot}}=Ud(t)$.
For half-filling Hubbard Hamiltonian ($n_{\uparrow}=n_{\downarrow}=1/2$) with
a semielliptic density of state $\rho_{hf}=\sqrt{4\mathcal{P}^{2}-\epsilon_{k}^{2}}/(2\pi\mathcal{P}^{2})$,
the kinetic energy per lattice site\cite{Eckstein M2} is $T_{{\rm kin}}=2\int d\epsilon_{k} \rho_{hf}(\epsilon_{k})n(\epsilon_{k},t)\epsilon_{k}$,
where the band energy $\epsilon_{k}$ here which obey the Dyson equation in lattice model with Green's function $G_{k}(t,t')$
\begin{equation} 
\begin{aligned}
G_{k}(t,t')(i\partial_{t}+\mu-\epsilon_{k}-\Sigma)=1,\ t=t'
\end{aligned}
\end{equation}
where the convolution product of local self-energy $\Sigma$ with $G_{k}$ yields the equal time double occupation in the homogeneity phase
and the self-consistency local Green function has\cite{Eckstein M} $G_{k}(t,t')=\int d\epsilon_{k}\rho(\epsilon_{k})G_{k}(t,t')$, where $G_{k}(t,t')$ is diagonal.
The approximation of Hartree-Fock which works well for the single-particle problem, 
affect the chemical potential $\mu$ which have a zero mean, by the particle number in canonical ensemble
\begin{equation} 
\begin{aligned}
\overline{\langle n_{\uparrow}n_{\downarrow}\rangle}=\frac{1}{N^{2}}\sum_{k,k'}\langle n_{k\uparrow}n_{k'\downarrow}\rangle=\frac{n^{2}}{4},
\end{aligned}
\end{equation}
and it contribute to the self-energy by the diagonalized Hartree-Fock Hamiltonian and provide the precise result in half-filling case,
but since the Hartree-Fock is sensitive to the spin-correlations\cite{Gebhard F2},
it fails when the spin degrees of freedom disappear.
In this case, one gives the second-order contribution to the self-energy by the form of\cite{Eckstein M4}
\begin{equation} 
\begin{aligned}
\Sigma(t,t')=-U(t)U(t')G_{0\sigma}(t,t')G_{0,\overline{\sigma}}(t',t)G_{0,\overline{\sigma}}(t,t'),
\end{aligned}
\end{equation}
here the unperturbed Green's function $G_{0\sigma}$ can be replaced by the full interacting one $G_{\sigma}$,
and the interaction $U$ can be viewed as a evolution propagator here.

Since the fact\cite{Gebhard F2} that the phase transition of metal-to-insulator in half-filling $1/r$ Hubbard chain occurs when the $U=W$,
which we set the bandwidth $W=4$ here,
i.e., $U_{c}=4$.
Note that the band energy $\epsilon_{k}$ is closely related to the continuity of momentum distribution,
e.g., it's discontinuity when $\epsilon_{k}=0^{-}$ and $0^{+}$ in the each side of critical value $U_{c}$.
When quench approaches to critical value $U_{c}$, $d(0)=1/8$,
and since we set the $n=1$ and the critical value is $U_{c}=4$,
the one-dimension half-filling $1/r$ Hubbard model have the double occupation as
\begin{equation} 
\begin{aligned}
d_{hf}(t)=\frac{1}{8}-\frac{(4-U)^{2}}{16U}-\frac{(16-U^{2})^{2}}{16U^{2}}{\rm ln}\left|\frac{4-U}{4+U}\right|-\frac{{\rm cos}(Ut){\rm cos}(4t)}{2Ut^{2}},
\ {\rm for\ quench\ from\ 0\ to\ U};\\
d_{hf}(t)=\frac{1}{8U}+\frac{(4-U)^{2}}{16U^{2}}+\frac{(16-U^{2})^{2}}{16U^{3}}{\rm ln}\left|\frac{4-U}{4+U}\right|+\frac{{\rm cos}(Ut){\rm cos}(4t)}{2U^{2}t^{2}},
\ {\rm for\ quench\ from\ \infty\ to\ U},
\end{aligned}
\end{equation}
while for the quench to reach $U_{c}$,
the behavior of double occupation is described by
\begin{equation}
\begin{aligned}
d_{c}(t)=\frac{1}{8}-\frac{1}{512}\left[\frac{48{\rm sin}(8t)}{t^{3}}+(\frac{6-32t^{2}}{t^{4}})({\rm cos}(8t)-1)\right]-\frac{3}{32t^{2}},
\ {\rm for\ quench\ from\ 0\ to\ U};\\
d_{c}(t)=\frac{1}{32}+\frac{1}{2048}\left[\frac{48{\rm sin}(8t)}{t^{3}}+(\frac{6-32t^{2}}{t^{4}})({\rm cos}(8t)-1)\right]+\frac{3}{128t^{2}}.
\ {\rm for\ quench\ from\ \infty\ to\ U}.
\end{aligned}
\end{equation}
Fig.11 shows the graphs of $d_{hf}(t)$ of Mott insulator for quenches from $0$ to $U$ and from $\infty$ to $U$ (according to Eq.(61)),
we can see that the later one is roughly the inverse version of the former one, and a significant features is the fast-saturation.
The larger the interaction $U$, the faster the curve tends to saturated.
Note that the double occupation here is indeed related to the realistic physical quantity of global correlation for bosons system,
and the disscution above is for a prediction for the behavior of long-time limit,
i.e., the stationary result, which consistent with the thermal values\cite{Kollar M2} :$1/4$ for interaction quenches from 0 to $\infty$,
$1/6$ for interaction quenches from $\infty$ to 0,
$1/8$ for interaction quenches from 0 or $\infty$ to $U_{c}$, (we set $n=1$ here).
The collapse of oscillations are scale as $1/\sqrt{g_{{\rm red}}}$,
i.e., the amplitude are continually decaying along the long-time scale limit cover the phase transition,
and $d(t)$ will shows strictly periodic behavior in the noninteracting regime with $g_{{\rm red}}=0$ (not shown in the Fig.11).
         For quenches from 0 to finite $U$, the prethermalization regime also shows large agreement with the stationary values of $d(t)$ in long-time limit.
The effect of damping on the amplitude of collapse-and-revival oscillations is always exist in the long-time scale,
and has important influence on the relaxation.
It produce the ``overdamp'' in the regime of sufficiently large $U$,
which nearly reduce the amplitude to 0 after instantly tends to saturate.
The process of damping is related to the velocity of spin wave in Goldsone model that
for zero frequency Goldstone mode is followed by a additional standing spin waves\cite{Plihal M,Polkovnikov A}.
By setting a list of interaction in Fig.11, we found that, for quench from 0 to a infinite interaction $U$, 
the closer the quenches to critical value $U_{c}$, the closer the $d_{hf}(t)$ to quasistationary value
which is obtained from the Fig.12 as 0.125 (see the bottom inset of Fig.12(a)):
the $U$ which close to $U_{c}$ in Fig.11(a) is setted as 3.299, and the long-time result for quench to this $U$ is 0.12499,
which is very close to the stationary prediction $1/8$,
and it's reasonably differ from the thermal prediction of $0.098$ by the equilibrium result\cite{Eckstein M3}.
While for the quench from $\infty$ to $U$, we obtain the same conclusion:
the result of quench to $U=3.299$ is 0.032 which is very close to the stationary value 0.0312 which is shown in the bottom inset of Fig.12(b).
That is the long-time behavior of nonequilibrium system which show agreement with the result of quasistationary value in phase transition point
(this conclusion will always exist in the time scale of $1/|\mathcal{P}|\ll t\ll U/\mathcal{P}^{2}$),

While for the anharmonicity case,
the couping $g_{{\rm red}}$ is still usable by the form of a symmetrical anharmonic term (see Setc.10),
the bare action of quantum system with $N$-component bosonic field $\phi_{\alpha}$ in $\phi^{4}$ field theory,
when the $g_{{\rm red}}$ close to the critical value with $U_{c}$, is\cite{Sachdev S,Polkovnikov A}
\begin{equation} 
\begin{aligned}
S=\int d^{d}r d\tau \frac{1}{2}[(\nabla_{r}\phi_{\alpha})^{2}+\frac{(\partial_{\tau}\phi_{\alpha})^{2}}{c^{2}}-(r_{c}+r)\phi_{\alpha}^{2}+\frac{\lambda r^{4}}{N}\phi_{\alpha}^{4}],
\end{aligned}
\end{equation}
where $\alpha=1\cdot\cdot\cdot N$, $c$ is the velocity, $\lambda r^{4}$ is the quartic nonlinear coupling term,
and the critical $r_{c}$ is reach in the $r=0$.
For the case of quenches from large $U$ to a small one which is close to zero,
i.e., from the Mott insulator initial state to the superfluid or metallic state, 
we introduce the vectors $k_{1}=2\pi n_{1}/N$ and $k_{2}=2\pi n_{2}/N$ which obey periodic boundary condition (see Appendix.C)
and have $n_{1}\neq n_{2}<N$,
then when the couping is close to zero, the time-dependent nearest-neighbor correlation in the bath with harmonic potential is given as\cite{Polkovnikov A}
\begin{equation} 
\begin{aligned}
\langle n_{r}(t)n_{r+1}(t)\rangle=\frac{2g_{{\rm red}}}{N}\sum^{N-1}_{r}\frac{{\rm sin}^{2}\mathcal{G}t}{\mathcal{G}},
\end{aligned}
\end{equation}
where the periodic correlator $\mathcal{G}=1+{\rm cos}k_{1}-{\rm cos}k_{2}-{\rm cos}(k_{1}-k_{2})$.
This utilize the periodicity of harmonic oscillators in superfluid regime and exclude the high-frequency part due to the periodic boundary condition,
i.e., keep the stable low-frequency only.

For many-body system,
the dispersion relation $\kappa$ of this bosonic model is oscillate as a function of $k$ with the period $\pi$ (see Fig.13).
From Fig.14, it's obvious to see that the periodic dispersion relation resulting in the degeneracy of energy.
In the process of relaxation of correlation,
the relevant parameter is assumed change linearly.
By setting the dispersion relations $\kappa$ before and after quench, the corresponding relaxation of correlations between the bosons is shown in the Fig.14,
we see that the oscillations approach to quasisteady state with small (non-zero) frequency,
and with the increasing of dispersion relation,
the amplitude of correlation is decreased and the required-relaxation time is shorter.
In fact this conclusion is always exist for all the many-body system in phase-space.

\section{Investigation of Relaxation of Chain Model to Gaussian State By the Transfer Matrices}
We then define the transfer matrix 
\begin{equation}
\begin{aligned}
t(x)={\rm Tr}(\prod_{l}^{a} T_{l}(x)), 
\end{aligned}
\end{equation}
where $T_{l}(x)=R^{l}_{n-1}(x)R^{l}_{n-2}(x)\cdot\cdot\cdot R^{l}_{0}(x)$ is the monodromy matrix with $n$-site R-matrices and $x$ is the spectral parameter.
Employing this transfer matrix representation, the initial state can be written as 

\begin{equation}   
\begin{aligned}
\mathcal{F}_{0}(x)=\lim_{n\rightarrow\infty}\frac{1}{n-1}\frac{\partial}{\partial x}\langle \psi(0)|t(x)t^{\dag}(x)|\psi(0)\rangle
\end{aligned}
\end{equation}
where the total number of particles $N$ is a integer multiple of number of transfer mtrices ${\rm num}(t_{1}(x))$.
Based on this, the localed free energy of per spin (or grid point in the network) is

\begin{equation}   
\begin{aligned}
E_{{\rm free}}=-\frac{{\rm num}(t_{1}(x))}{N}\frac{1}{\beta}\lim_{M\rightarrow\infty}{\rm ln}\lambda_{max}
\end{aligned}
\end{equation}
where $M$ is the number depends on how many parts temperature divided into (i.e., the Trotter number), 
and $\lambda_{max}$ is the maximum eightvalue of transfer matrix and in the limit of $N\rightarrow\infty$, it has
\begin{equation}   
\begin{aligned}
\lambda_{max}^{N}=\lim_{M\rightarrow\infty}{\rm Tr}\ t_{1}^{{\rm num}(t_{1}(x))}(x),
\end{aligned}
\end{equation}
i.e., in the case of infinity-system-size the maximum eightvalue is equal to the trace of transfer matrices.
Further, we deduce that
\begin{equation}   
\begin{aligned}
\lim_{N\rightarrow\infty}\frac{{\rm ln}(\lambda_{max}^{N})}{N}=\lim_{M\rightarrow\infty}\frac{{\rm ln}\lambda_{max}}{N}\cdot{\rm num}(t_{1}(x)),
\end{aligned}
\end{equation}
which can be easily confirmed by numerical methods.
In the framework of auxiliary space which estabished in above, one can define the matrix $A_{i}$ which acting on the auxiliary space\cite{Fagotti M},
then the wave function of ground state can be redefined as 
\begin{equation}   
\begin{aligned}
|\psi(0)\rangle=\sum_{s_{i}}{\rm Tr}(\prod_{i=0}^{n-1}A_{i})|\prod_{i=0}^{n-1}s_{i}\rangle
\end{aligned}
\end{equation}
where $|\prod_{i=0}^{n-1}s_{i}\rangle$ denotes a normalized computational basis state\cite{?nidari? M},
while the set of unnormalized part form a projective space $\mathbb{P}$ with dimension $d_{i}d_{j}-1$\cite{Cubitt T}.

Since in normalization case the expectation value of initial state is $\langle \psi(0)|\mathcal{J}_{i}|\psi(0)\rangle$ with $\langle\psi(0)|\psi(0)\rangle =1$,
the transfer matrices in two subspaces can be obtained by the algebraic Bethe ansatz\cite{Fagotti M3}
\begin{equation}   
\begin{aligned}
t(i+\mathcal{R})={\rm Tr}(A_{n-1}(\mathcal{R})A_{n-2}(\mathcal{R})\cdot\cdot\cdot A_{0}(\mathcal{R})),\\
t^{\dag}(i+\mathcal{R})={\rm Tr}(A^{\dag}_{n-1}(\mathcal{R})A^{\dag}_{n-2}(\mathcal{R})\cdot\cdot\cdot A^{\dag}_{0}(\mathcal{R})),
\end{aligned}
\end{equation}
where $\mathcal{R}$ is a constans of motion and
the matrices $A$ and $A^{\dag}$ are isomorphic with the bipartite space of $\mathbb{C}^{d_{i}}\otimes\mathbb{C}^{d_{j}}$.
In convex hull construcsion for nuclear norm, a direction of subgradient is consist of the orthogonal set $\{ s_{i}\}$ and $\{ s_{i}\}_{\bot}$\cite{129},
and it's well know that the Schmidt rank $R$ is invariant by local operations and classical communication (LOCC) when the 
but variable when bipartite state is mixed\cite{Cubitt T,Terhal B M}.
For localized quantum communication, Eqs.(43,44) give the exponential suppression for transfer which reflected as the the exponentially fast quantum propagation in
branched tree graph and the exponentially slow down of latter-time motion in the quantum graph\cite{Keating J P} 
for which the information flow toward the random path in local relaxation process.

In the above Bose-Hubbard model, using the Wigner representation which is generally negative definite\cite{Drummond P D}
we also have the characteristic function of density matrix $\mathcal{J}_{i}$ as\cite{Cramer M}
\begin{equation}   
\begin{aligned}
{\rm Tr}[\mathcal{J}_{i}e^{\alpha b^{\dag}_{i}-\alpha^{*}b_{i}}]=e^{-\frac{|\alpha|^{2}}{2}}\prod_{d_{r}}L_{m}(|\alpha|^{2}e^{2it\mathcal{P}}(d_{r})),
\end{aligned}
\end{equation}
where the translation operator $e^{\alpha b^{\dag}_{i}-\alpha^{*}b_{i}}=e^{\alpha b^{\dag}_{i}}e^{-\alpha^{*}b_{i}}e^{-|\alpha|^{2}/2}$ where 
the state of c-number variable $|\alpha\rangle=e^{-|\alpha|^{2}/2}(\alpha b_{i}^{\dag}-\alpha^{\dag} b_{i})$\cite{Steel M J},
and $L_{m}$ is the Laguerre polynomial,
which is noniterative and utilized to express the boundary conditions of parameter space.
Here the density matrix $\mathcal{J}_{i}={\rm Tr}(|\psi\rangle\langle\psi|)$ and
$b_{i}^{\dag}b_{i}=-(\frac{\partial}{\partial\alpha}+\frac{\alpha^{*}}{2})(\frac{\partial}{\partial\alpha^{*}}+\frac{\alpha}{2})$,
$b_{i}b_{i}^{\dag}=(\frac{\alpha}{2}-\frac{\partial}{\partial\alpha^{*}})(\frac{\partial}{\partial\alpha}-\frac{\alpha^{*}}{2})$.
After the local relaxation (dephasing) to a steady state ensemble
with stationary state $\overline{\rho}_{i}$,
the Eq.(72) tends to the Gaussian form with 
$e^{-(\overline{\rho}_{i}+1/2)\alpha^{\dag}\alpha}$\cite{Cramer M}
where $\overline{\rho}_{i}$ is the average of initial states for finite system
and reach the maximum entanglement related to the second moments.
The Hamiltonian has $\lim_{t\rightarrow\infty}\langle\psi(0)|e^{\tau H}H_{\tau}e^{-\tau H}|\psi(0)\rangle={\rm Tr}(\overline{\rho}H_{\tau})$.
For integrable homogeneous system (like the one we present in the Sect.6), the translation invariance in transition states and it's also meaningful in the investigation of relaxation of degrees of freedom,
the small displacement of coordinates due to the local potential produce a negative Hessian eigenvalue\cite{Morgan J W R}, 
and since the site-shift invariance has been broken by the local conservation law\cite{Fagotti M2},
The result of Ref.\cite{Cramer M} shows that the local relaxation is always preserves the full information of initial state, 
which shows that the information of initial state
is not or at least not only be recorded by the factors of Lagrange multipliers\cite{Fagotti M2},
and this is consistent with the above result in Gaussian form which contain the term about initial states.
While for inhomogeneous case (like most of the damped or polarized model), since the translation invariance is broken, 
the thermal behaviors and scattering is very different compare to the homogeneous one,
and the prediction of GGE to the final state is also inadequate\cite{Lancaster J}.
Further, the relaxed result for nonequilibrium system can be constructed as the sum of Gaussians which is associated to the related collective variables\cite{Bussi G}
or canonical variables which can be utilized to diagonalize the inhomogeneous model\cite{Suttorp L G}.
Note that this Gaussian state is quasifree and contains only second moments,
i.e., the redistribution by the scattering.
We will further represent this process by matrix method in the next section.
When the system have already relax to the equilibrium distribution, the dynamic is well described by a stochastic partial differential equation,
e.g., the quantum Langevin equation\cite{Gardiner C W}.
For this equilibrium state under large time evolution, the diffusion have a non-negligible influence to system and
produce the recurrences which occur in a time scale larger that the relaxation time
(i.e., the diffusion time is larger than the relaxation time),
and the recurrences period is also depends on the transfer velocity of information.


\section{Matrices Processing}

The density matrices of Eqs.(14,15) can be represented by the Schmidt decomposition of bipartite state
\begin{equation}   
\begin{aligned}
|\psi\rangle=\sum_{R}^{R}\sqrt{\lambda_{R}}|\mathcal{J}_{iR}\rangle\otimes|\mathcal{J}_{jR}\rangle
\end{aligned}
\end{equation}
where $\lambda_{R}$ is the maximum eightvalues of density matrix for each $R$. If we set the the maximum rank is $\mathds{R}$, 
then it have $\sum_{R}^{\mathds{R}}\lambda_{R}=1$ and $(\sum_{R}^{\mathds{R}}\sqrt{\lambda}_{R})^{2}\le\mathds{R}$,
Definition\cite{Terhal B M} 
shows that the Schmidt rank is just $\mathds{R}$ under the condition $\mathds{R}-1<(\sum_{R}^{\mathds{R}}\sqrt{\lambda}_{R})^{2}\le\mathds{R}$.
In the case of $(\sum_{R}^{\mathds{R}}\sqrt{\lambda}_{R})^{2}<\mathds{R}$, only the eighnvector which has maximum rank $\mathds{R}$ is needed,
that also explain why the singular values decomposition reserved only the largest singular value (Eq.(12)).
The set spaces $\mathbb{S}$ with convex constrution always have $\mathbb{S}_{R}\subset\mathbb{S}_{\mathds{R}}$.
In the zero-entanglement case, 
the square root of eightvalue of $\mathcal{J}\mathcal{J}^{*}$ have $\sqrt{\lambda_{R}}=(V\mathcal{J} V^{\dag})_{ij}$ with another index $j$ when $(V\mathcal{J} V^{\dag})_{ij}$ is diagonal,
and in another expretation is $\langle A_{i}|\sigma_{y}A_{j}^{*}\rangle=\lambda_{R}\delta_{ij}$,
where $\sigma_{y}=\begin{pmatrix} 0&-i\\ i&0 \end{pmatrix}$ and $A$ is the matrix-product state. 
Here we conside the spin flip in the term $\sigma_{y}A_{j}^{*}$, and it also have
$\langle A_{i}|\sigma_{y}A_{j}^{*}\rangle={\rm Tr}[(\sigma_{y}A_{j}^{*})^{\dag}A_{i}]={\rm Tr}[(A_{j}^{T}\sigma_{y})A_{i}]$ (not the scalar product).
In such a flip in tilted state scheme\cite{Fagotti M} we let the eightvalue $\lambda_{i}=e^{S_{i}^{z}}$,
and $A=e^{i\theta\sum_{i}S_{i}^{z}}$, i.e., spin flip when the $\theta=\pi$. It's found that $\sum_{i}e^{2i\theta}\lambda_{i}=0$ in the zero-entanglement case\cite{Wootters W K}.

A prerequisite to satisfy this formula is zero-entanglement, i.e., the two subsystem $i$ and $j$ is separable (or distillable).
The density matrix $\mathcal{J}$ here is assume have the eightvalue $\lambda_{R}$ and it diagonalized
by matrix $V$ when $\mathcal{J}$ is symmetry,
and in this case, the eightvalues of $\mathcal{J}\mathcal{J}^{*}$ is non-negative.
Assuming $V$ is a $m\times n$ matrix with $n$ orthonormal columns and $m<n$, 
thus $V$ acts periodic with period of square of number of column $n$.
Let $\Sigma$ is the $m\times m$ diagonal matrix which made up of singular values of $\mathcal{J}$,
then its nuclear norm can be expressed as $||\mathcal{J}||_{*}={\rm Tr}(V\Sigma V^{T})$.

If here $V$ is the matrix $A$ which appear in Eq.(71), then the trace norm of $A$ which equals the sum of square root of eightvalues
$\sum_{i}\sqrt{\lambda_{i}}$
have $||A||_{{\rm Tr}}={\rm Tr}\sqrt{AA^{\dag}}$ and $||A\otimes A^{\dag}||_{{\rm Tr}}=||A||_{{\rm Tr}}\cdot||A^{\dag}||_{{\rm Tr}}=||A||^{2}_{{\rm Tr}}$.
In normalized case with $\langle\psi|\psi\rangle=1$, the operator norm of $A$ has the similar property with Hermitian conjugate matrices:
$||A^{*}\otimes A||_{{\rm op}}=||A||^{2}_{{\rm op}}=1$.
This corresponds to a absolute value of the maximal eightvalue which is normalized and it's found nondegenerate for $S=1/2$ Heisenberg model\cite{Xiang T}.
For separable case, $\sum_{R}^{\mathds{R}}\lambda_{R}\le 1$ due to the convertibility and the decomposition of entangled state into unentangled pure states
in the case that the maximum eightvalue is smaller than the sum of rest eightvalues\cite{Wootters W K}, 
i.e., $\lambda_{1}<\lambda_{2}+\lambda_{3}+\cdot\cdot\cdot\lambda_{n}$ (here set the $i=1,2,\cdot\cdot\cdot,n$).
For pure state we have
\begin{equation}   
\begin{aligned}
\frac{n-1}{n}\ge 1-\sum_{i}\lambda_{i}^{2}\ge\frac{4}{n(n-1)}(\sum_{i<j}\sqrt{\lambda_{i}\lambda_{j}}).
\end{aligned}
\end{equation}
A general bound of dimension of subspace is that
the largest dimension of space is almost $d_{i}\times d_{j}$ and the smallest one is $(d_{i}-\mathds{R}+1)(d_{j}-\mathds{R}+1)$, 
and these subspaces which dimension within this range, i.e., the rank $R<\mathds{R}$ can be represented by the affine variety\cite{Cubitt T}.
Since a precondition of increase of the Schmidt rank is incresing the dimension of subspace, and the degree of entanglement is also 
reaches maximally when it grows into
the largest subspace, 
we can obtain that in most case the largest subsystem which almost is full rank have the almost maximal entanglement, except somecase for the pure state
which is unmixed\cite{Terhal B M}.
The largest subspace form the largest-probability set with the constants of motion which proportional to the dimension of corresponding Hilbert space
or projector onto its eightvalues, or its integer powers of Hamiltonian\cite{Rigol M}.

Without losing general, for distillable state, the upper bound of entanglement entropy formed by the logarithmic negativity \cite{Plenio M B}
$S_{N}={\rm ln}||\mathcal{J}^{\Gamma}||_{{\rm Tr}}$, where $\mathcal{J}^{\Gamma}$ is the partial transpose of density matrix $\mathcal{J}$
and the corresponds covariance matric is $\gamma^{\Gamma}=P\gamma P$ where $\gamma$ is covariance matric and 
the diagonal matrix $P=(-{\bf I}_{i})\oplus{\bf I}_{j}$ with the diagonal identities matrices ${\bf I}$.
Let $V$ is the nonsingular and skew-sysmetric column vector, and it's real.
Then we have $V^{T}\mathcal{J}V=\mathcal{J}$,
(i.e., $\mathcal{J}$ is diagonal)
so the nonincreasing ordered symplectic eightvalues $\lambda^{\Gamma}$ with symplectic matrix 
$\Omega=\begin{pmatrix} 0&{\bf I}_{d}\\ -{\bf I}_{d}&0 \end{pmatrix}$ 
which describe the reduced Gaussian state
\cite{Plenio M B,Eisert J} have

\begin{equation}   
\begin{aligned}
{\rm ln}||\mathcal{J}^{\Gamma}||_{{\rm Tr}}=\sum_{i}{\rm ln}({\rm max}[1,(\lambda_{i}^{\Gamma})^{-2}])\le\sum_{i}((\lambda_{i}^{\Gamma})^{-2}-1),
\end{aligned}
\end{equation}
while the normal eightvalue of $\mathcal{J}$ is $\lambda_{i}$ which equal to $(\lambda_{i}^{\Gamma})^{2}$ (see Appendix.B for the detail).


\section{Relaxation of Nonequilibrium System With Stochastic Dynamical Variables}

Since for mixed system, if the initial state is homogeneous, the second moments is conserved and it prevent the system to relax to the thermal state\cite{Cramer M},
so the effective disentanglement is impossible in this case,
and thereforce some microstates are inaccessible since the final state is onstrained by the conserved constans of motion
no matter the system is integrable or not.
Like the integrable system which guided by the corresponding GGE with maximal entropy $S_{ij}=-{\rm Tr}(\rho_{ij}{\rm ln}\rho_{ij})$, 
will reach nonthermal steady states and share the similar propertice with the prethermalization plateaus in the long time limit\cite{Kollar M},
which also called the prerelaxation in the time evolution of GGE,
and this has been founded in the isolated or open quantum system\cite{Gring M,Marino J},
while for the nonintegrable system it's thermalize directly\cite{Rigol M}.
In the inhomogeneous case like the most damping model, the conserved law is no more exist and then the thermal state is achievable directly.
The local minima free energy which separated by barriers in free energy surface is connected by along the steepest descent path 
in the scenario of discretized evolution\cite{Laio A} and thereforce update the collective coordinates.
This is a powerful way to obtain the symmetric tensor in the flattened space macroscopically, and even the supersymmetry system with gloabal minimal potential energy.
In these special points, 
the gradient of free energy as well as the potential energy vanish, and the energy is rised by the little displacement of coordinates\cite{Morgan J W R}.

Defining $\mathcal{Z}$ as a collective rariable with the coordinate $x$, 
then for harmonic oscillators with mass $m$, in the free energy surface, the disturbution of Gaussians
can be described by the biasing potential which is guided by the differece of free energy $E(\mathcal{Z})-E_{G}(\mathcal{Z},\tau)$\cite{Laio A2}

\begin{equation}   
\begin{aligned}
V_{{\rm bias}}=w\sum_{{\rm Gaussians}}{\rm exp}(-[\mathcal{Z}(x)-\mathcal{Z}(x_{G})]^{2}/2(\delta\mathcal{Z})^{2}),
\end{aligned}
\end{equation}
where $w$ and $\delta\mathcal{Z}$ are the height (amplitude) and width of the Gaussians and $x_{G}$ is the positon of Gaussians,
and in the limit of $w\rightarrow 0$, it have $\int d\mathcal{Z}e^{-\beta E(\mathcal{Z})}=e^{-\beta E_{G}(\mathcal{Z},\tau)}$.
Such a biasing potential is indeed a history-dependent term which is appear in the non-Markovian dynamics equation and as a biased estimator for the free energy,
and while the unbiased estimate require a Markovian one\cite{Bussi G}.
A experiment done recently\cite{Yang B} of one-dimension Tomonage-Luttinger liquid model 
that the Gaussians propagation, which are adjusted by microwave, are along the one-dimension trajectorys (``tubes'')
and accompanied by a negative perturbation in the time evolution of $w$ and $\delta\mathcal{Z}$ which shows stability in the chaotic scenario.
This biasing potential is indeed a bias estimator for the quantum states with multiple phases,
and we can see that it follows the Gaussian decay.
Here the summation symbols is used due to the discretized evolution. 
Note that this expression is for harmonic oscillators, i.e., the system is linear response.
While for anharmonicity oscillators, which produced by, .eg., the detuned Gauusian laser\cite{Arlt J,Bretin V} (blue-shift or red-shift)
or the (two-photon) Raman detuning\cite{Zhang J Y},
this potential need to modified by adding some variational parameter describing the asymmetry (three-order term) or symmetry (quartic term) anharmonic\cite{Li G Q,Hou J X}
to the exponent part of Eq.(76).
The coupling in this case is nonlinear, like the scenario in FPU theorem.
The free energy $E_{{\rm free}}=-V_{{\rm bias}}$, and it's govern by the force $F=-\partial_{G}E_{{\rm free}}$\cite{Laio A}.
After the flatting process on free energy surface (for a intuitive schematic view, see, e.g., the Ref.\cite{Odagaki T}), 
the change of the distribution makes the new Gaussains which goven by new Hamiltonians,
and hence the new equilibrium states,
that can only happen in the inhomogeneous situation. 
After the local minimums of differece of free energy were mostly eliminated, 
then the probability distribution is nearly uniform,
and the remaining corrugations are independents of the $E(\mathcal{Z})-E_{G}(\mathcal{Z},\tau)$.
The action describing this dynamic of evolution in complex time scale is ($\tau$ is the complex time here)

\begin{equation}   
\begin{aligned}
S(\mathcal{Z})=\frac{m}{2}\int^{\tau'}_{\tau} d\tau[\dot{\mathcal{Z}(\tau))}^{2}-\mu^{2}\mathcal{Z}^{2}(\tau)],
\end{aligned}
\end{equation}
where $\mu$ here is the natural frequency.
Note that for macroscopic model, the actions of harmonic oscillators which are viewed as matter fields coupled with the reservoir or the external electric field
is not stationay and therefore belongs to the nonequilibrium dynamic, and the corresponding kernel functions are also in a nonequilibrium form,
(see Ref.\cite{Eckstein M4}). 

The correlation matrix $\Gamma_{G}$ which obey the Gaussian disturbution is

\begin{equation}   
\begin{aligned}
\Gamma_{G}(\tau)=\langle \mathcal{Z}(\tau')\mathcal{Z}(\tau)\rangle_{G}=\frac{\delta\mathcal{Z}}{2}\langle R_{i}(\tau')R_{j}(\tau)\rangle,
\end{aligned}
\end{equation}
where $R$ is the coupling operators between the states with dissipation scenario (e.g., the reservoir),
and the evolution is $\Gamma_{G}(\tau)=e^{-\tau H_{G}}\Gamma(0)$. The coupling is fadeout in damping system through this evolution.
Then we have the action function 
\begin{equation}   
\begin{aligned}
S(\mathcal{Z})=\int^{\tau'}_{\tau} d\tau g(\mathcal{Z},\mathcal{Z}(\tau)),
\end{aligned}
\end{equation}
which contain the non-Markovian kernel $g(\mathcal{Z},\mathcal{Z}(\tau))$.
In the classical limit approximately, the harmonic motion can be described by

\begin{equation}   
\begin{aligned}
M\ddot{\mathcal{Z}}+s\dot{\mathcal{Z}}=-\frac{d}{d\mathcal{Z}}V(\mathcal{Z})+F_{n}(\tau),
\end{aligned}
\end{equation}
where $s$ is a friction parameter and $s=M\int^{\tau'}_{\tau}d\tau \mathcal{S}(\tau'-\tau)$ where $\mathcal{S}$ is the friction kernel,
$F_{n}$ is the noise force.
Since for Markovian noise which obey the Markovian evolution and can be well fitted to the master equation Eq.(33),
we then need to replace the history-independent potential term which mentioned above
by the form of Eq.(78), i.e., taking the bath coupling $R$ as the noise sourse which are real and Gaussian, 
and then it have $\langle R_{i}(t')R_{j}(t)\rangle=\delta_{ij}\delta_{t'-t}$.
Here is because that for the harmonic oscillator, using Wick theorem, the density matric can be diagonalized with a quadratic Gaussian potential (see Ref.\cite{Karrasch C}),
and then the Green's function with infinite imaginary-time becomes\cite{Caldeira A O}

\begin{equation}   
\begin{aligned}
G(\mathcal{Z}_{i},\mathcal{Z}_{j};\tau)=\int^{\mathcal{Z}(\tau')}_{\mathcal{Z}(\tau)}Ds(t){\rm exp}(-S_{{\rm eff}}(\mathcal{Z}(t))/\hbar),
\end{aligned}
\end{equation}
where the Euclidean effective action 

\begin{equation}   
\begin{aligned}
S_{{\rm eff}}(\mathcal{Z}(t))=\int^{\tau'}_{\tau}(\frac{1}{2}M\dot{\mathcal{Z}}(t)+V(\mathcal{Z}))dt-\int^{\tau'}_{\tau}dt\delta(\mathcal{Z}-\mathcal{Z}(t))+V_{0}
\end{aligned}
\end{equation}
where $V_{0}$ is the time-independent potential. In this expression, the state in next time step is only depends on the state in this time,
i.e., variables satisfy the Marcovian evolution, and more important, the contributions of noise in the imaginary axis is vanish,
that's also match the real noise sourse, so we only need to consider the noise in real part.
Then the time derivative of $\mathcal{Z}$ has the form

\begin{equation}   
\begin{aligned}
\frac{d}{dt}\mathcal{Z}=A(t)+B(t)F_{n}(t),
\end{aligned}
\end{equation}
with the $2d\times 2d$ positive definite diffusion matrix $D=BB^{T}$ which is symmetry in Wigner representation and both
$A$ and $B$ are positive and real matrix.
By the way, in this case, the quantum Fisher information matrix satisfy its saturation condition\cite{Pezzè L}.
This Markovian stochastic evolution can be expressed by the second-order Fokker-Planck equation in a stochastic description

\begin{equation}   
\begin{aligned}
\frac{\partial}{\partial t}E=\left[-\sum_{i}\frac{\partial}{\partial \mathcal{Z}}A(t)+\frac{1}{2}\sum_{ij}\frac{\partial}{\partial \mathcal{Z}_{i}}\frac{\partial}{\partial \mathcal{Z}_{j}}D_{ij}\right]E
\end{aligned}
\end{equation} 
where $E$ is the free energy of the system influenced by the noise variables.
Indeed this expression for the anharmonic case is due to the truncation which discard the asymmetry or symmetry anharmonic terms (see above). 
While in a probabilistic description, a Laplacian operator equal to the second time-derivative of non-Markovian kernel which is
negative definite is contained in the
Markovian form Fokker-Planck equation (see Ref.\cite{Bussi G}).

Now that in macroscopic system the observables are usually represented by thermal 
states directly since the error of statistical prediction is negligible\cite{Kollar M}.
We then investigate the rate of variance of the statistical prediction of observable $P_{i}$ which belong to the canonical ensemble, 
i.e., the relate to the decay rate of Liouvillean relaxation\cite{Prosen T2}.
Writting its statistical prediction as ${\rm Tr}(\rho P_{i})$ where $\rho$ is the canonical ensemble.
As we discussed above, the damp-out process is associate with the decoupling with the dissipation,
and therefore we also can define the Hamiltonian here as the damping spectrum of the observable, which classified discussion here for bosons and fermions,
i.e., decompose the $P_{i}$ into real part and imaginary part. 
Consider a bath with space $\mathbb{C}^{2d}\otimes\mathbb{C}^{2d}$,
then for bosons, the communication relation is $[b_{i},b_{j}^{\dag}]=\delta_{ij}$
and for linear bath Hamiltonian which is in a quadratic form (even sector) is $H=u^{T}H_{b}u$ where $H_{b}$ is symmetry, and for fermions $[f_{i},f_{j}^{\dag}]_{-1}=\{f_{i},f_{j}^{\dag}\}=\delta_{ij}$ with
Hamiltonian $H=w^{T}H_{f}w$ where $H_{f}$ is antisymmetry, where $u$ and $w$ are real vectors. 
Since the real part of prediction can be represented by the covariance matric\cite{Eisert J} $(\gamma_{b})_{ij}=\frac{1}{2}{\rm Tr}\rho P_{b}$ where
$P_{b}=\{u_{i},u_{j}\}$ and the imaginary part $(\gamma_{f})_{ij}=\frac{i}{2}{\rm Tr}\rho P_{f}$ where $P_{f}=[w_{i},w_{j}]$,
and here always have $\gamma_{b}\ge \sigma_{y}$.
Writting the bath matrix as $M=\sum_{i}l_{i}\otimes l_{i}^{\dag}$ with $l_{i}$ the vector with dimension $2d$ describing the bath coupling,
then we have\cite{Eisert J,Diehl S}

\begin{equation}   
\begin{aligned}
\partial_{t}\gamma=X^{T}\gamma+\gamma X-Y
\end{aligned}
\end{equation}
where for fermions $X=2{\rm Re}M$ and $Y=4{\rm Im}M$
while for bosons $X=2{\rm Im}M$ and $Y=4{\rm Re}M$,
This Sylvester matrix equation also clarify the FDR.

Through this, the materials of bulk-edge-coupling type like the topological insulators or topological superconductors with the quantum spin Hall effect,
have the full pairing gap inside the bulk and the gapless state which protected by the time-reversal invariance in the edge \cite{Chung S B}
can decoupling with the bulk part,
i.e., without dissipation at the sample boundary\cite{Liu C C} and the subspaces of edge and bulk will separate throught the long-enough time evolution
(The closing of gap is due to the effect of off-diagonal term here and often leads to the phase transition, e.g., which follow the power law decay with system size $N$
in the a spinor condensate system\cite{Zhang Z}).
For example, the chiral superconductor with $d+id'$ pairing phase\cite{Liu F} which break the time-reversal symmetry
(it's realizable by, e.g., applying a strong magnetic field\cite{Brandner K}), 
or the non-Abelian statistical in the Majorana zero model\cite{Sun H H}.
For most the bulk-edge-coupling type model which is the spinless fermions model, the time evolution is presented in the Appendix.C.

We already know that the integrable system in the homogeneous phase can only relaxes to the nonthermal steady state,
but there are some models which can't find the thermalization (e.g., can only to the generalized canonical),
like the soft-core bosons model (e.g., the Mott insulator\cite{Kollath C}), spinless fermions model, integrable Luttinger model\cite{Cazalilla M A}, etc.
This kind of model can't be effectively predicted by the form of Eq.(24).
While for the models which nearly integrable (like the Hubbard model) or nonintegrable, the expectation will relax to thermal equilibrium finally,
the resulting quasistationary state of this kinds of model is nonthermal\cite{Manmana S R}.
The final state which not be thermalized is quasisteady due to the off-diagonal contribution.
But there are still some integrable system
which have the features of thermalization for some specific variables whose final state is described by the Gibbs ensemble,
like the hard-core bosons system\cite{Manmana S R,Rigol M2},
so the integrablility is not the only criterion of the thermodynamic behavior,
the varied or conservative observables which have nonnegligible effect
and their off-digonal contribution as well as the integrability broken (broken of integrals of motion\cite{Yurovsky V A})(see Appendix.C) are also important to consider.
The required distance to the nonthermal steady state is in a infinite time average,
and the required distance away from integrable point for thermalization occur is infinitesimal\cite{Kollar M},
while for a nonintegrable system, the thermalization will gradually (``smoothly'') broken when approaches to an integrable point\cite{Rigol M3}
with a infinite time scale.

\section{Conclusion}
This work mainly investigate time evolution of quantum many-body system as well as the thermodynamics of macroscopical system with the non-Markovian processes
in the free-energy surface for which the steepest descent is used to find the minimal coupling (similar to the method of covariant derivatives).
The condition of the presence of thermalization in a relaxation process of quantum many-body system
is discussed in this work as well as the entropy and entanglement in the harmonic and anharmonic system.
The main model of our investigation is the nonisolated system and so that the degrees of freedom can be traced out from the discussed canonical ensembles
(or the microcanonical one), and thereforce the ergodic is suppressed which the detial investigation is presented in the above.
Althought the integrable system which governed by the corresponding GGE 
keeps the expectation value of observables in initial state while the chaotic one keeps the initial memory little
and it helps to understand the quenches towards the stationary state in the ordered phase or disordered phase in thermodynamic limit or scaling limit respectively,
the required numberical computation is more demanding and the eigenstate thermalization hypothesis is failure\cite{Rigol M3}.
We also obtain that,
the integrability is not only affected by the constants of motion, but some other important considerable factors
which constitute the integrability breaking term (see Appendix.C).

To investigate the approaching to Gaussian state with maximum local entropy within the relaxation,
a estimator in terms of trace norm is presented in the Setc.8 which related to the matrix method.
The open quantum system is discussed in depth in the above sections,
while for a closed quantum system which begin with a pure state with ${\rm Tr}\rho^{2}=1$ ($\rho$ is the square root of eigenvalue of the density matrix),
will never relax to the thermal state with ${\rm Tr}\rho^{2}<1$ which corresponds to $\sum_{R}^{\mathds{R}}\lambda_{R}\le 1$ which is discussed in Sect.9.
For the diagonal Hamiltonian which make the observables tend to diagonal form with the infinite time average
can be implemented by the methods like Bogoliubov transformation and a fast relaxation to diagonal ensemble
(reach a quasisteady state) required the system spectrum is nondegenerate\cite{Rigol M3} where we exclude the accidental degeneracies
of diagonal ensemble.
In this case, the eightenergies is linear like the one which mentioned in Sect.4,
and the globally observable follows the relation Eq.(126) in long-time limit.
Note that such a nondegenerate will not long-live since the irregular dispersion in boundaries or the nonlinear waveguide will generate the degeneracies.

\section{Appendix A : Deduction of $\beta$-function and the coupling in perturbed system}

The $\beta$-function can be defined as $\beta=\mu\frac{\partial}{\partial\mu}g=\frac{d}{d({\rm In}\lambda)}g$, where ${\rm ln}\lambda=\frac{1}{2}\mu^{2}$. 
When $\lambda\rightarrow +\infty$, the 
$g\rightarrow 0$\cite{Caswell W E}.
Since the bare coupling $g^{b}$ is independent of the mass, so $\frac{d}{d\mu}g^{b}=0$, according to the relation given in the Ref.\cite{McKeon D G C}

\begin{equation}   
\begin{aligned}
\mu\frac{d}{d\mu}g^{b}=(\mu\frac{\partial}{\partial \mu}+\mu\frac{d}{d\mu}g\frac{\partial}{\partial g})g^{b}.            
\end{aligned}
\end{equation}
We can deduce that $-\partial _{\mu}g^{b}=\frac{d}{d\mu}g\frac{\partial}{\partial g}g^{b}\ne 0$, 
according to the asymptotic series expansion which given in the Ref.\cite{McKeon D G C}

\begin{equation}   
\begin{aligned}
\mu\frac{d}{d\mu}g^{b}=\varepsilon g^{b}-\varepsilon g\frac{\partial}{\partial g}g^{b}+(b_{3}g^{2}+b_{5}g^{4}+b_{7}g^{6}+O(g^{8}))g\frac{\partial}{\partial g}g^{b},          
\end{aligned}
\end{equation}
when the $\varepsilon\rightarrow 0$, i.e., dimension $n\rightarrow n_{c}$, 

\begin{equation}   
\begin{aligned}
\mu\frac{d}{d\mu}g^{b}\rightarrow\mu\frac{d}{d\mu}g\frac{\partial}{\partial g}g^{b}.                     
\end{aligned}
\end{equation}

The coefficient of Eq.(7) is \cite{Caswell W E,Banks T}

\begin{equation}   
\begin{aligned}
\beta_{0}&=\frac{11}{3}C^{(2)}_{ij}-\frac{4}{3}T_{ij}\\
\beta_{1}&=\frac{34}{3}(C^{(2)}_{ij})^{2}-\frac{20}{3}C^{(2)}_{ij}T_{ij}-4C^{(2)}_{F}T_{ij}\\   
\beta_{2}&=\frac{2857}{54}(C^{(2)}_{ij})^{3}-\frac{5033}{162}C^{(2)}_{ij}T_{ij}+\frac{2925}{864}C^{(2)}_{F}T_{ij}^{2},  
\end{aligned}
\end{equation}
where $C^{(2)}_{ij}$ is the quadratic Casimir operator acting on the adjacent nodes, which equal to $N$ for SU(N) system\cite{Caswell W E},
$C^{(2)}_{F}$ is the quadratic Casimir operator acting on fermions,
and has the relation with mass as $\frac{1}{4}C^{(2)}_{ij}{\rm dim}(T_{ij})$=m\cite{Caswell W E,Banks T}, where $m$ is the number of fermion multiplets\cite{Gross D J}.
With the increase of $m$, there will be a lot of novel nature in fermion stand model which we don't discuss here, for a reference can see the Ref.\cite{Picek I}.

According to the supersymmetry SU(3) Yang-Mills theory in Ref.\cite{Elias V,Jones D R T},
the quadratic Casimir operator which have $C^{(2)}_{ij}=F^{\mu\nu}F_{\mu\nu}$ where $F^{\mu\nu}$ is the field strength tensor 
or the SU(N) generate meta (here is the group generator of SU(3)) which have the below relation with the coupling $g$

\begin{equation}   
\begin{aligned}
 \frac{\beta(g)}{g}F\tilde{F}=-\frac{11}{4}\partial_{\mu}(\psi^{\dag}(x)\gamma^{\mu}\gamma_{5}\psi(x)),
\end{aligned}
\end{equation}
where $F\tilde{F}=\varepsilon_{\mu\nu\rho\sigma}F^{\mu\nu}F^{\rho\sigma}$, $\varepsilon_{\mu\nu\rho\sigma}$ is the Levi-Civita symbol.
Note that this relation is correct for $l$-loop order where $l\ge 2$ since it's gauge-independent for $\beta(g)$ in one-loop order. It's easy to obtain that

\begin{equation}   
\begin{aligned}
F\tilde{F}&=-\frac{11}{4}\frac{g}{\beta(g)}\partial_{\mu}(\psi^{\dag}(x)\gamma^{\mu}\gamma_{5}\psi(x)),\\
C_{ij}^{(2)}&=\frac{16\pi^{2}}{g}[-\frac{8}{33}(\frac{\beta(g)}{g})^{2}-\frac{1}{3}\frac{\beta(g)}{g}],
\end{aligned}
\end{equation}
where $\frac{\beta(g)}{g}$ in SU(3) system obeys \cite{Elias V}.

\begin{equation}   
\begin{aligned}
\frac{\beta(g)}{g}=\frac{-3C_{ij}^{(2)}}{16\pi^{2}-2C^{(2)}_{ij}}=\frac{-9}{16\pi^{2}-6},
\end{aligned}
\end{equation}
here utilize the virtue of invariance of $\gamma_{5}$ as $\Lambda_{\frac{1}{2}}\gamma_{5}\Lambda_{\frac{1}{2}}^{-1}=\gamma_{5}$.

 
\section{Appendix B : The Supplement of Covariance Matrix}
Firstly we consider the Minkowski space function\cite{Rahi S J}
\begin{equation}   
\begin{aligned}
Z={\rm Tr}e^{-iHt}=\int D P_{i}e^{-iS(t)},
\end{aligned}
\end{equation}
where $A$ is the vector potential,
and the partition function $Z(\beta)=Z(-i\beta)={\rm Tr}e^{-\beta H}$.
Consider the canonical ensemble
\begin{equation}   
\begin{aligned}
\rho(\beta)=\frac{e^{-\beta H}}{Z(\beta)}.
\end{aligned}
\end{equation}
We take the Hamiltonian of components of decomposed covariance matrix $\gamma=(H_{1}\oplus H_{2})/2$,
where $H_{1}=V^{-1/2}$ and $H_{2}=V^{1/2}$ and $V$ is the potential matrix, into the blocks of $1/\beta$.
Then the free energy in entropy ensemble is\cite{Audenaert K}
\begin{equation}   
\begin{aligned}
E(\beta)&={\rm Tr}H_{2}(\beta)\\
&=\sum {\rm ln}\gamma(\beta)\\
&=\sum {\rm ln}\frac{H_{1}(\beta)\oplus H_{2}(\beta)}{2}
\end{aligned}
\end{equation}
where $H_{1}(\beta)=V^{-1/2}[{\bf I}_{d}+2({\rm exp}(\beta H_{2})-{\bf I}_{d})^{-1}]$
and $H_{2}(\beta)=V^{1/2}[{\bf I}_{d}+2({\rm exp}(\beta H_{2})-{\bf I}_{d})^{-1}]$.
Then the Eq.(75) can be represented as
\begin{equation}   
\begin{aligned}
{\rm ln}||\mathcal{J}^{\Gamma}||_{{\rm Tr}}=\sum_{i}{\rm ln}({\rm max}[1,\lambda_{i}^{-1}])\le ||\lambda_{i}^{-1}-1||_{{\rm Tr}}\le 2(e^{\beta H_{2}}-{\bf I_{d}})^{-1}
\end{aligned}
\end{equation}
here $e^{\beta H_{2}}=-\Omega^{-1/2}\gamma^{\Gamma}\sigma_{y}\gamma^{\Gamma}(-\Omega)^{1/2}$ is the blocks of $H_{2}$ and
indeed it play a key role in the coupling between the target region with the rest.
The maximal $l_{1}$-norm\cite{Yu L W} of $(e^{\beta H_{2}}-{\bf I_{d}})$ is linear bounded\cite{Plenio M B} by the size of target region,
(linear with the number of degrees of freedom of boundary of $\rho$),
but it's independent of size of the total size (contain the nontarget-region).

Then we take the equation of stochastic-description dynamics (Eq.(83)) into consider and let the $A$, $B$, $F$ be the matrices.
For quantitative analysis, we form a new potential matrix $Q=\begin{pmatrix} A&B\\ B&A \end{pmatrix}$.
Through the mathematical method, we have\cite{Audenaert K}
\begin{equation}   
\begin{aligned}
S^{-1}\begin{pmatrix} A&B\\ B&A \end{pmatrix}S=(A+BF)\oplus(A-BF)
\end{aligned}
\end{equation}
where $S=(P+F)/\sqrt{2}$ and $S^{-1}=S$, and we have $A+BF=(A-BF)^{-1}$.
The Hamiltonian which describe the conserved observable becomes 
\begin{equation}   
\begin{aligned}
H={\rm Tr}(F{\rm In}Q)={\rm Tr}({\rm In}\frac{A+BF}{A-BF}),
\end{aligned}
\end{equation}
then the determinant ${\rm det}[A+BF]={\rm exp}(-p{\rm Tr}(F{\rm In}Q))$,
where $p$ is the probability within the canonical ensemble $\rho=\sum p|\psi\rangle\langle\psi|$.
Through Jacobi's formula, we have
\begin{equation}   
\begin{aligned}
\partial_{t}{\rm det}[A+BF]&={\rm Tr}({\rm adj[A+BF]\cdot\partial_{t}[A+BF]})\\
&={\rm exp}(-p{\rm Tr}(F{\rm In}Q))\cdot\partial_{t}Q,
\end{aligned}
\end{equation}
where ${\rm adj}[\cdot]$ denotes the adjoint matrix.

\section{Appendix C : The Perturbation Theory Applied to Diagonalized Ising Chain Hamiltonian and The Discuss of Off-Diagonal Contribution Term}

We next taking the Ising chain model $H=-J\sum^{N-2}_{i=0}\sigma^{x}_{i}\sigma^{y}_{i+1}-gJ\sum^{N-2}_{i=0}\sigma^{z}_{i}$ with $g<1$
as a example to detect the effect of perturbation theory in diagonalization.
For fermion quasiparticles with quasimomentum\cite{Gebhard F} which have even parity, have even fermion number $N_{{\rm even}}$ and obey the 
antiperiodic boundary conditions $\psi(r+N)=-\psi(r)$ with essential vectors $k=\pi (2n-1)/N$ ($n$ is a integer), while
for the odd parity one which have a odd fermion number $N_{{\rm odd}}$ is obey the 
periodic boundary conditions $\psi(r+N)=\psi(r)$ with essential vectors $k'=2\pi n/N$.
Note that these two sectors can well describe the stationary phase-space probability distribution
by the WKB spectrum\cite{Polkovnikov A,Bloch I}.
Then the WJ fermions $c^{\dag}_{r}$ satisfy
\begin{equation}   
\begin{aligned}
\sigma^{+}_{r}=\frac{\sigma_{r'}^{x}+i\sigma_{r}^{y}}{2}=c^{\dag}_{r}e^{i\pi N},
\end{aligned}
\end{equation}
\begin{equation}   
\begin{aligned}
\sigma^{-}_{r}=\frac{\sigma_{r'}^{x}-i\sigma_{r}^{y}}{2}=c^{\dag}_{r}e^{i\pi N},
\end{aligned}
\end{equation}
with $r$ and $r'$ satisfy the anticommute relation $\{c_{r},c_{r'}\}=\delta_{r,r'}$ \cite{Dziarmaga J}.
We introduce the Guassian white noise to this in this model, then the conserved observebles 
follow the Guassian distribution after the quench, which with the Gaussian amplitude $\omega=1/(\delta\mathcal{Z}\sqrt{2\pi})$ (see Eq.(76)). 

For the currents which is proportional to the diagonalization\cite{Suttorp L G},
the antiperiodic boundary conditions which also called the Neveu-Schwarz sector\cite{Calabrese P} corresponds to the left current $J^{c}_{L}$,
and the periodic boundary conditions corresponds to the right current $J^{c}_{R}$, which are
\begin{equation}   
\begin{aligned}
J^{c}_{R}(k)=\sum_{k}\psi^{\dag}_{R}(k+k')\psi_{R}(k')+\psi_{R}^{\dag}(k+k')\psi^{\dag}_{R}(k')+{\rm H.c.}
\end{aligned}
\end{equation}
\begin{equation}   
\begin{aligned}
J^{c}_{L}(k')=\sum_{k'}\psi^{\dag}_{L}(k+k')\psi_{L}(k)+\psi_{L}^{\dag}(k+k')\psi^{\dag}_{L}(k)+{\rm H.c.}
\end{aligned}
\end{equation}
The lagerest current is appear in the ground state, i.e., the $J^{c}(0)$, and the
net current $J_{{\rm net}}=N_{R}-N_{L}$ which is conserved.
The observable $A$ in long-time limit has
\begin{equation}   
\begin{aligned}
\lim_{t\rightarrow\infty}\langle \psi(t)|A|\psi(t)\rangle=\lim_{t\rightarrow\infty}\frac{\langle \psi_{R}(t)|A|\psi_{R}(t)\rangle+\langle \psi_{L}(t)|A|\psi_{L}(t)\rangle}{2},
\end{aligned}
\end{equation}
and
\begin{equation}   
\begin{aligned}
\frac{\langle \psi_{R}(t)|\psi_{R}(t)\rangle}{\langle \psi_{L}(t)|\psi_{L}(t)\rangle}=1+O(e^{-nt}),
\end{aligned}
\end{equation}
where $n$ is a constant associate with the $J_{{\rm net}}$,
i.e., the wave function in the pictures of left current and right current are nearly equivalence if $J_{{\rm net}}$ is small enough.

Mapping the fermi field into the Fourier space for simplicity through the transformation
$\sigma^{z}_{r}=1-2c_{r}^{\dag}c_{r}$ and $\sigma_{r}^{x}=-\prod_{r'=0}^{r-1}(1-2c_{r'}^{\dag}c_{r'})(c_{r}+c_{r}^{\dag})$, we have
$\psi_{r}(k)=\frac{1}{\sqrt{N}}\sum_{k}\psi_{k}e^{ikr}$ for even parity, 
and 
$\psi_{r}(k')=\frac{1}{\sqrt{N}}\sum_{k'}\psi_{k'}e^{ik'r}$ for odd parity,
we obtain the quadratic Hamiltonian (but no diagonalized)
\begin{equation}
\begin{aligned}
H=2\sum_{k>0}{\bf c}^{\dag}_{k}{\bf H}_{k}{\bf c}_{k},
\end{aligned}
\end{equation}
where Nambu vector  
${\bf c}^{\dag}_{k}=\begin{pmatrix} c_{k}^{\dag}\\c_{-k}\end{pmatrix}$, 
and ${\bf H}_{k}={\bf H_{0}}+R(t,k)\sigma_{z}$ where $R(t,k)\sigma_{z}$ is the term associate to the noise and
${\bf H_{0}}$ is the Hamiltonian without the noise which is
\begin{equation}   
\begin{aligned}
{\bf H_{0}}=\begin{pmatrix} 2J(g-{\rm cos}k)&-2Ji{\rm sin}k\\ 2Ji{\rm sin}k& -2J(g-{\rm cos}k)\end{pmatrix}
\end{aligned}
\end{equation}
To make the Hamiltonian diagonal in a nonperturbative treatment,
we use the Bogoliubov transformation (rotation) to obtain the expression of Bogoliubov quasiparticles with Bogoliubov angle $\theta(k)$ 
(assuming the lattice spacing $\bar{a}=1$)
\begin{equation}   
\begin{aligned}
c(k)={\rm cos}\theta(k)c_{0}(k)+i{\rm sin}\theta(k)c_{0}^{\dag}(-k),
\end{aligned}
\end{equation}
\begin{equation}   
\begin{aligned}
c^{\dag}(k)=i{\rm sin}\theta(k)c_{0}(-k)+{\rm cos}\theta(k)c_{0}^{\dag}(k),
\end{aligned}
\end{equation}
with the gap is $\Delta=\epsilon_{0}=2J|1-g|$ which vanish in the phase transition point (quantum critical point $k_{c}$=1)
where the interactions of quasiparticle become more effective.
The excitation probability of quasiparticles becomes $\langle \psi(0)|c^{\dag}(k)c(k)|\psi(0)\rangle={\rm tan}^{2}[(\theta(k)-\theta(0))/2]$
and obey the nonthermal disturbution.
When $g\gg 1$, the ground state is strictly a paramagnetic,
while when $g\ll 1$, the ground states are two degenerate ferromagnetic.
If we ignore the noise term, the diagonalized Hamiltonian after the transformation is
\begin{equation}   
\begin{aligned}
H_{0}=2\sum_{k}\epsilon_{k}(c_{0}^{\dag}(k)c_{0}(k)-c_{0}(-k)c_{0}^{\dag}(-k)-1),
\end{aligned}
\end{equation}
where the linear dispersion $\epsilon_{k}$ which dependents on ${\bf H}_{0}$, and $\epsilon_{k}=\sqrt{|{\bf H}_{0}|}=2J\sqrt{g^{2}-2g{\rm cos}k+1}$.
This a noninteracting Hamiltonian and has the accidental degeneracies due to the periodic dispersion which has mentioned above.
This procedure is also avilable for the phonon field operators,
which the Hamiltonian can be exactly diagonalized in harmonic-oscillator\cite{Cazalilla M A,Mattis D C}.  
If we consider the noise term, the density matrix of diagonalized Hamiltonian which satisfy the master equation (Eq.(33)) can be written as
\begin{equation}   
\mathcal{J}(k)=\begin{pmatrix} c_{0}^{\dag}(k)c_{0}(k)&c_{0}^{\dag}(k)c^{\dag}_{0}(-k)\\ c_{0}(-k)c_{0}(k)&c_{0}(-k)c_{0}^{\dag}(-k) \end{pmatrix},
\end{equation}
where the two elements in the main diagonal stands for the number of levels in momentum space which is invariant under the time evolution,
and the two elements in the vice diagonal describe the coherence which will decay exponentially under time evolution and finally lead the system to
the mixed state with decoherence superposition.
For example, we denote the element $c_{0}(-k)c_{0}(k)$ as $c_{10}$, then $c_{10}(t)=e^{-\mathcal{K}t}c_{10}(0)$,
i.e., it vanish when $t\gg 1/\mathcal{K}$, this result is obey the thermal Glauber dynamics\cite{Marino J}.
So it has $\partial_{t}\mathcal{J}(k)\neq 0$.
Base on the Bogoliubov transformation introduced above,
the initial state beforce the quench can be written as\cite{Marino J2}
\begin{equation}   
\begin{aligned} 
|\psi(g_{0})\rangle=N\prod_{k,k'>0}[1+i{\rm tan}\Delta\theta\ c^{\dag}(k)c^{\dag}(-k)]|\psi(g)\rangle,
\end{aligned}
\end{equation}
where the difference of Bogoliubov angle $\Delta\theta(k)=\theta(k;g)-\theta(k;g_{0})$ for the left current regime or
$\Delta\theta(k)=\theta(k';g)-\theta(k';g_{0})$ for right current regime, and $N={\rm exp}[-\frac{1}{2}\sum_{k,k'>0}{\rm ln}(1+\Delta\theta(k))]$.
More parameterized, the difference of Bogoliubov angle $\Delta\theta(k)$ has\cite{Calabrese P}
\begin{equation}   
\begin{aligned} 
{\rm cos}\Delta\theta(k)=\frac{\epsilon_{k}^{2}(g_{0}g)}{\epsilon_{k}(g_{0})\epsilon_{k}(g)},
\end{aligned}
\end{equation}
where $\epsilon_{k}(g_{0}g)=2J\sqrt{g_{0}g-(g_{0}+g){\rm cos}k+1}$.

Since the WJ fermions is exist here, it's spinless and thereforce the thermalization can't be found in this model, which it's similar to the one mentioned in the 
Ref.\cite{Manmana S R}. For the setups of model mentioned in Sect.10 which have a damping model with damping spectrum,
the result is different with what disscussed above.
In integrable case for this Majorana fermions setup,
the Hamiltonian can be simplified as $H=-i\mathcal{P}_{f}(i\gamma_{L}+\gamma_{R})$ where $\gamma$ are the Majorana models and $\mathcal{P}_{f}$ is the hopping of nearest-neighbor fermions.
The Majorana model in the edge of sample is nonlocal and decoherence,
the total edge localized  model is 
\begin{equation}   
\begin{aligned} 
c_{M}(k)=\frac{1}{2}(i\gamma_{L}(k)+\gamma_{R}(k)),
\end{aligned}
\end{equation}
i.e., the conserved currents coupling to the Majorana models.
This combination process cost energy $2\mathcal{P}_{f}$ and form a dissipative gap with the bulk
(this gap requires that the on-site interaction $U<2\mathcal{P}_{f}$ \cite{Alicea J}).
Since the damping feature, the bulk part of density matrix (not the Eq.(111)) is decay with time evolution, and its time derivative have the same form with Eq.(85),
while the edge part is not,
i.e., the both the main diagonal and vice diagonal are decay with time exponentially,
so the final state is become a pure state ($\mathcal{J}=|\psi\rangle |\langle\psi|$) with coherence superposition 
(in a similar way to Eq.(112)).

In perturbation theory, with the variables driven by time-dependent white noise,
the correlation matrix becomes 
$\Gamma(t)=\frac{\mathcal{K}}{2}\langle R_{i}(t')R_{j}(t)\rangle=\frac{\mathcal{K}}{2}\delta_{ij}\delta_{t'-t}$,
i.e., the coupling strength $\mathcal{K}$ is associate with the dephasing effect of noise
which accelerate the relaxation in a time\ scale of order $1/\mathcal{K}$\cite{Marino J} while the diverging length scale is $1/\Delta$.
We take the approximation $H=H_{0}+gH_{1}$,
where $H_{0}=\sum_{k}\epsilon_{k}c^{\dag}(k)c(k)$ and $H_{1}=\sum_{k}\frac{\delta_{k}}{2}c^{\dag}(k)c^{\dag}(k)c(k)c(k)$
where $H_{1}$ is second quantized and $\delta_{k}$ is a nonlinear two-body interaction potential unlike the linear eigenenergy $\epsilon_{k}$.
Then we introduce the anti-Hermitian operator $s$ as $s=gs_{1}+\frac{1}{2}g^{2}s_{2}+O(g^{3})$ where $g$ is time-dependent parameter
and diagonalize the Hamiltonian through canonical transformation have been presented in the Ref.\cite{Kollar M}
\begin{equation}   
\begin{aligned}
H_{d}&=H_{0}+gH_{d}^{(1)}+g^{2}H_{d}^{(2)}+O(g^{3})\\
&=H_{0}+g(H_{1}+[s_{1},H_{0}])+g^{2}(\frac{1}{2}[s_{2},H_{0}]+[s_{1},H_{1}]+\frac{1}{2}[s_{1},[s_{1},H_{0}]])+O(g^{3})
\end{aligned}
\end{equation}
then the conserved observable $P_{i}$ have $[H_{d},P_{i}]=O(g^{3})$. In this way, the diagonalized quasiparticles are
$c^{\dag}(k,t)=e^{iH_{d}t}c^{\dag}(k)e^{-iH_{d}t}$ and $c(k,t)=e^{iH_{d}t}c(k)e^{-iH_{d}t}$.
In the range of $1/|g|\ll {\rm time\ scale}\ll 1/g^{2}$ \cite{Kollar M},
the pure state have the same expectation value with the mixed state,
i.e., the main diagonal and vice diagonal of diagonalized Hamiltonians' density matrix have the same degree of decaying.

In the case of $g^{2}\ll 1$, the $s$ can be viewed as $gs_{1}$, then since $H_{d}(t)=e^{gs_{1}}He^{-gs_{1}}$,
we obtain
\begin{equation}   
\begin{aligned}
\frac{d}{dg}H_{d}(t)=e^{gs_{1}}[s_{1},H]e^{-gs_{1}},
\end{aligned}
\end{equation}
\begin{equation}   
\begin{aligned}
\frac{d^{2}}{dg^{2}}H_{d}(t)=e^{gs_{1}}[s_{1},[s_{1},H]]e^{-gs_{1}},
\end{aligned}
\end{equation}
$$ \cdot\cdot\cdot$$,
then we further obtain
\begin{equation}   
\begin{aligned}
\frac{d}{dg}H_{d}(t)=e^{s}[\frac{s}{g},H]e^{-s},
\end{aligned}
\end{equation}
\begin{equation}   
\begin{aligned}
\frac{d^{2}}{dg^{2}}H_{d}(t)=e^{s}[\frac{s}{g},[\frac{s}{g},H]]e^{-s},
\end{aligned}
\end{equation}
$$ \cdot\cdot\cdot$$.
For a globally conserved observeble $A=\prod_{i}P_{\alpha_{i}}$,
apply $H_{d}$ to it with the GGE average, we have\cite{Kollar M2}
\begin{equation}   
\begin{aligned}
\langle A\rangle_{GGE}=\sum_{\alpha_{1}\cdot\cdot\cdot\alpha_{n}}\tilde{A}_{\alpha_{1}\cdot\cdot\cdot\alpha_{n}}\prod_{i=1}^{n}\langle P_{\alpha_{i}}\rangle_{GGE},
\end{aligned}
\end{equation}
where $\tilde{A}_{\alpha_{1}\cdot\cdot\cdot\alpha_{n}}$ is the perturbation-averaged matrix elements which is utilized to diagonalize the $P_{\alpha_{i}}$ here
and it have the property of
\begin{equation}   
\begin{aligned}
\langle A\rangle_{GGE}=\langle \prod^{n}_{i=1}P_{\alpha_{i}}\rangle_{{\rm GGE}}=\prod^{n}_{i=1}\langle P_{\alpha_{i}}\rangle_{{\rm GGE}}=
\langle \prod^{n}_{i=1}P_{\alpha_{i}}\rangle_{0}=\prod^{n}_{i=1}\langle P_{\alpha_{i}}\rangle_{0}+O(g^{3})
\end{aligned}
\end{equation}
we have\cite{Kollar M}
\begin{equation}   
\begin{aligned}
\langle A(t)\rangle&=\langle \psi(0)|e^{iHt}\mathcal{F}e^{-iHt}|\psi(0)\rangle\\
&=\langle \psi(0)|e^{-s}e^{iH_{d}t}e^{s}\mathcal{F}e^{-s}e^{-iH_{d}t}e^{s}|\psi(0)\rangle,
\end{aligned}
\end{equation}
which is diagonalized, and
with $s(t)=e^{iH_{d}t}se^{-iH_{d}t}$.
This transformation use the formula $e^{iHt}=e^{-s}e^{iH_{d}t}e^{s}$,
we define the $e^{-s}e^{iH_{d}t}e^{s}=e^{e^{R_{-s}}}e^{iH_{d}t}$ where the real linear map $R_{-s}:=ad_{-s}$\cite{Hall B},
and have\cite{Hall B2}
\begin{equation} 
\begin{aligned}
-s\cdot(iH_{d}t)=-s+\frac{R_{-s}(iH_{d}t)}{1-e^{R_{-s}}},
\end{aligned}
\end{equation}
then it's easy to obtain
\begin{equation} 
\begin{aligned}
{\rm In}(e^{-s}e^{iH_{d}t})\approx -s+\frac{s^{-1}e^{R_{-s}}s}{e^{R_{-s}}-1}.
\end{aligned}
\end{equation}

A estimator for the integrability breaking is given by the Ref.\cite{Yurovsky V A} that add the integrability broken term to the expression of observable
\begin{equation} 
\begin{aligned}
A(t)\approx\mu A_{{\rm initial}}+(1-\mu)A_{{\rm thermal}},
\end{aligned}
\end{equation}
for which the system in a completely integrable case when $\mu=1$,
the system expectation value is the same as the initial one in this case,
and it's different from the thermal expectation value of microcanonical ensemble in the completely chaotic case (nonintegrable)
which can be well described by the standard statistical mechanics\cite{Rigol M3}. The later case appear in the case $\mu\ll 1$ 
and average over the initial states 
which equal to the thermal one, as $\langle \psi(0)|A_{{\rm thermal}}|\psi(0)\rangle=\langle \psi(t)|A_{{\rm thermal}}|\psi(t)\rangle$,
and all these eightstate are within the relevant energy windows with different weights\cite{Cassidy A C}.
That allow the precise prediction for thermal state in long-time limit with the energy close to the initial one.
So the thermalization require a large number of coarse-grained observables\cite{Kollar M2}.
As predicted in the classical system by KAM theorem, it's a crossover of regular and chaotic regime\cite{Yurovsky V A},
and the achievement of thermalization require enough integrability breaking
(otherwise the ergodicity is ineffective and the thermalization is suppressed) and a long-time process ($\sim 1/g^{3}$ in our limit) 
or as a infinite time average to the diagonal ensemble and fluctuate around it in the latter time\cite{Rigol M3}, which shown as
(not consider the possible degeneracies here)
\begin{equation} 
\begin{aligned}
\langle A(t)\rangle=&\lim_{t\rightarrow\infty}\frac{1}{t}\int^{t}_{0}dt{\rm Tr}(A\rho(t))=\langle \psi(t)|A|\psi(t)\rangle_{{\rm diag}}\\
=&\sum_{\alpha}|\langle \alpha|\psi(0)\rangle|^{2}\langle \alpha|A|\alpha\rangle,
\end{aligned}
\end{equation}
where $|\alpha\rangle=\sum_{b}[(|b\rangle\langle b|gH_{1}|\alpha\rangle)/(E_{\alpha}-E_{b})]$.
This equation gives the long-time average, and keeps the diagonal term only.
This long-time average will equal to the GGE expectation value or not which dominated by the conserved $P_{i}$.
For Eq.(122), when the state $\rho$ which can be described by the Hamiltonian $H=H_{0}+gH_{1}$ is nondiagonal while the observable $A$ is diagonal
(i.e., $[A,H_{d}]=0$), it becomes\cite{Kollar M}
\begin{equation} 
\begin{aligned}
\langle A(t)\rangle&=-\langle\psi(0)|(s(t)-s)A(s(t)-s)|\psi(0)\rangle +O(g^{3})\\
&=-2(\langle \psi(0)|sAs|\psi(0)\rangle-{\rm Re}\langle \psi(0)|sAs(t)|\psi(0)\rangle),
\end{aligned}
\end{equation}
where the term $-{\rm Re}\langle \psi(0)|sAs(t)|\psi(0)\rangle$ is due to the off-diagonal contribution as
\begin{equation} 
\begin{aligned}
-{\rm Re}\langle a|sAs(t)|a\rangle={\rm Re}\sum_{b}\frac{|\langle a|gH_{1}|b\rangle|}{(E_{a}-E_{b})^{2}}\langle b|A|b\rangle e^{-i(E_{a}-E_{b})t}+O(g^{3}),
\end{aligned}
\end{equation}
where we simplify the initial state $\psi(0)$ as $a$ and the quenched state $\psi(t)$ ($t>0$) as $b$.
But in the case of both $\rho$ and $A$ are off-diagonal, this off-diagonal contribution term becoms
\begin{equation} 
\begin{aligned}
-2{\rm Re}\sum_{b}\frac{(|\langle a|gH_{1}|a\rangle|-|\langle a|gH_{1}|b\rangle|-|\langle b|gH_{1}|b\rangle|)^{2}}{(E_{a}-E_{b})^{2}}\langle a|A|b\rangle e^{-i(E_{a}-E_{b})t}+O(g^{3}).
\end{aligned}
\end{equation}
While the diagonalized state is
\begin{equation} 
\begin{aligned}
\rho_{{\rm diag}}(|b\rangle)=\sum_{a}P_{a}\rho_{0}P_{a},
\end{aligned}
\end{equation}
where the prejector $P_{a}=|a\rangle\langle a|$ which project onto the subspace of initial state $|a\rangle$.

\end{large}
\renewcommand\refname{References}
\bibliographystyle{unsrt}

\begin{thebibliography}{unsrt}
\bibitem{92}
For quantum chaotic non-integrable system, the Kolmogorov entropy $S_{k}=1.1$ and it's indenpendent of $n$ and error\cite{Prosen T}.
Since the Kolmogorov entropy can be written as 
\begin{equation}   
\begin{aligned}
S_{k}=-\sum_{i+1}\lambda_{i+1}{\rm ln}\lambda_{i+1}+\sum_{i}\lambda_{i}{\rm ln}\lambda_{i}
\end{aligned}
\end{equation}
So we can obtain that for the non-integrable chaotic system, the eightvalue $\lambda_{i}=1.82632$ with $i=0,1,\cdot\cdot\cdot,n-1$ here.
\bibitem{35}The quadratic Casimir operator $C^{(2)}_{ij}$ here is the first-order one, i.e., the bilinear relation and fit for the two-body model.
For the three- or more-body model we need to introduce the second Casimir operator $C^{(3)}_{ijk}=\gamma_{abc}F^{a}F^{b}F^{c}$ (see, for example, Ref.\cite{Dmitra?inovi? V})
\bibitem{129}For finite dimension symplectic space, the searching of orthogonal set of state $|s_{i}\rangle$ 
often use the Darboux basis $\omega_{n}=|\partial_{n} |S_{i}\rangle +|s_{i}\rangle \langle\partial_{n} s_{i}|s_{i}\rangle$
which have $\langle s_{i}|\omega_{n} \rangle=\partial_{n}\langle s_{i}|s_{i}\rangle=0$ in the Gram-Schmidt process which presented in the Ref.\cite{Pezzè L}.
\end{thebibliography}

\clearpage
\section{Figure captions}
\begin{large}
Fig.1:(Color online)$\beta(g)$ as a function of $g$ in SU(3) system (i.e. $C^{(2)}_{ij}=3$ (see Appendix.A) with the number of fermion multiplets 
$m=0,1,2,3,5,8,10,15,20$, i.e., the 0-plet, 1-plet,$\cdot\cdot\cdot$, 20-plet fermion configration.


Fig.2:(Color online)
     (a) Energy difference between the excited state and initial state as a function of staggered magnetic field $h_{s}$ for different dimension of matrices.
(b) Probability of excitation $P_{ex}$ as a function of temperature for different dimension.

Fig.3:(Color online)(a)Inverse spin correlation length (square) and spin correlation (triangle) for $S=1$ Ising spin chain at different site $i$.
(b)Inverse spin correlation length and spin correlation for $S=1/2$ Heisenberg spin chain at different site $i$ for different $J_{2}$.
The $J_{1}$ here is setted as 0.7.

Fig.4:(Color online)Spin correlation for $S=1/2$ spin chain as a function of temperature for different next-nearest neighbor coupling $J_{2}$.

Fig.5:Spin correlation for $S=1$ Ising spin chain and $S=1/2$ Heisenberg chain as a function of temperature.

Fig.6:(left) Spin correlation as a function of temperature by the method of Bethe ansantz; 
(right) Comparison of the results of spin correlation under low temperature between 
Bethe ansatz and renormalization group (RG).

Fig.7:(Color online)Graph of Eq.(47) with phase $\alpha=1,2,3$. It's obviously to see that the contours is bounded by a power function.

Fig.8:The graphs of $\langle n^{2}(t)\rangle$ as a function of $t$ (Eq.(49)). The reduced coupling 
$g_{{\rm red}}=0.01,0.1,1,1.5,2$ from (a) to (e), respectively.

Fig.9:The large time behavior of $\langle n^{2}(t)\rangle$ with coupling $g_{{\rm red}}=0.1,0.05,0.01,0.001$ from left to right (Eq.50).


Fig.10:(Color online) Double occupation for half-filling Mott insulator $d_{hf}(t)$ quenches from $U=0$ to $U$ (a) and from $\infty$ to $U$ (b).
The insets show the enlarged views of the $d_{hf}(t)$ for quenches to $U=1$.

Fig.11:Double occupation for quenches from 0 to critical value $U_{c}$ (a) and from $\infty$ to $U_{c}$. 
The top insets show the enlarge views on short-time scale, and the bottom insets show the large-time behavior in more detial.

Fig.12:Kinetic energy of $1/r$ Hubbard chain as a function of $U$ with different $n$ and bandwidth $W=1/2,1,4$.
The bandwidth-dependent hopping constants of $1/r$ Hubbard chain as a function of distance is shown in the inset.

Fig.13:(left)The dispersion relation in $k$ space with different regulatory paramater (0 to 1 from bottom of graph to the upper);
(right)The upper, lowest, and ground state energy in a same space according to
Ref.\cite{Xiang T}.

Fig.14:(Color online) Correlations
as a function of distance $r$ for different quench of dispersion relation are shown,
the curves with different colors
from outside to inside corresponds to $\kappa_{1}\ {\rm to}\ \kappa_{2}$,$\kappa_{2}\ {\rm to}\ \kappa_{3}$,$\kappa_{4}\ {\rm to}\ \kappa_{5}$,
$\kappa_{6}\ {\rm to}\ \kappa_{7}$,
and
$\kappa_{7}\ {\rm to}\ \kappa_{8}$, respectively.
The dispersion relations are
$\kappa_{1}=0.191820018,
\kappa_{2}=0.331662479,
\kappa_{3}=0.45825757,
\kappa_{4}=0.5,
\kappa_{5}=0.619656837,
\kappa_{6}=0.866025404,
\kappa_{7}=1.118033989,
\kappa_{8}=1.322875656.$

\end{large}

\clearpage

\section{Tables}

Table.A:
\begin{table}[!hbp]
\centering
\resizebox{\textwidth}{!}{
\begin{threeparttable}
\begin{spacing}{1.19}
\begin{tabular}{|c|c|c|c|}
\hline
Model                &  Time scale of relaxation                   & Period of collapse and revival    &    Ref(s).               \\
\hline    
Falicov-Kimball      &  $\hbar$/bandwidth                          & $h/U$\tnote{\S}         &     \cite{Eckstein M} \\

Bose-Hubbard         &  1/$\mathcal{P}$                            & $h/U$         &     \cite{Kollath C}  \\   

Spin glasses         &  macroscopical and with a very broad range  &  -            & \cite{Lundgren L},\cite{Binder K} \\

Tomonaga-Luttinger   &  2$\sim$3 orders of time                    &$h/J$ ($J$ is the coupling of nearest-neighbor)          &     \cite{Karrasch C}\tnote{*},\cite{Manmana S R} \\   

Hubbard              & $\rho_{F}^{-1}U^{-2}\sim\rho_{F}^{-3}U^{-4}$\tnote{\dag}  &    -   & \cite{Moeckel M}\\  

One-dimension hard core bosons  & $1/\mathcal{P}_{{\rm f}}$\tnote{\ddag}   &$h/U$&\cite{Rigol M3},\cite{Manmana S R}\\
\hline       
\end{tabular}
\end{spacing}
\begin{tablenotes}
        \footnotesize
        \item{*} Here taking the decaying of time derivative of initial Hamiltonian as the criterion of relaxation.
         \item{\S} $h$ is the Planck constant and $U$ is the strength of nearest interaction (The belows are also follow this).
         \item{\dag} $\rho_{F}$ is the density of states at the Fermi level.
\item{\ddag} $\mathcal{P}_{{\rm f}}$ is the hopping of finial state after quench.
\end{tablenotes}
\end{threeparttable}}
\end{table}
\clearpage
\section{Figures}
Fig.1
\begin{figure}[!ht]
   \centering
   \begin{center}
     \includegraphics*[width=0.8\linewidth]{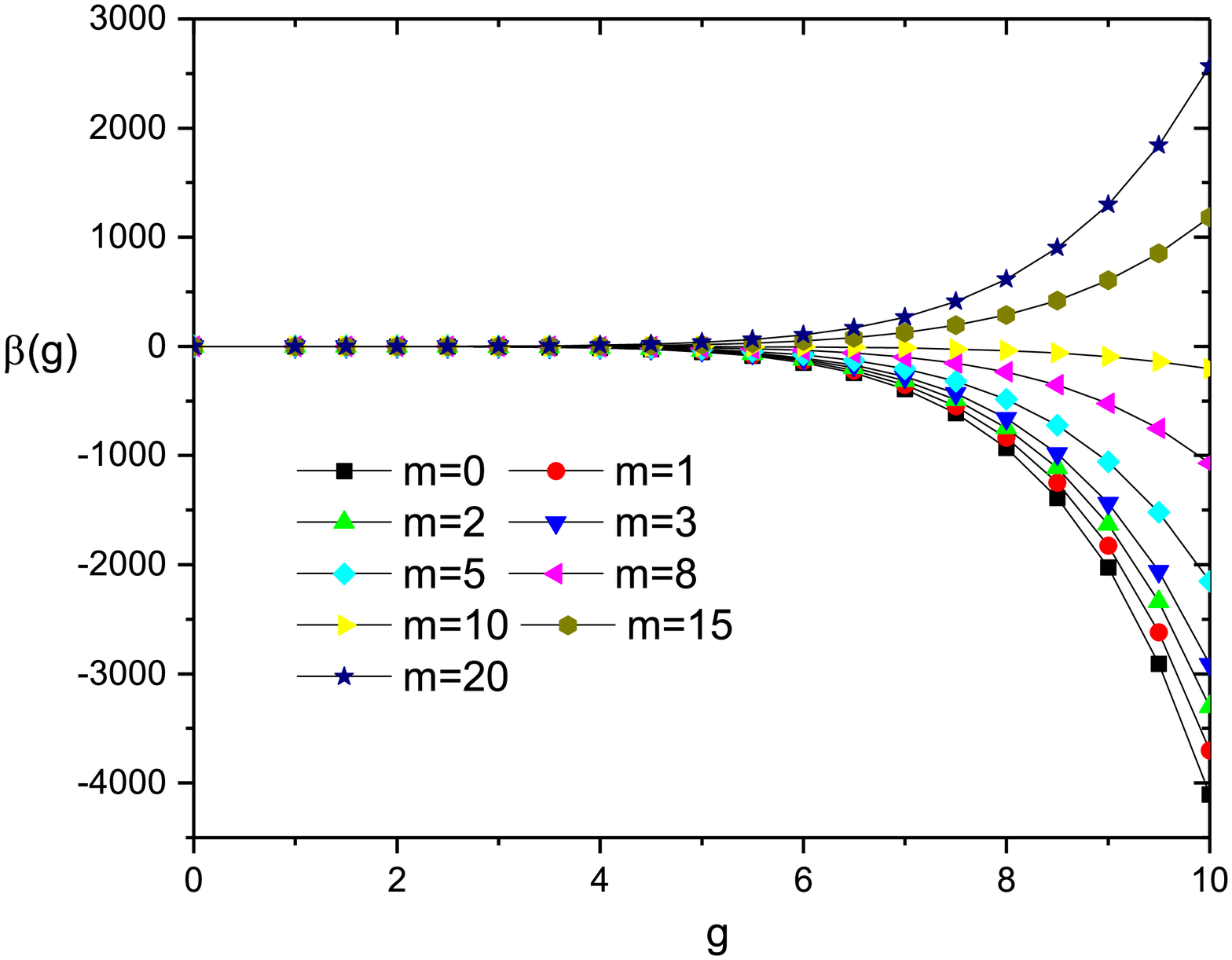}
   \end{center}
\end{figure}
\clearpage
Fig.2
\begin{figure}[!ht]
\begin{minipage}[t]{0.3\textwidth}
\centering
\includegraphics[width=1.5\linewidth]{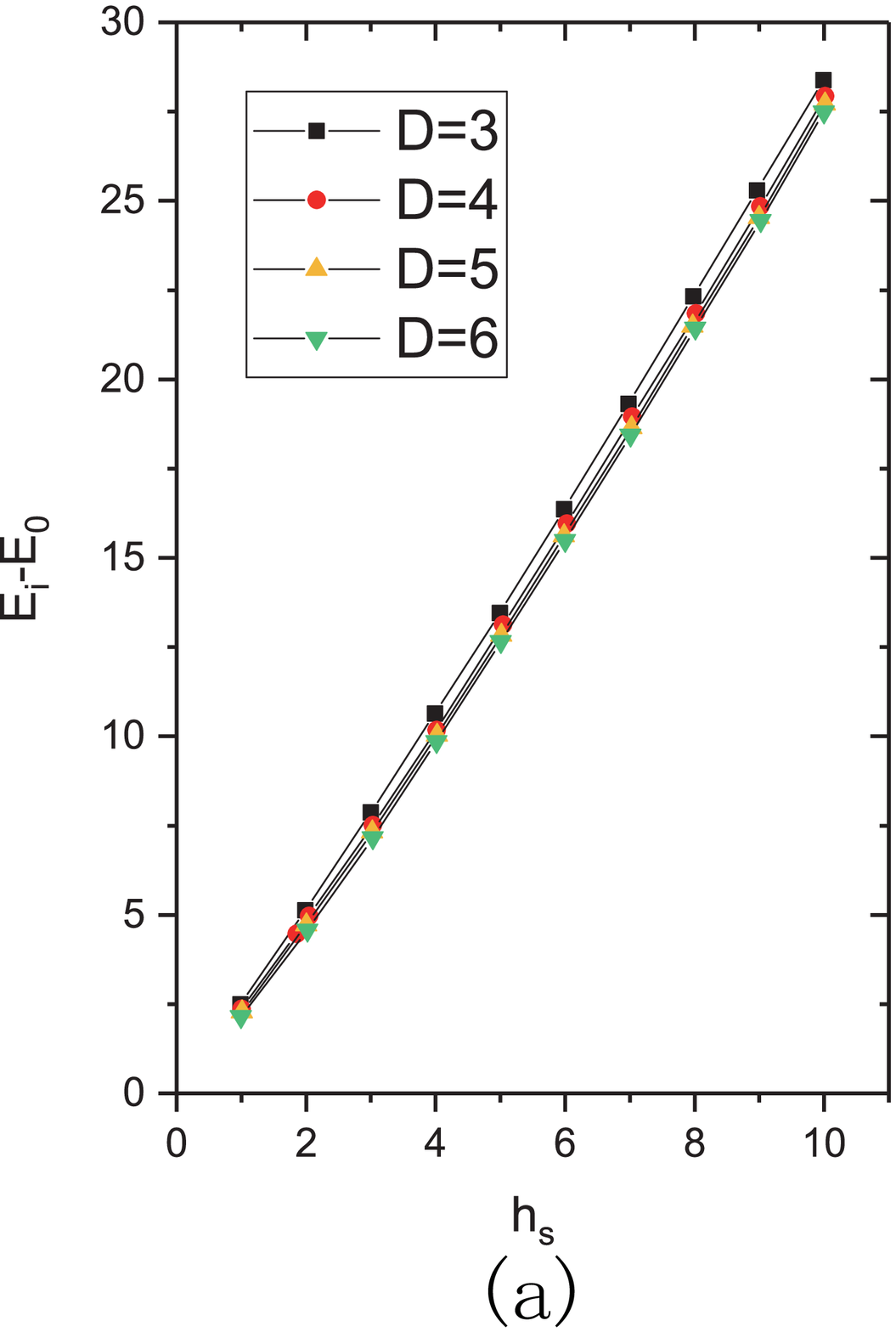}
\label{fig:side:a}
\end{minipage}\\
\begin{minipage}[t]{0.3\textwidth}
\centering
\includegraphics[width=1.5\linewidth]{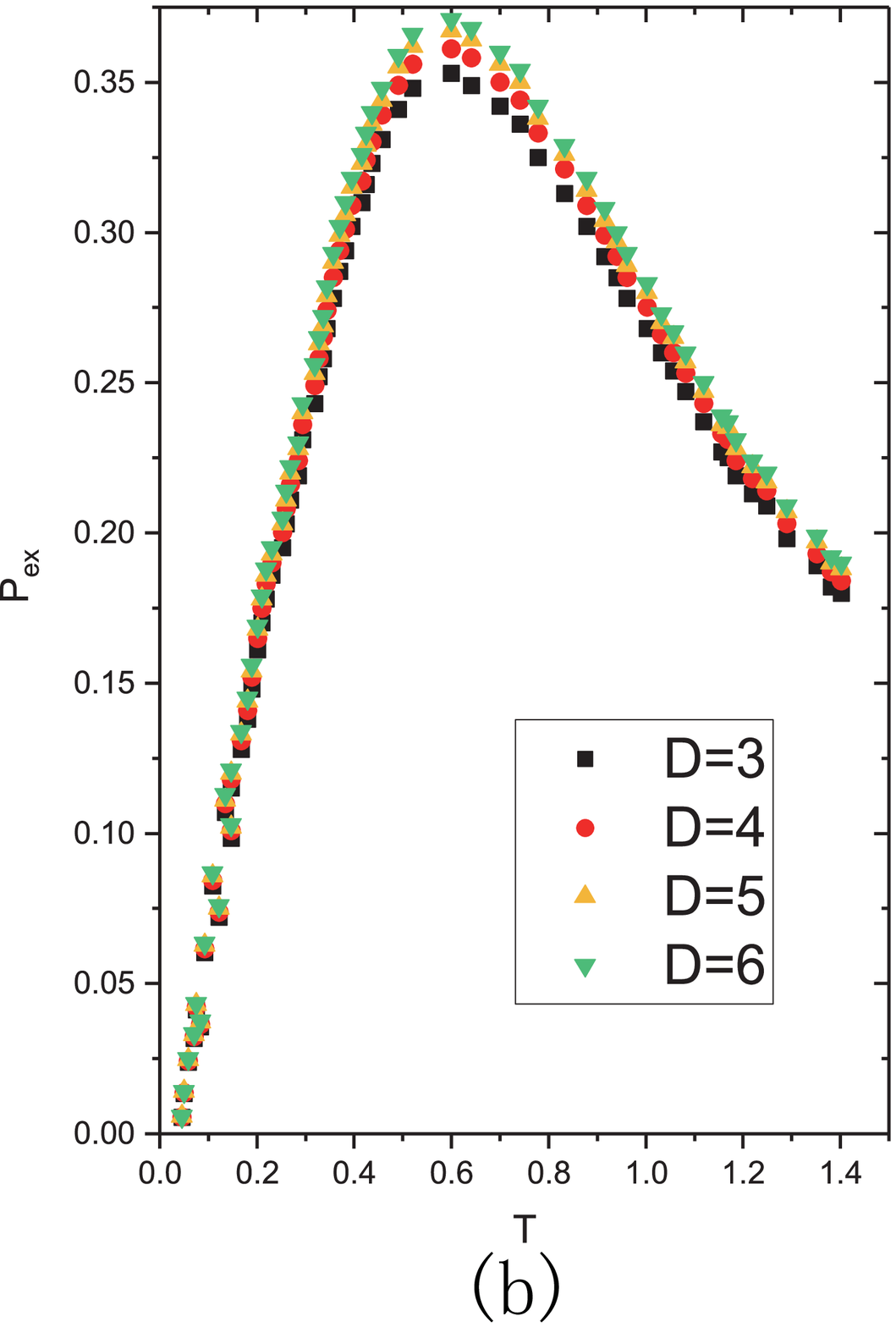}
\label{fig:side:b}
\end{minipage}
\end{figure}
\clearpage
Fig.3
\begin{figure}[!ht]
\begin{minipage}[t]{0.4\textwidth}
\centering
\includegraphics[width=1.5\linewidth]{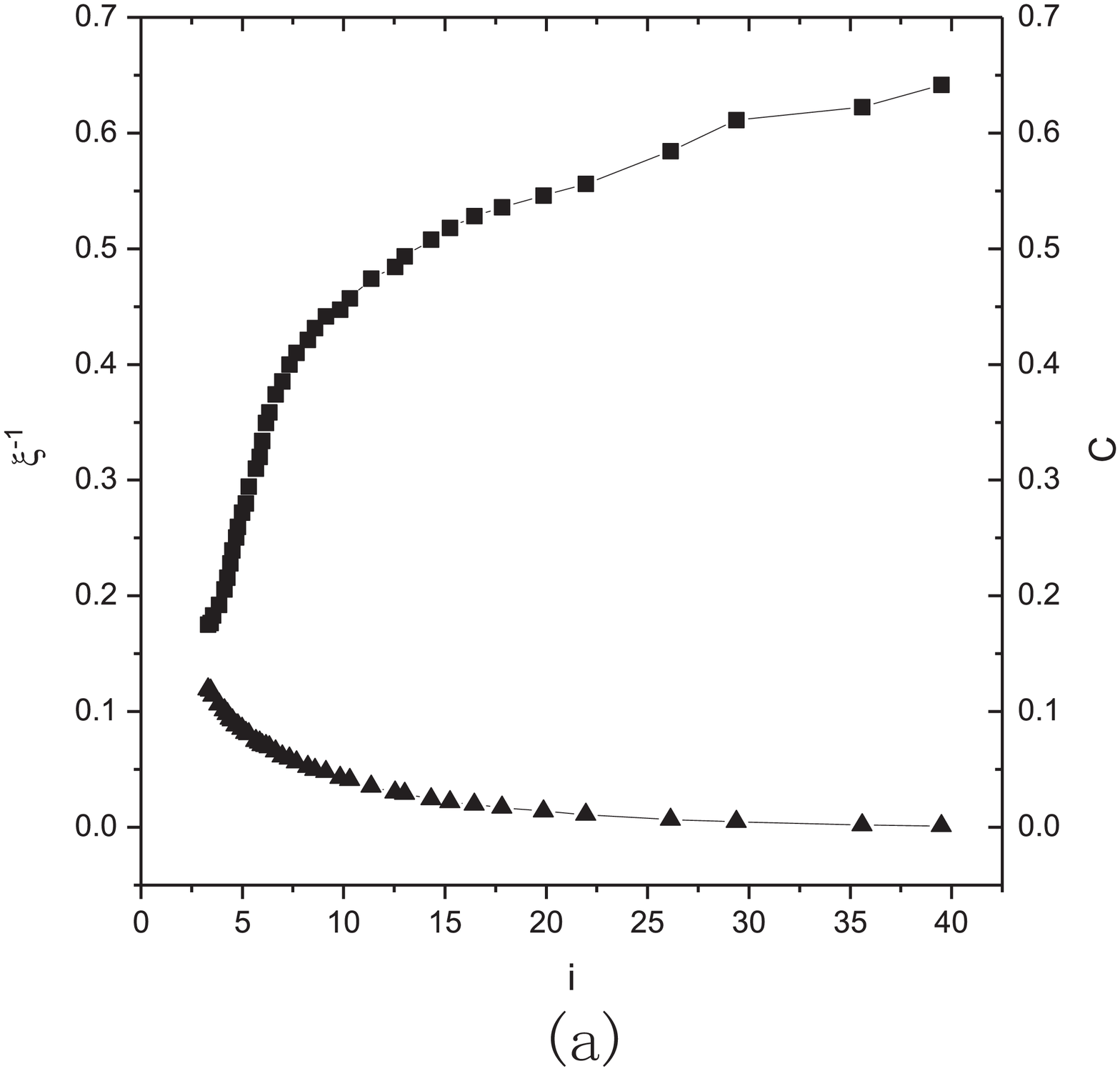}
\label{fig:side:a}
\end{minipage}\\
\begin{minipage}[t]{0.4\textwidth}
\centering
\includegraphics[width=1.5\linewidth]{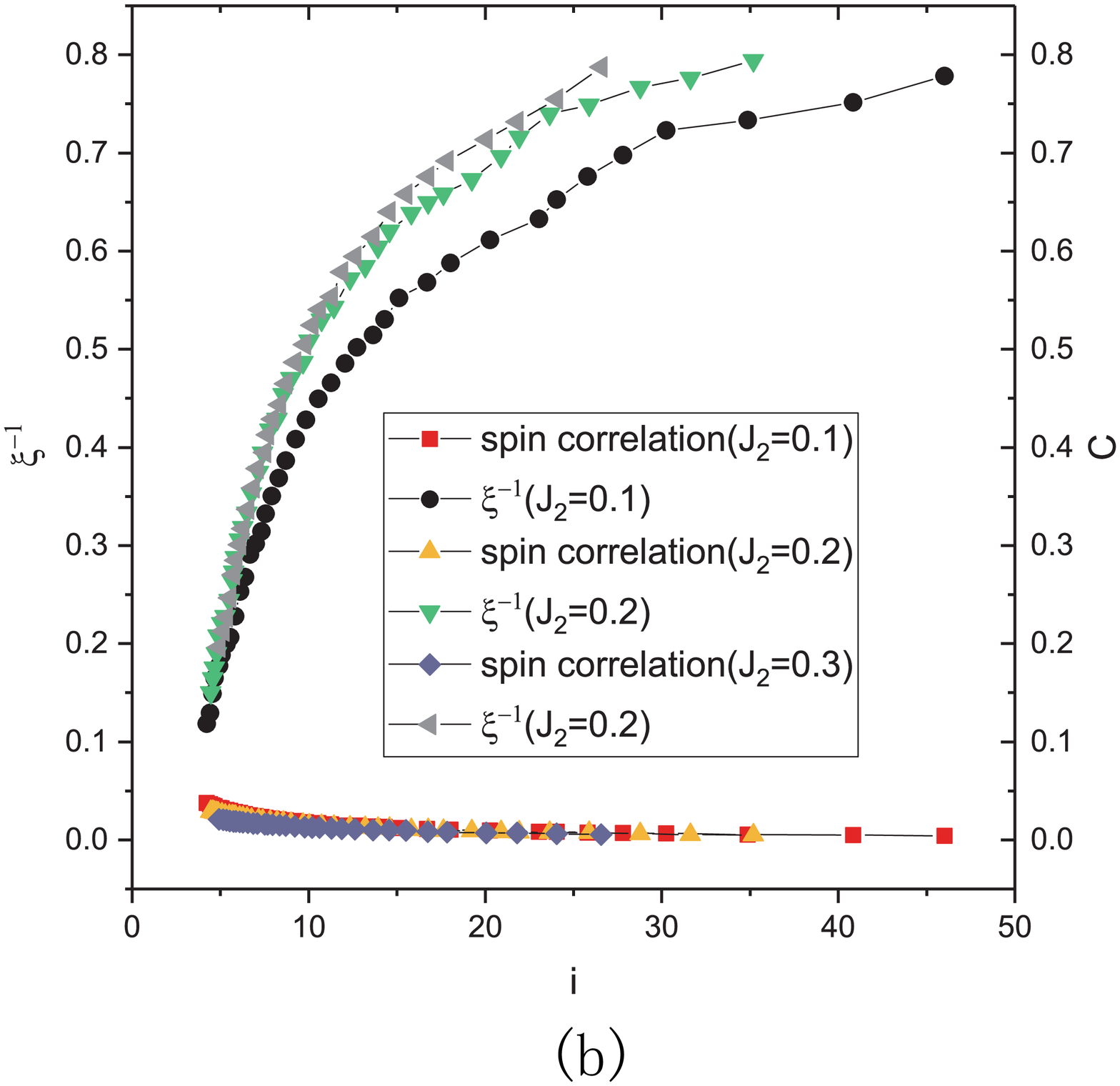}
\label{fig:side:b}
\end{minipage}
\end{figure}
\clearpage

Fig.4
\begin{figure}[!ht]
   \centering
   \begin{center}
     \includegraphics*[width=0.8\linewidth]{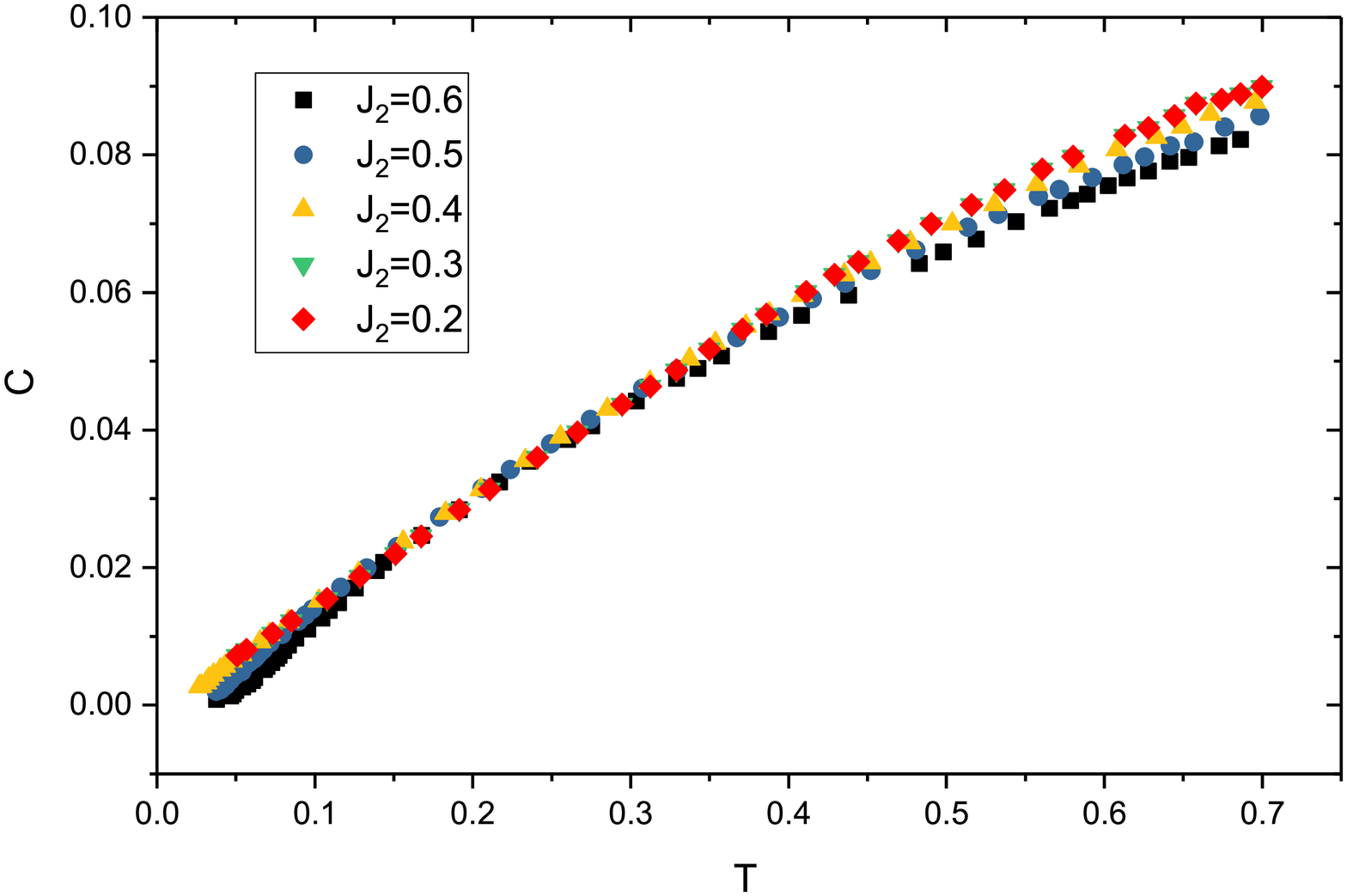}
   \end{center}
\end{figure}

Fig.5
\begin{figure}[!ht]
   \centering
   \begin{center}
     \includegraphics*[width=0.8\linewidth]{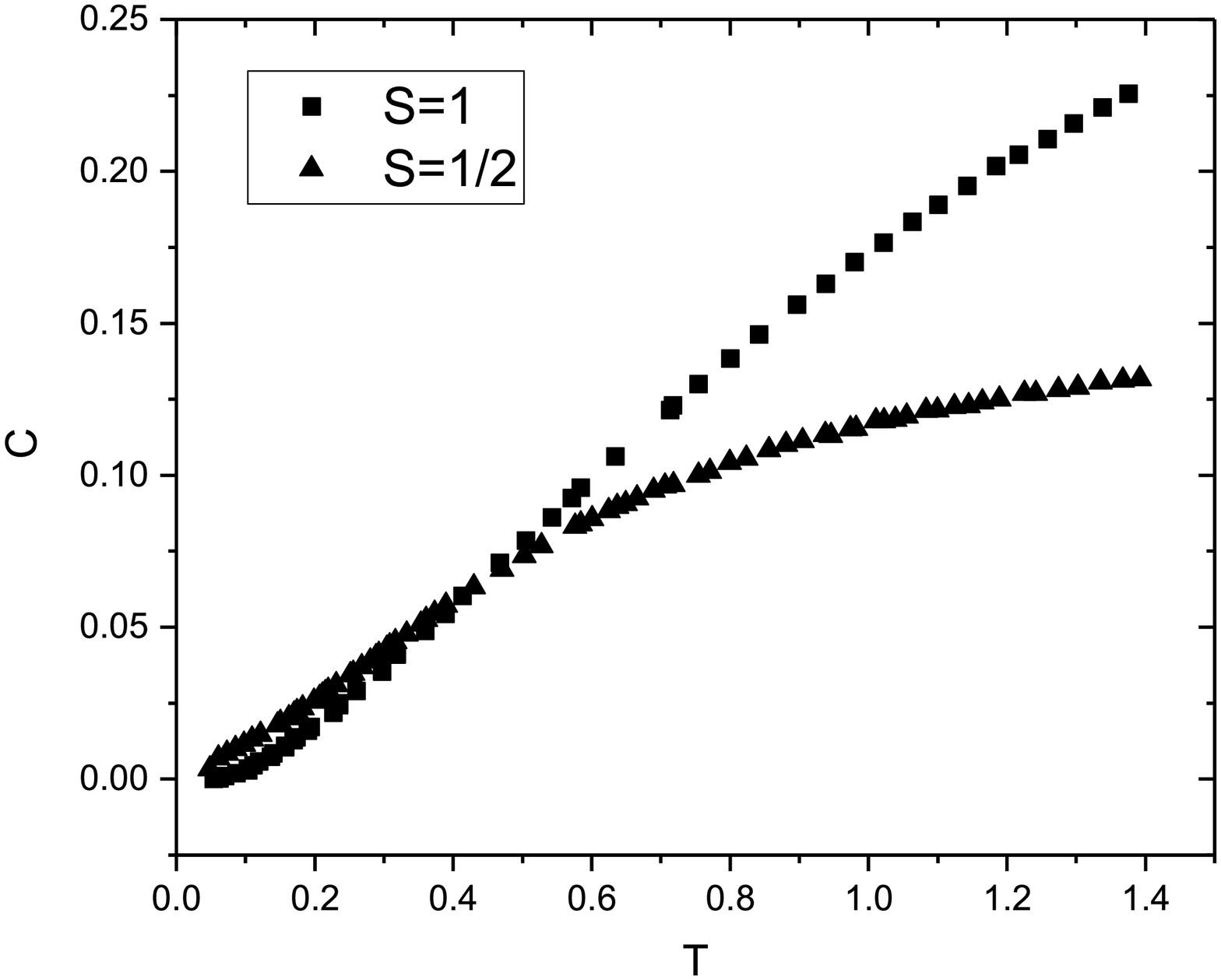}
   \end{center}
\end{figure}
\clearpage
Fig.6
\begin{figure}[!ht]
\begin{minipage}[t]{0.5\textwidth}
\centering
\includegraphics[width=1\linewidth]{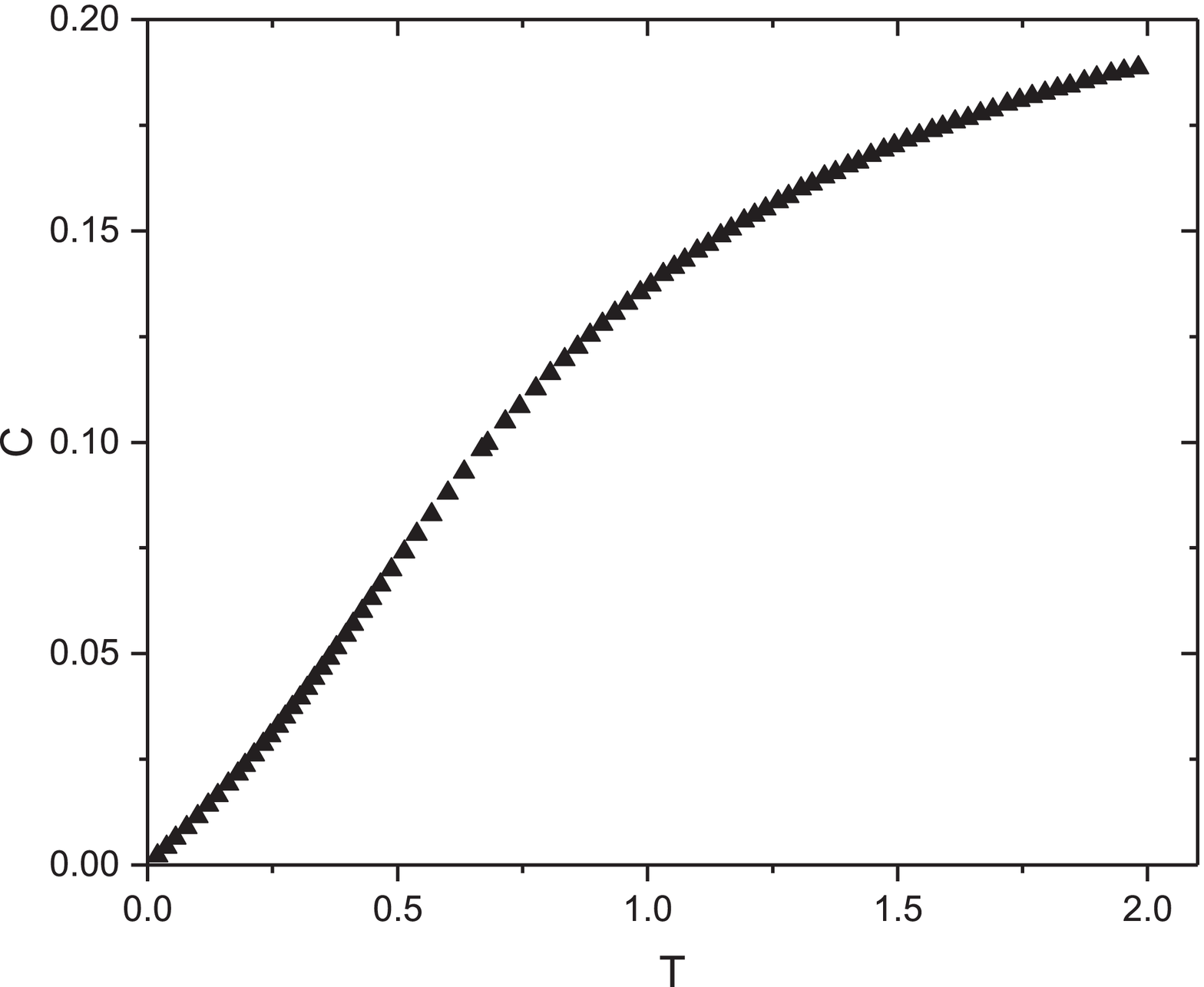}
\label{fig:side:a}
\end{minipage}
\begin{minipage}[t]{0.5\textwidth}
\centering
\includegraphics[width=1\linewidth]{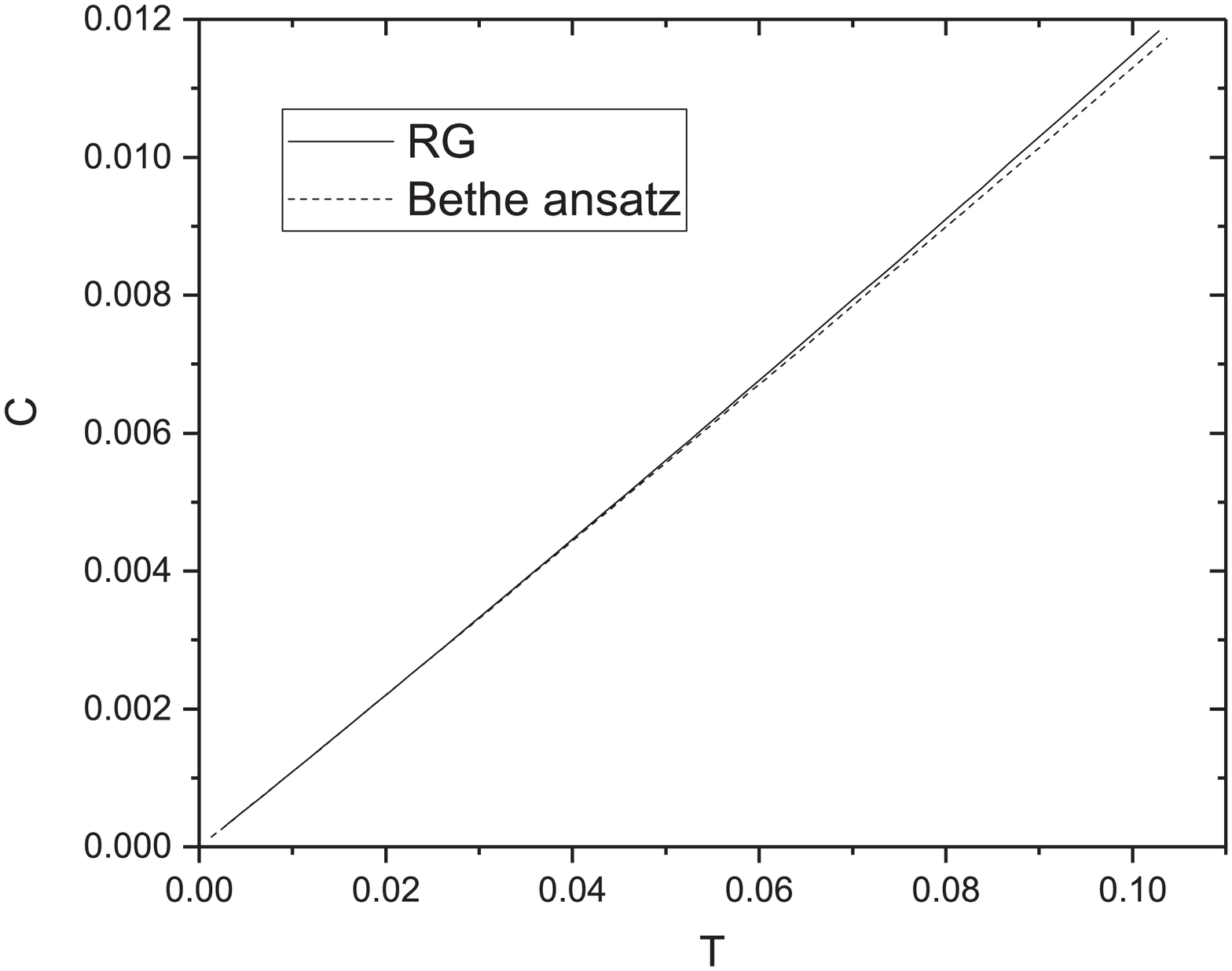}
\label{fig:side:b}
\end{minipage}
\end{figure}

Fig.7
\begin{figure}[!ht]
   \centering
   \begin{center}
     \includegraphics*[width=0.8\linewidth]{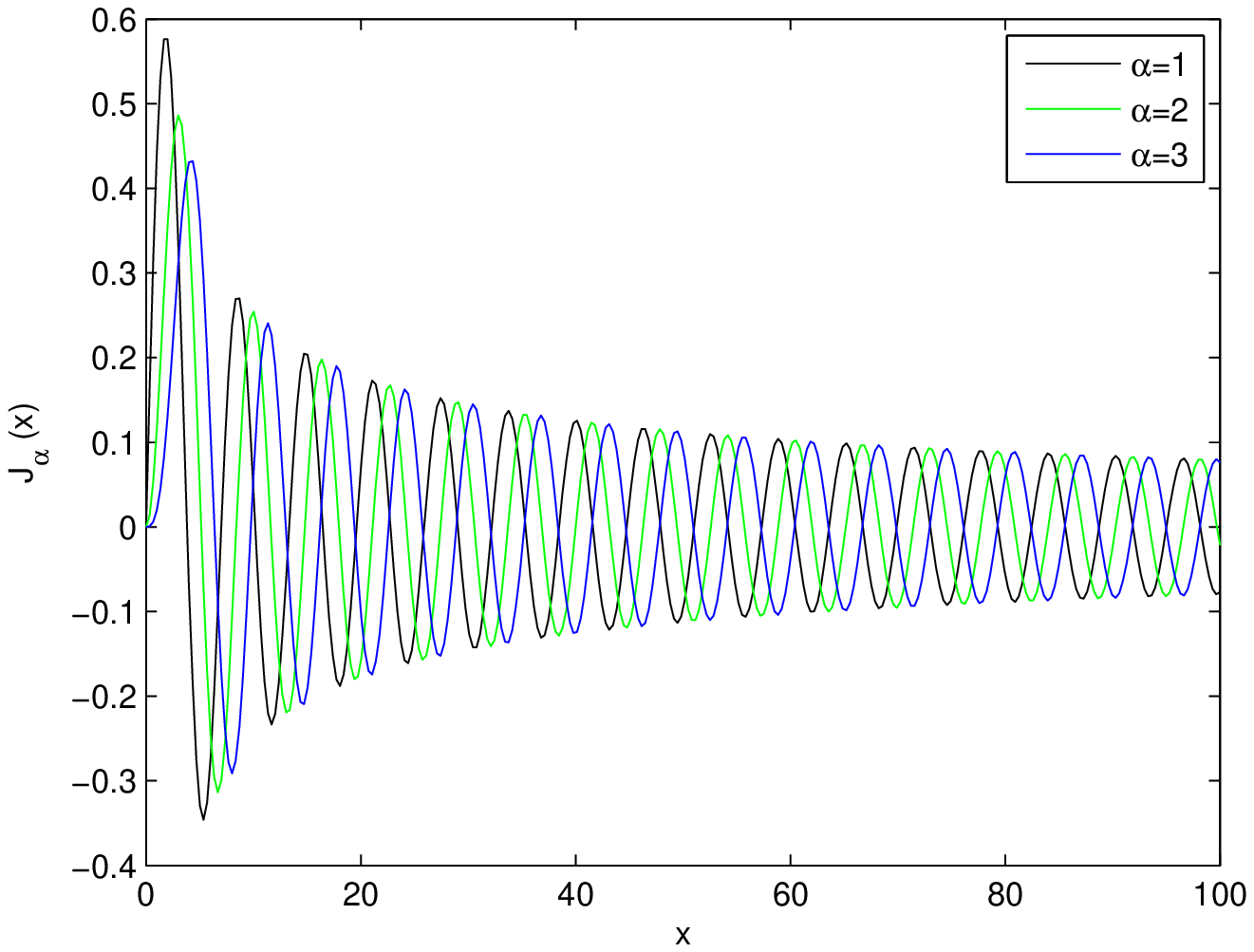}
   \end{center}
\end{figure}
\clearpage

Fig.8
\begin{figure}[!ht]
   \centering
     \subfigure[]{\includegraphics[width=0.2\linewidth]{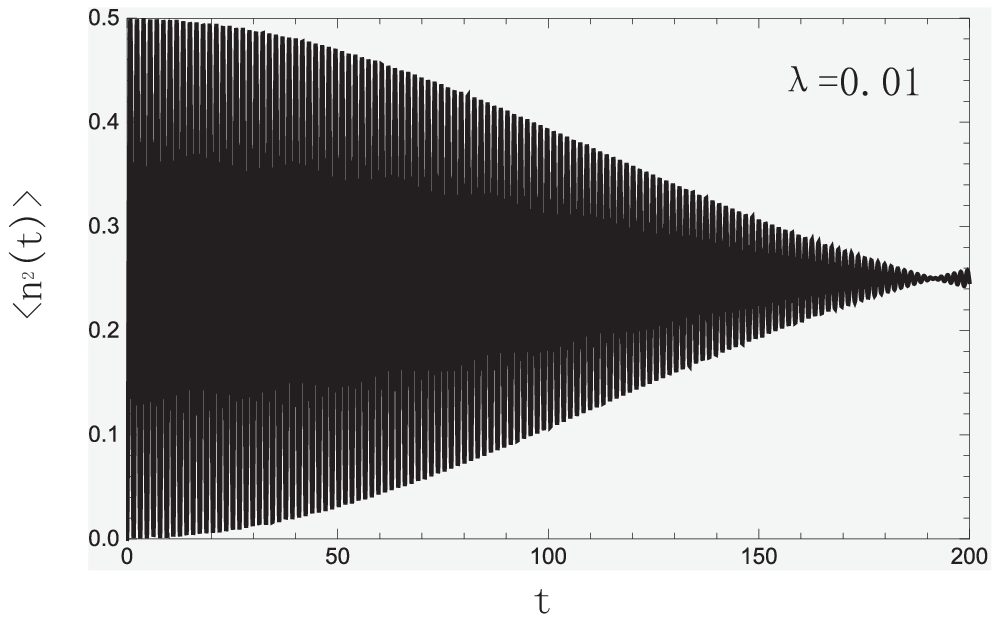}}
     \subfigure[]{\includegraphics[width=0.2\linewidth]{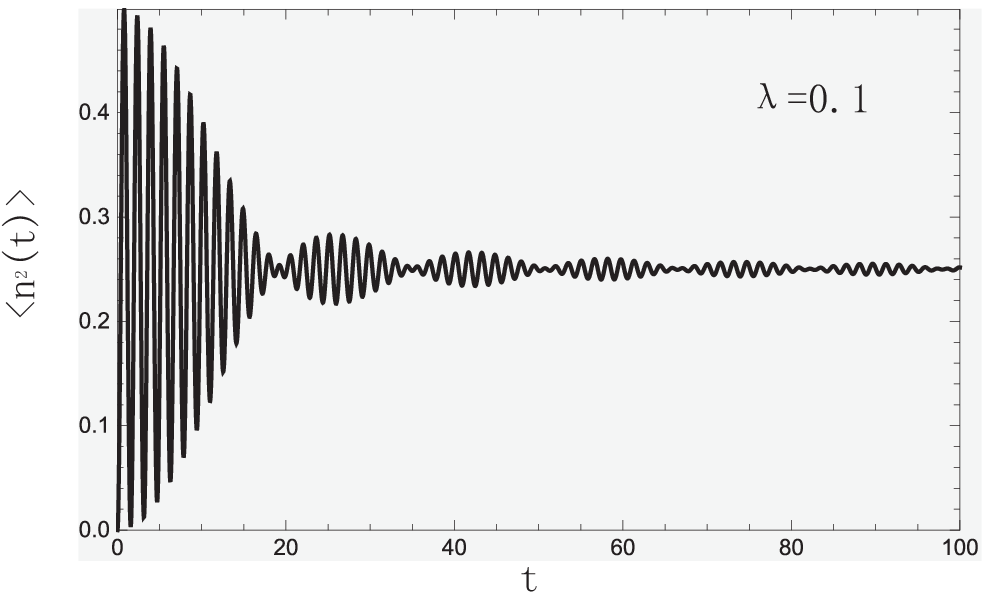}}
     \subfigure[]{\includegraphics[width=0.2\linewidth]{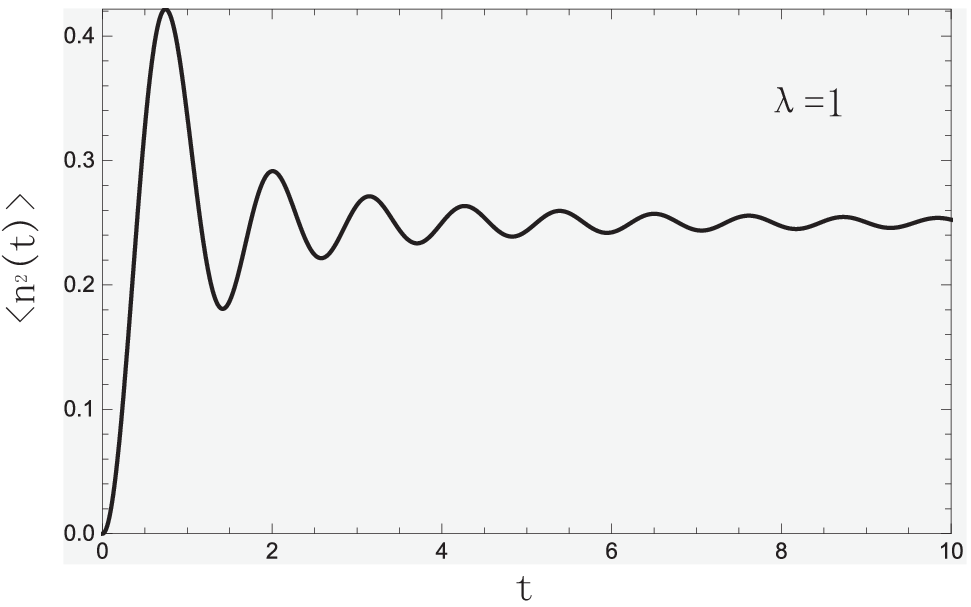}}
     \subfigure[]{\includegraphics[width=0.2\linewidth]{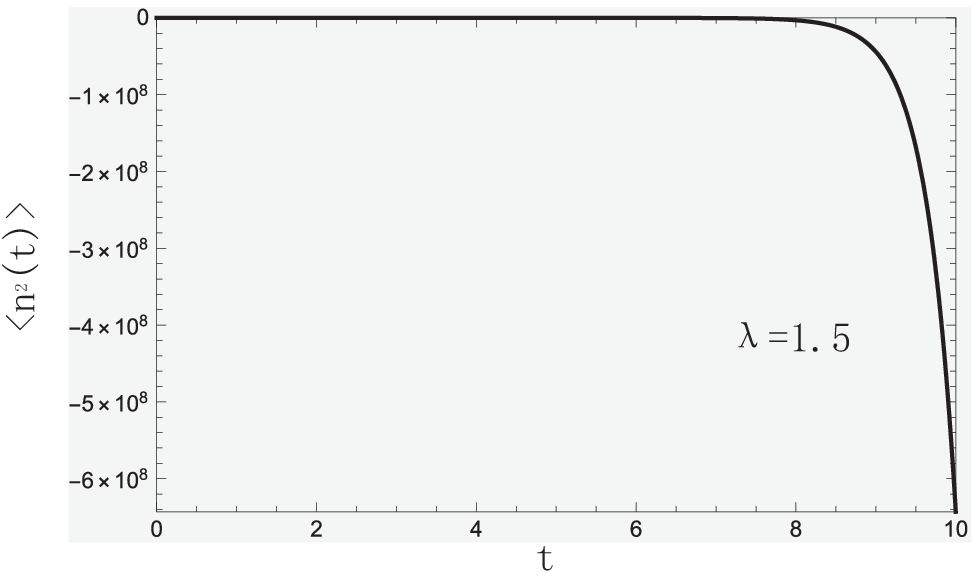}}
     \subfigure[]{\includegraphics[width=0.2\linewidth]{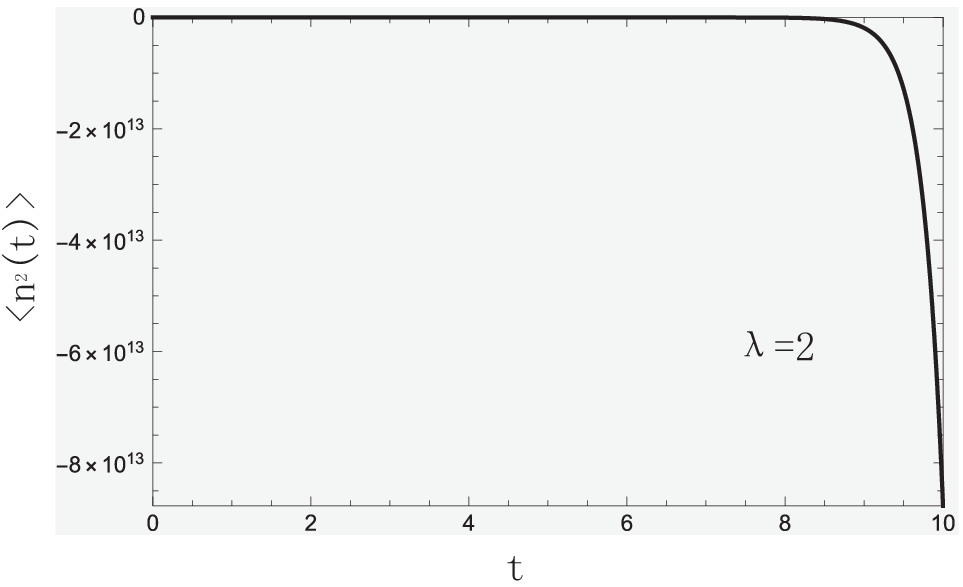}}
\end{figure}

Fig.9
\begin{figure}[!ht]
\begin{minipage}[t]{0.2\textwidth}
\centering
\includegraphics[width=1\linewidth]{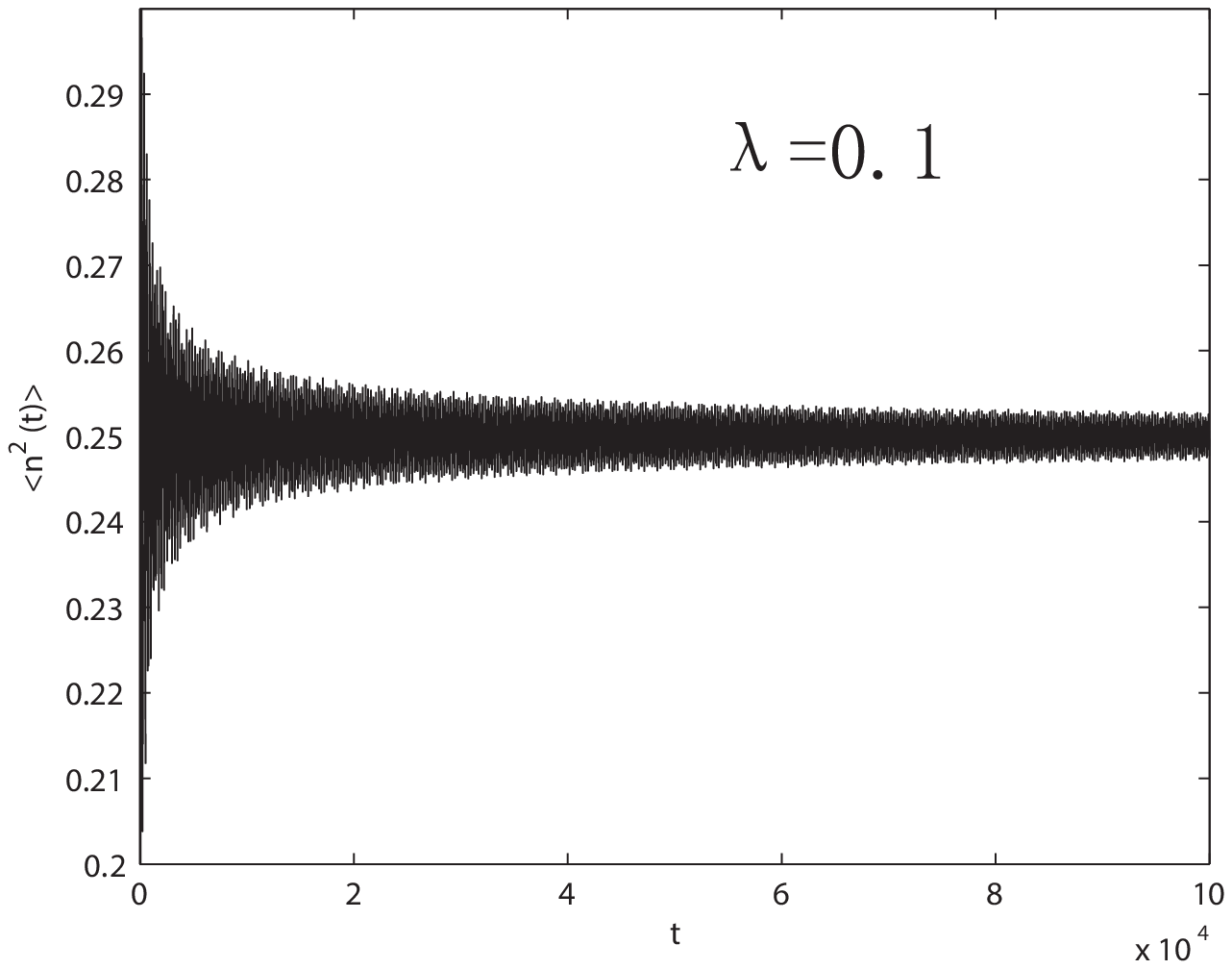}
\label{fig:side:a}
\end{minipage}
\begin{minipage}[t]{0.2\textwidth}
\centering
\includegraphics[width=1\linewidth]{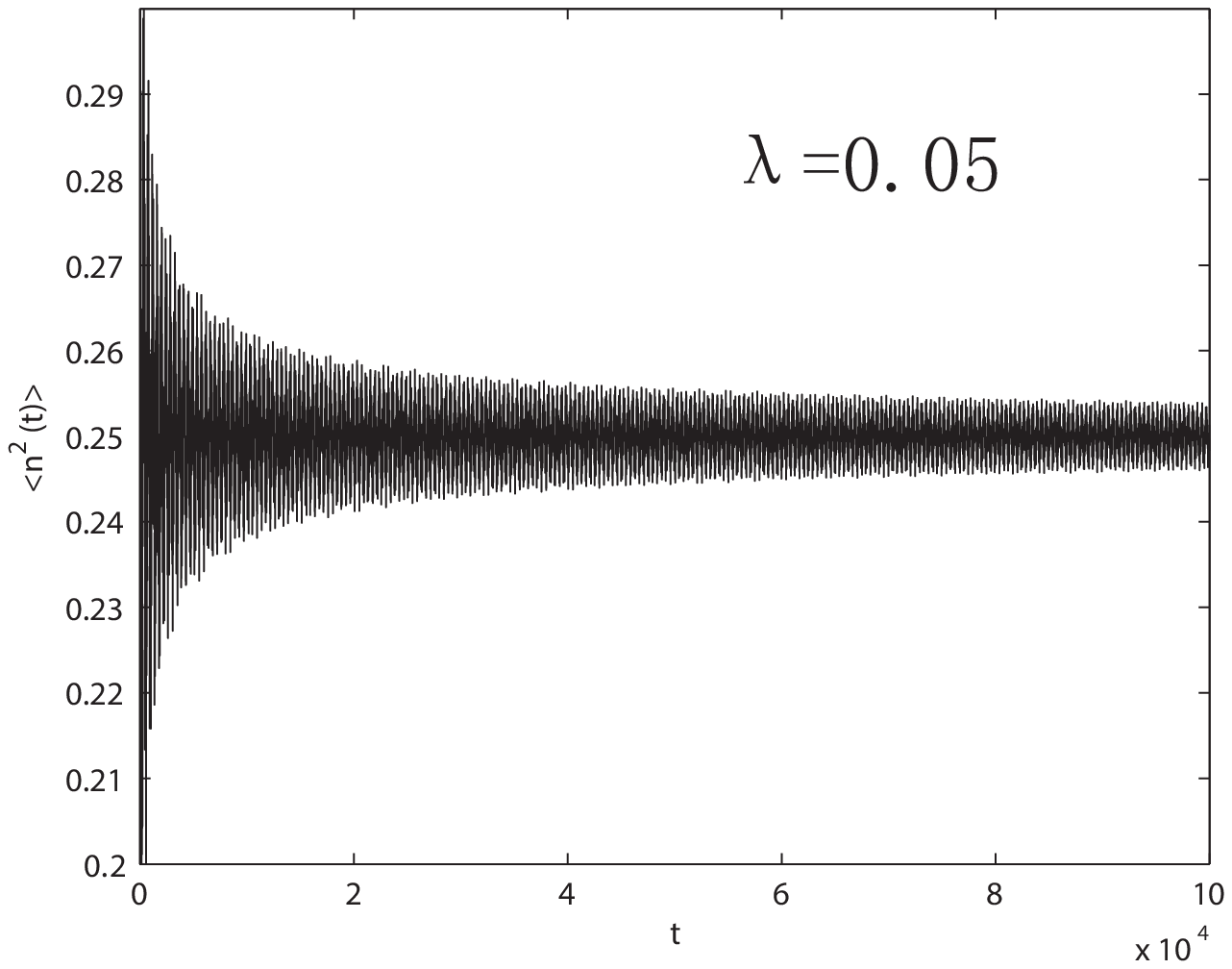}
\label{fig:side:b}
\end{minipage}
\begin{minipage}[t]{0.2\textwidth}
\centering
\includegraphics[width=1\linewidth]{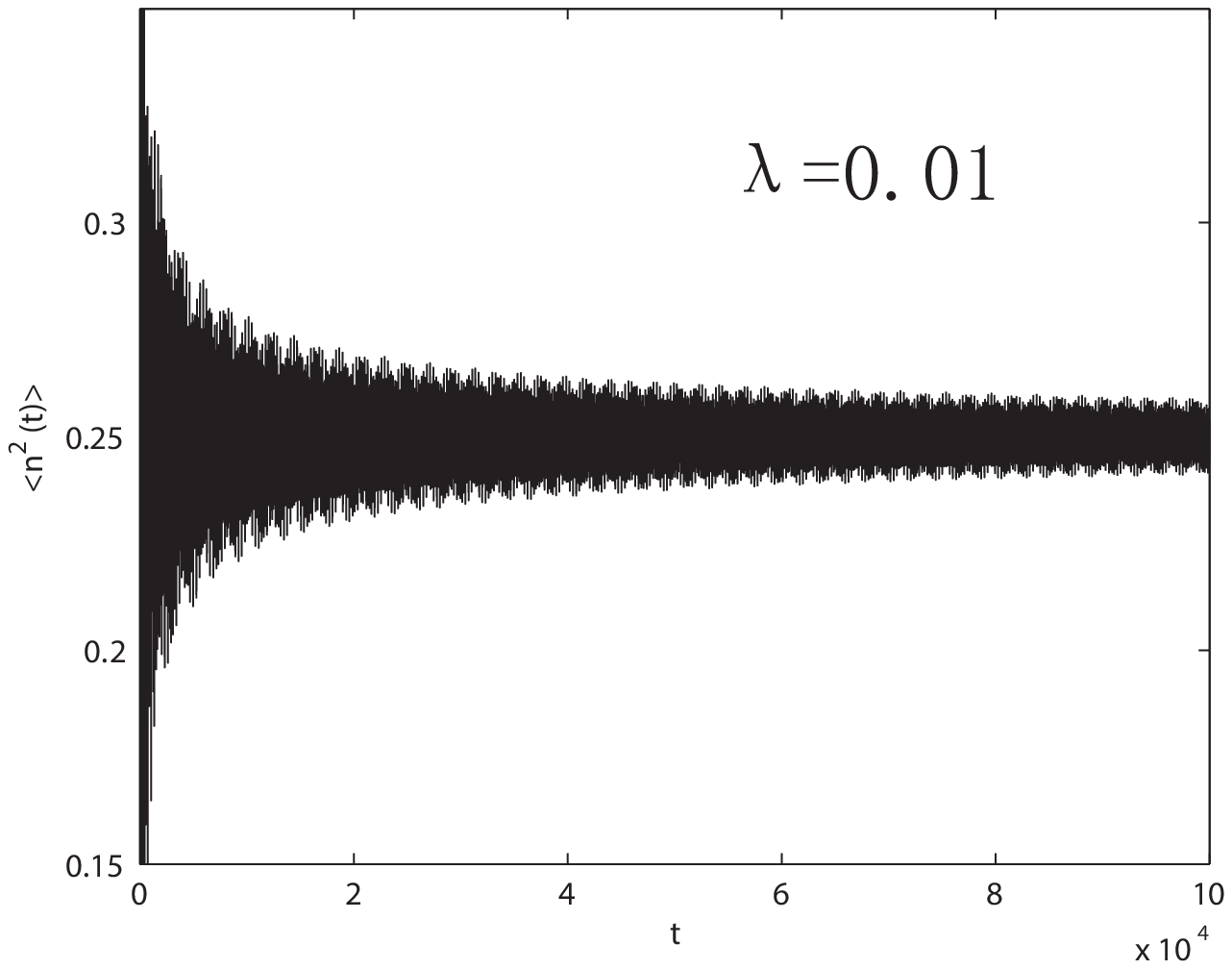}
\label{fig:side:a}
\end{minipage}
\begin{minipage}[t]{0.2\textwidth}
\centering
\includegraphics[width=1\linewidth]{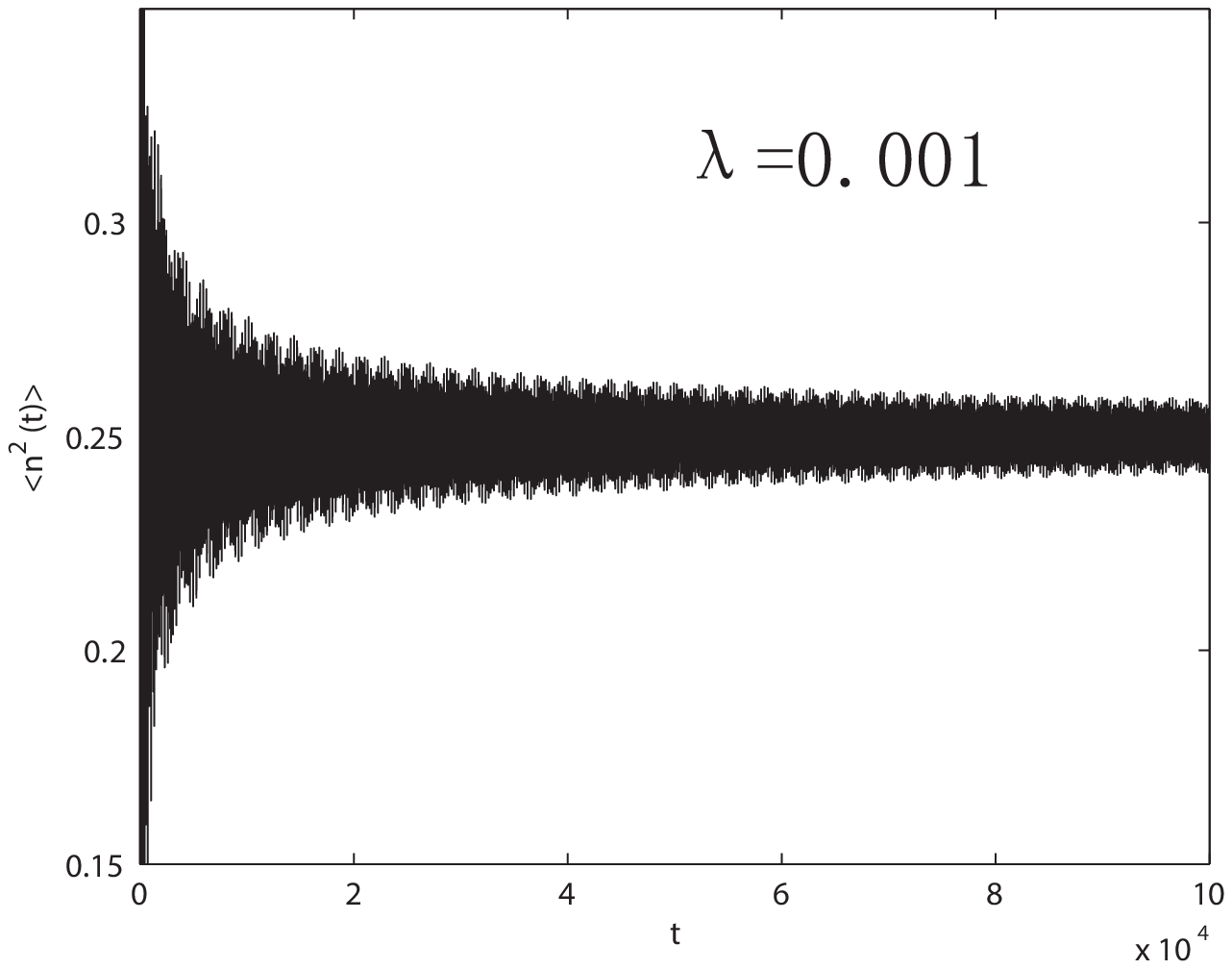}
\label{fig:side:b}
\end{minipage}
\end{figure}

\clearpage

Fig.10
\begin{figure}[!ht]
   \centering
   \begin{center}
     \includegraphics*[width=0.8\linewidth]{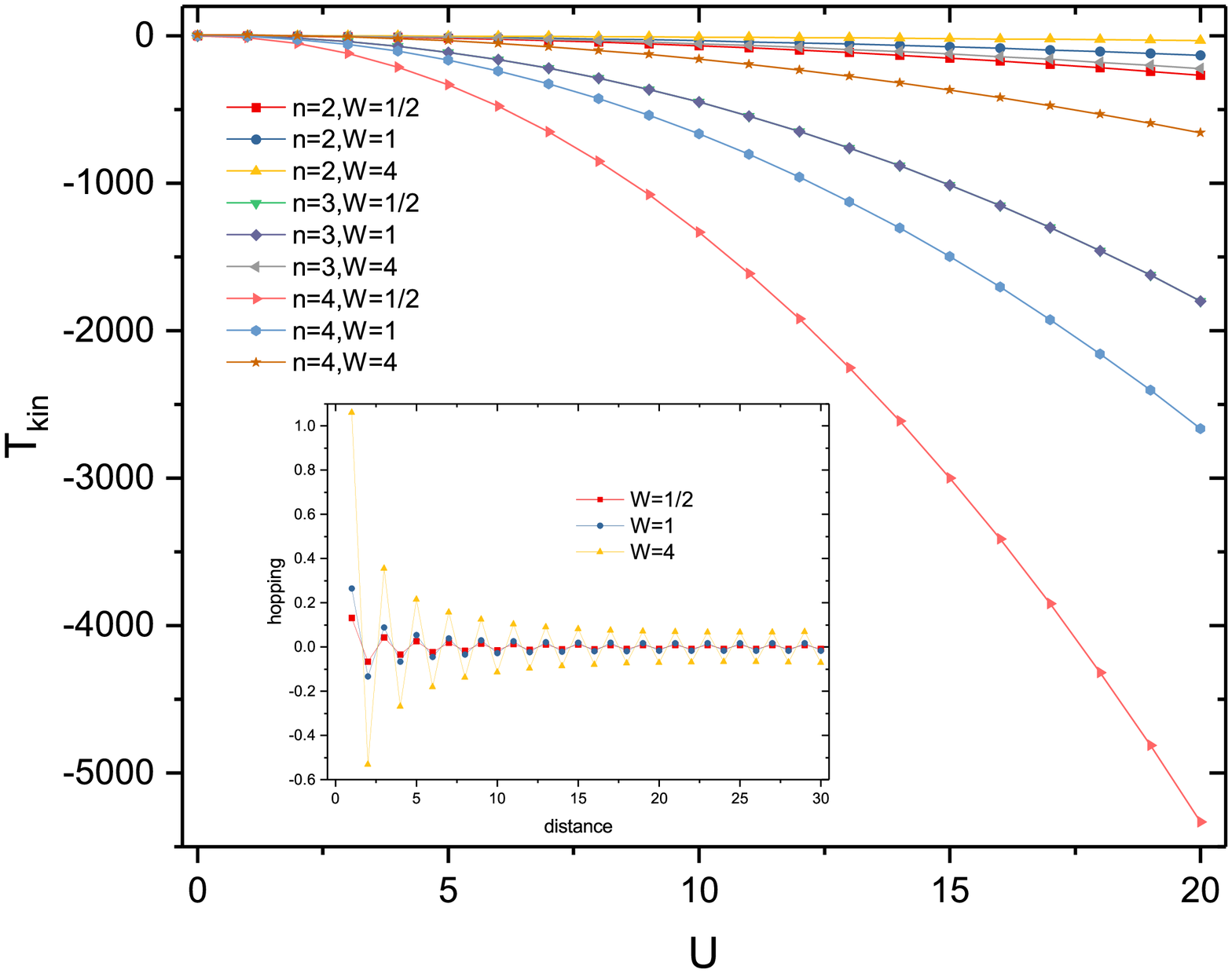}
   \end{center}
\end{figure}
\clearpage
Fig.11
\begin{figure}[!ht]
   \centering
   \begin{center}
     \includegraphics*[width=0.5\linewidth]{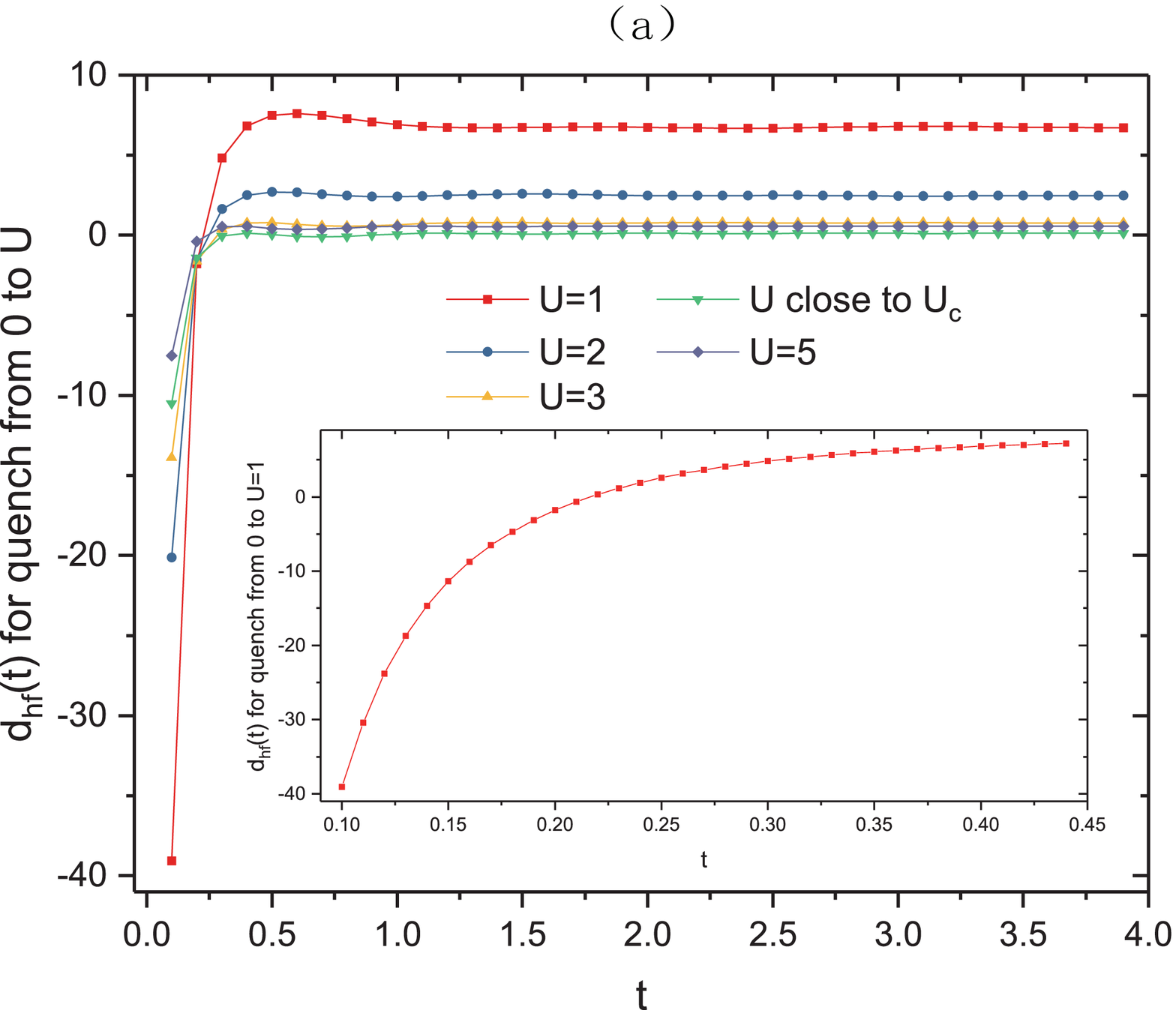}
   \end{center}
\end{figure}
\begin{figure}[!ht]
   \centering
   \begin{center}
     \includegraphics*[width=0.5\linewidth]{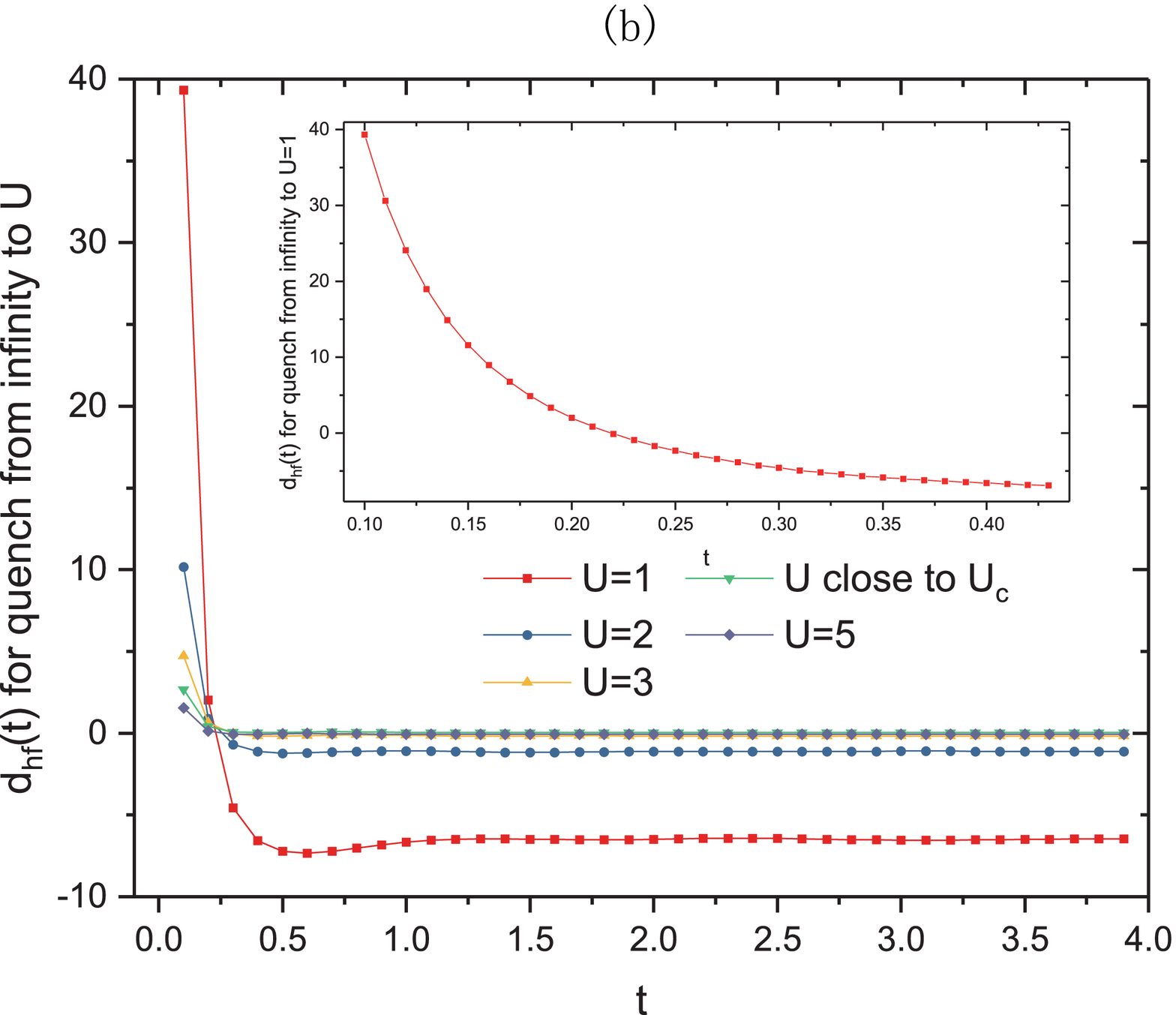}
   \end{center}
\end{figure}
\clearpage
Fig.12
\begin{figure}[!ht]
   \centering
   \begin{center}
     \includegraphics*[width=0.5\linewidth]{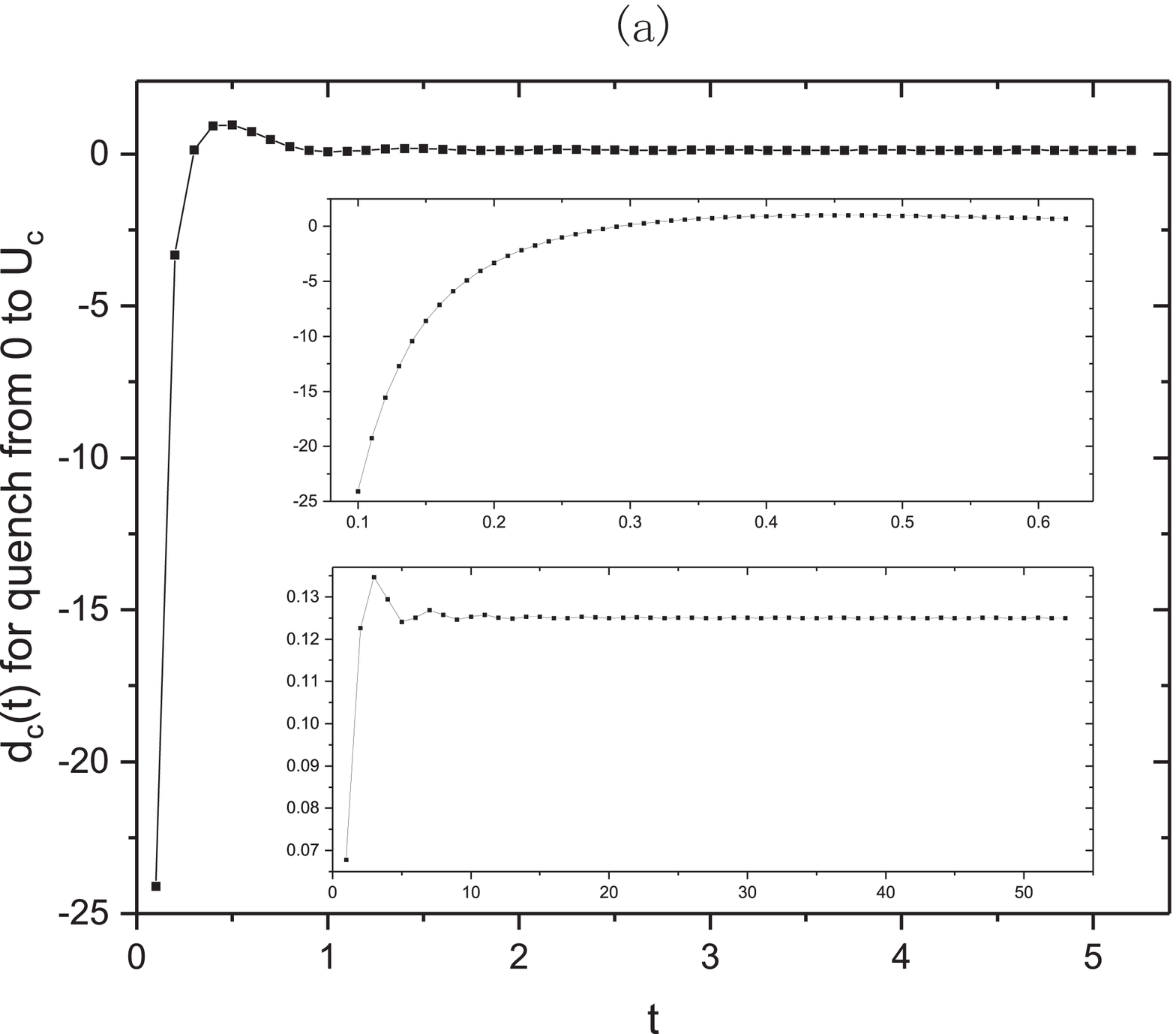}
   \end{center}
\end{figure}
\begin{figure}[!ht]
   \centering
   \begin{center}
     \includegraphics*[width=0.5\linewidth]{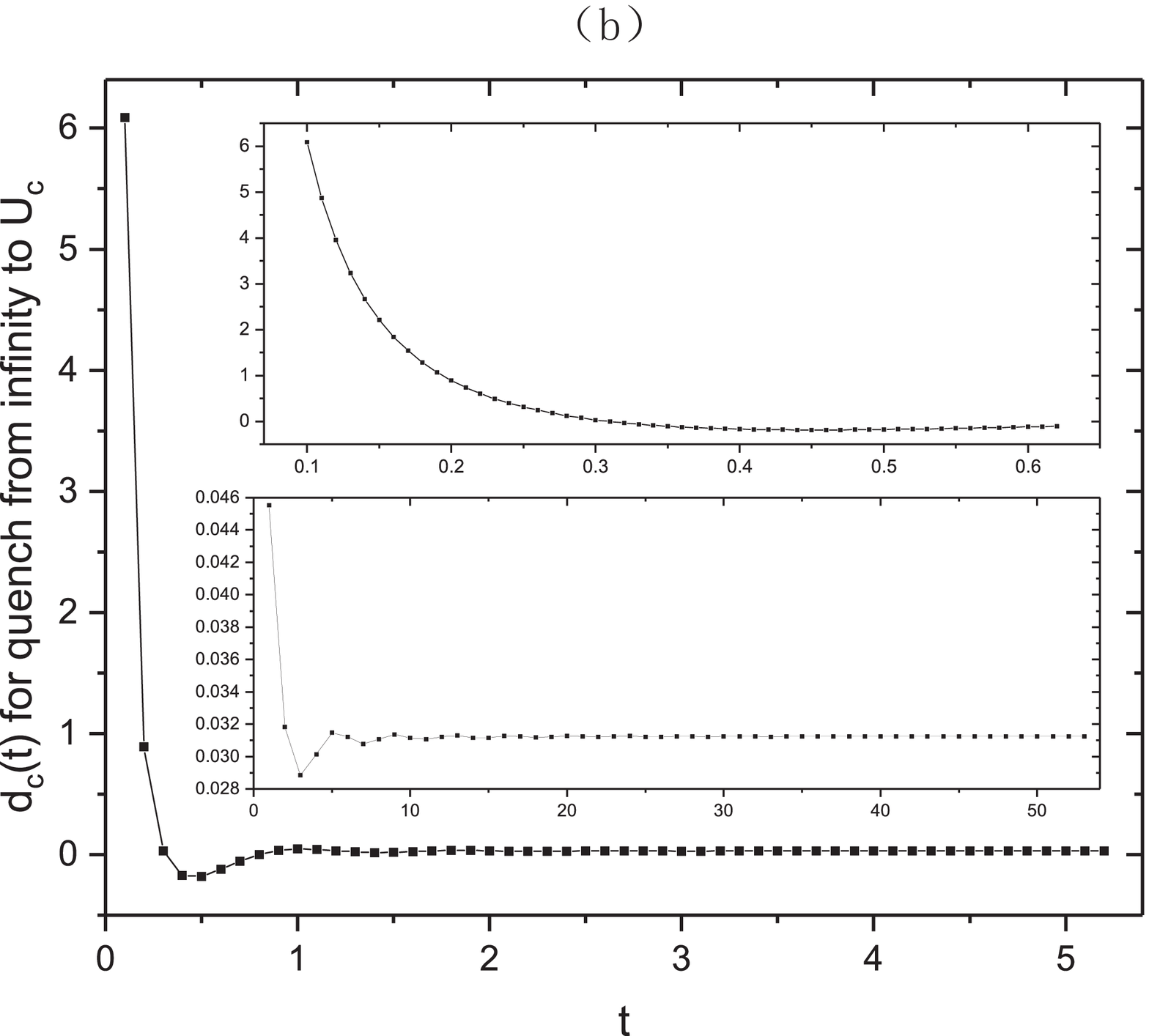}
   \end{center}
\end{figure}
\clearpage

Fig.13
\begin{figure}[!ht]
\begin{minipage}[t]{0.5\textwidth}
\centering
\includegraphics[width=1\linewidth]{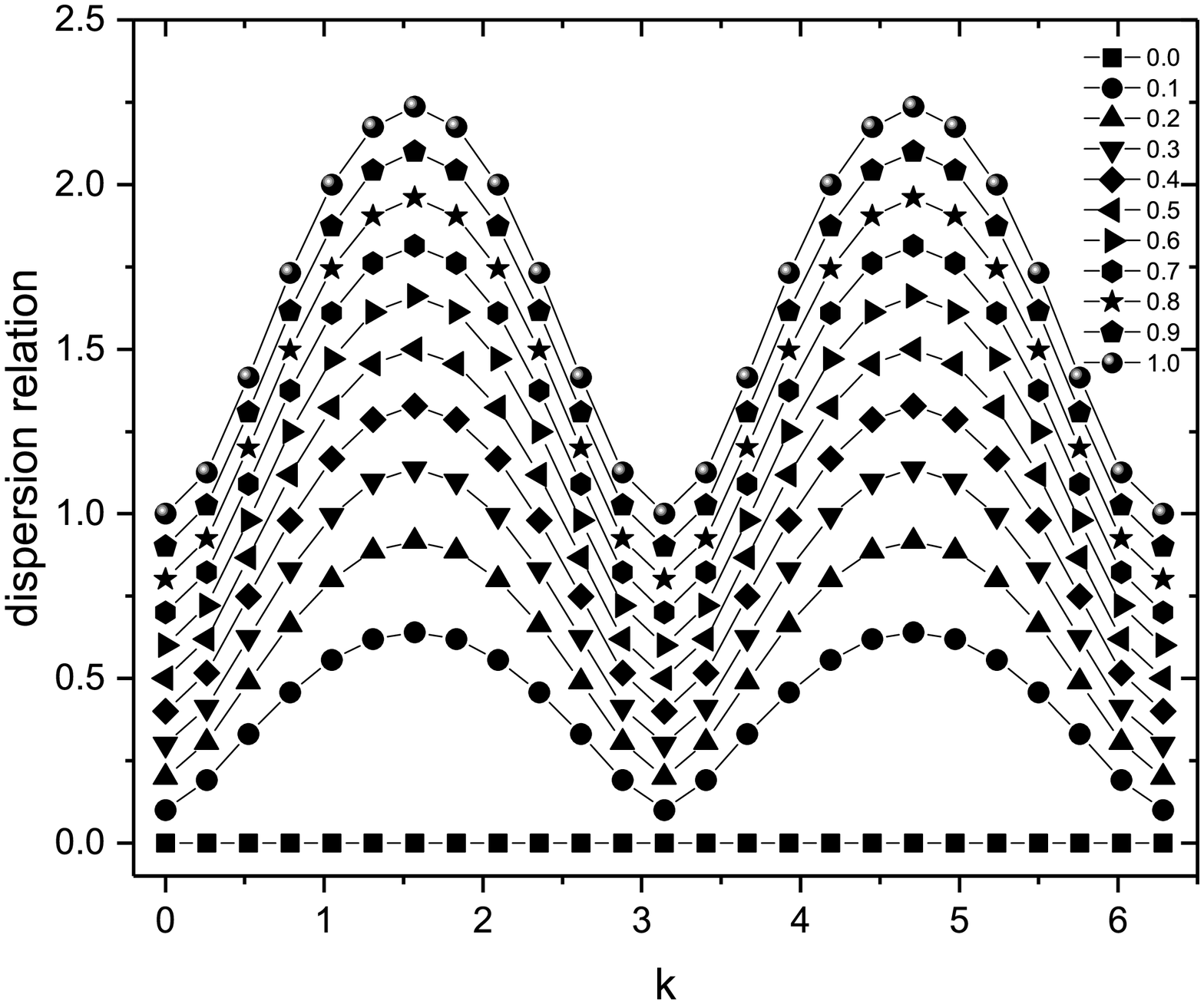}
\label{fig:side:a}
\end{minipage}%
\begin{minipage}[t]{0.5\textwidth}
\centering
\includegraphics[width=1\linewidth]{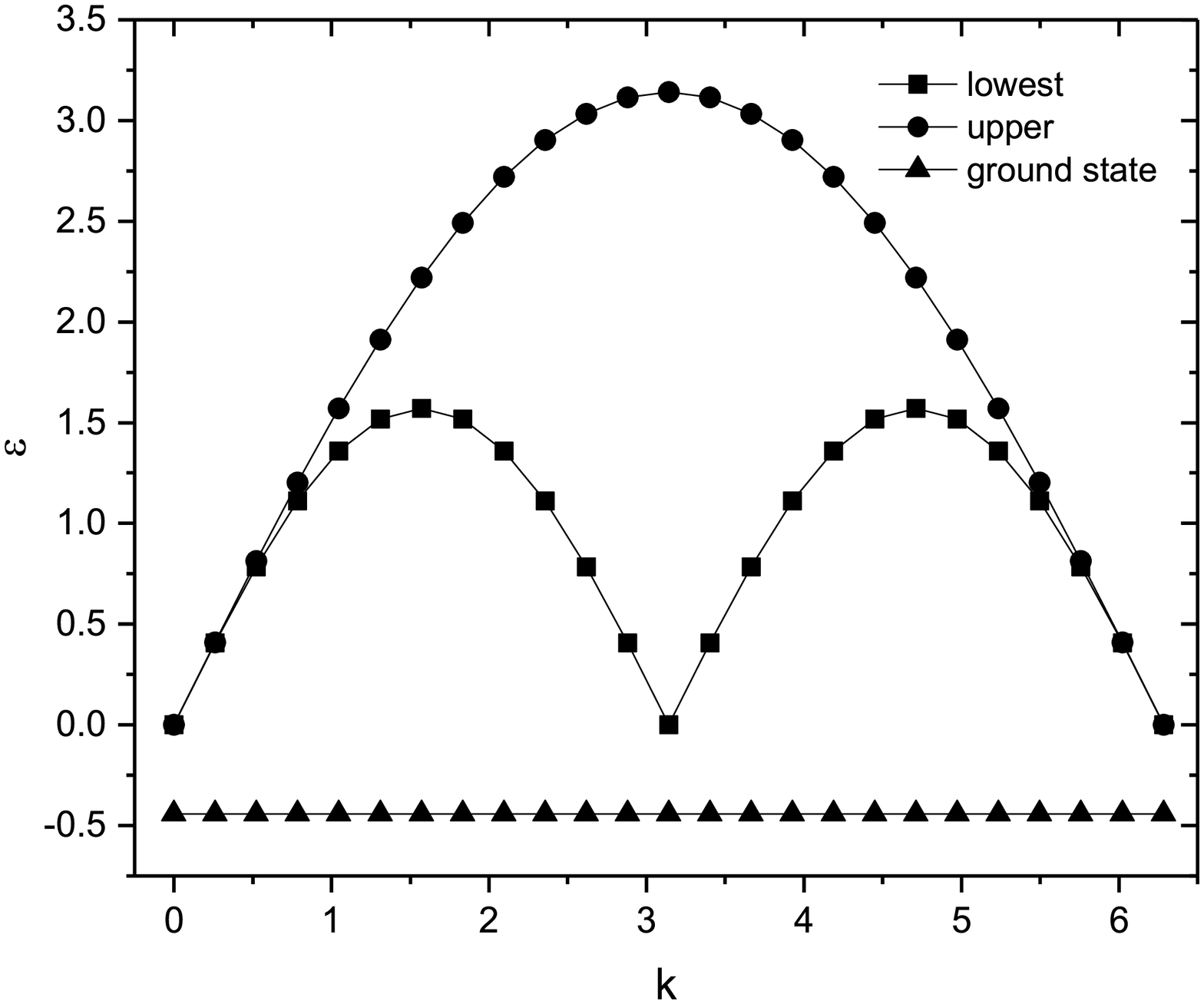}
\label{fig:side:b}
\end{minipage}
\end{figure}

Fig.14
\begin{figure}[!ht]
   \centering
   \begin{center}
     \includegraphics*[width=0.8\linewidth]{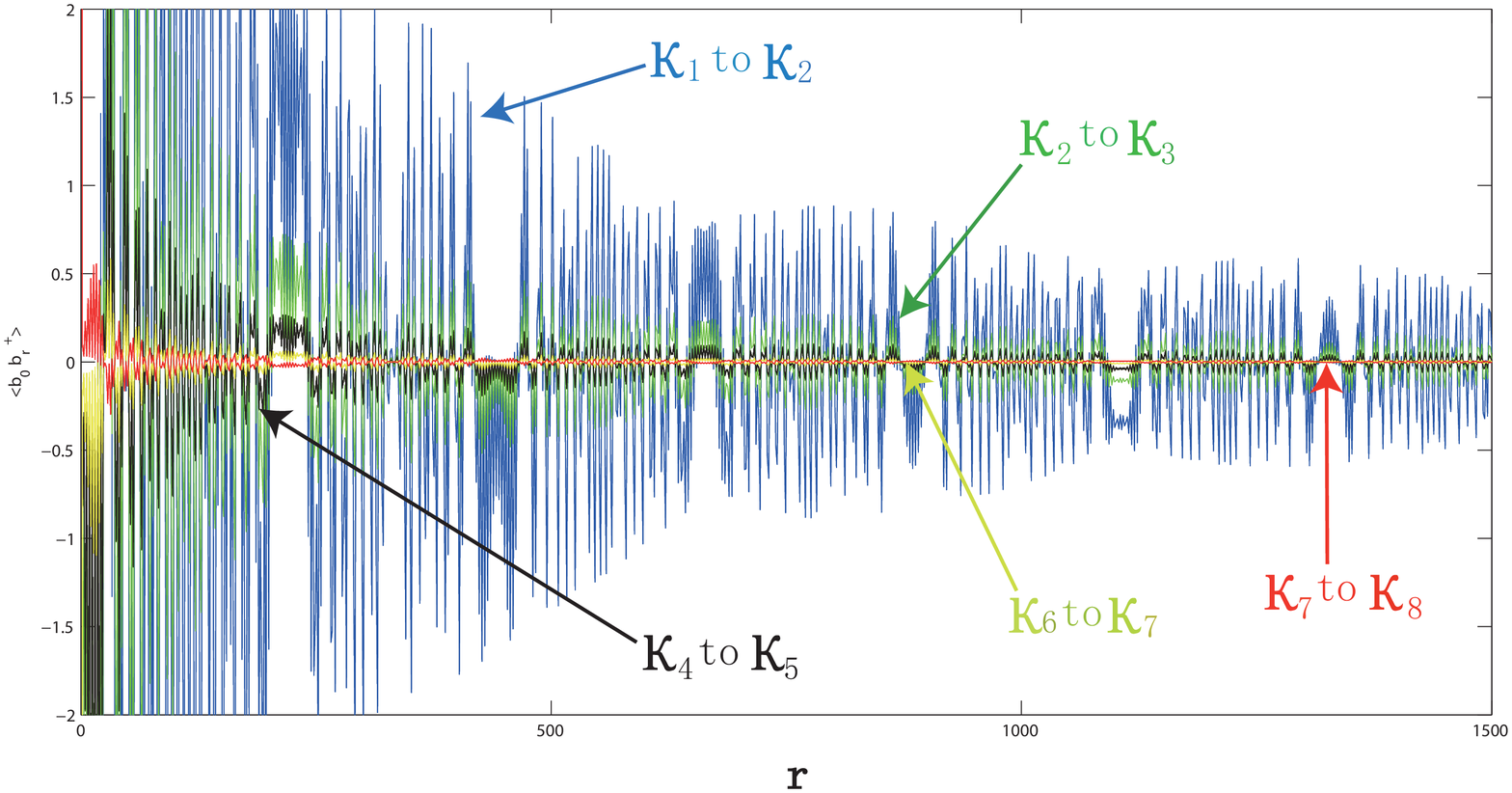}
   \end{center}
\end{figure}

\end{document}